\documentclass{article}
\usepackage{arxiv}
\usepackage[utf8]{inputenc}
\usepackage[T1]{fontenc}
\usepackage{amsmath,amssymb,amsfonts,amsthm,mathtools,bm,mathrsfs,fix-cm}
\usepackage{graphicx,rotating,float}
\usepackage[table]{xcolor}
\usepackage{booktabs,array,tabularx,threeparttable,adjustbox,longtable,ragged2e}
\usepackage{pifont}
\usepackage{enumitem,placeins,microtype,setspace}
\usepackage{caption}
\usepackage{etoc}
\usepackage[super,sort&compress]{natbib}
\usepackage{url}
\usepackage[hidelinks,bookmarksnumbered=true]{hyperref}

\allowdisplaybreaks[1]
\graphicspath{{Figures/}}
\DeclareGraphicsExtensions{.pdf,.png,.jpg}
\renewcommand{\shorttitle}{\small Doubly robust win statistics for hierarchical survival endpoints}
\renewcommand{\headeright}{}
\date{}

\theoremstyle{definition}
\newtheorem{theorem}{Theorem}

\newtheorem{proposition}{Proposition}

\newtheorem{assumption}{Assumption}
\newtheorem{condition}{Condition}

\newcommand{\E}{\mathbb E}
\newcommand{\Prb}{\mathbb P}
\newcommand{\ind}{\mathbf 1}
\newcommand{\Var}{\operatorname{Var}}
\newcommand{\Cov}{\operatorname{Cov}}
\newcommand{\IF}{\operatorname{IF}}

\newcommand{\op}{o_{\Prb}}
\newcommand{\Op}{O_{\Prb}}
\providecommand{\doi}[1]{\url{https://doi.org/#1}}

\title{Doubly robust win statistics for hierarchical survival endpoints in individual- and cluster-randomized trials}
\author{Xi Fang$^{1}$ \quad Fan Li$^{2,3,*}$ \\[4pt]
  \normalfont\small $^{1}$Data Science Institute, Medical College of Wisconsin, Milwaukee, WI, USA \\
  \normalfont\small $^{2}$Department of Biostatistics, Yale School of Public Health, New Haven, CT, USA \\
  \normalfont\small $^{3}$Center for Methods in Implementation and Prevention Science, \\
  \normalfont\small Yale School of Public Health, New Haven, CT, USA \\[3pt]
  \normalfont\small $^{*}$Corresponding author: Fan Li, \texttt{fan.f.li@yale.edu}}
\hypersetup{
  pdftitle={Doubly robust win statistics for hierarchical survival endpoints in individual- and cluster-randomized trials},
  pdfauthor={Xi Fang and Fan Li},
  pdfkeywords={composite survival endpoints, double robustness, estimands, informative cluster size, right censoring, U-statistics}
}
\begin{document}
\maketitle
\begin{abstract}
Prioritized composite endpoints compare trial participants according to a prespecified clinical hierarchy, and the win estimands are increasingly used to summarize treatment effects. When the component event times are right-censored, a pairwise comparison may remain unresolved, and treating unresolved pairs as ties yields a win statistic whose value depends on the censoring distribution. We define win, loss, and tie probabilities at a fixed restriction time for prioritized time-to-event outcomes in individual-randomized trials and extend them to cluster-randomized trials, where cluster-average and individual-average estimands assign equal weight to clusters and to participants, respectively. We propose doubly robust estimators that combine inverse probability of censoring weighting with joint outcome models to incorporate information recorded before censoring. Under covariate-dependent censoring, the resulting U-statistics are consistent when either the censoring models or joint outcome models are correctly specified for both treatment arms, but not necessarily both. We establish asymptotic normality and derive analytic variance estimators for computationally efficient inference. Simulation studies under both designs evaluate bias, precision, and confidence interval coverage under working model misspecification. Applications to two real-world studies demonstrate the use of the proposed methods under individual- and cluster-randomized trials.
\end{abstract}

\etocdepthtag.toc{main}

\section{Introduction}
\label{sec:introduction}

Composite endpoints are commonly used in cardiovascular trials, where death is combined with nonfatal events such as hospitalization to capture the overall burden of disease. Although the component endpoints typically differ in clinical importance, the standard analysis based on the time to the first event uses only the earliest event, so that a death following hospitalization does not contribute to the analysis. To address this limitation, Buyse \cite{buyse2010generalized} and Pocock et al.\ \cite{pocock2012win} proposed the generalized pairwise comparison approach, which compares participants from the two treatment groups on the most important outcome first and proceeds to the next outcome only when the pair cannot be distinguished. Typical summary measures including the win ratio, net benefit, and win odds, are different functions of the same win, loss, and tie probabilities and can complement one another.\cite{buyse2010generalized,brunner2021winodds,dong2023complement} We refer to the corresponding estimators of these summary measures collectively as win statistics. For right-censored time-to-event endpoints where win statistics are frequently used, however, a tie under the clinical hierarchy must be distinguished from a comparison that remains unresolved because of censoring. Comparing each pair over its common observed follow-up and treating unresolved pairs as ties yields quantities that depend on the censoring distribution, even under independent right censoring.\cite{oakes2016win,mao2024defining} This entangles the target clinical contrast with a nuisance component of the observation process, leaving the intended clinical estimand ambiguous, since the resulting quantity can change with the censoring mechanism even when the underlying outcome distributions remain unchanged. A natural remedy is to define the win and loss probabilities for a fixed target population at a prespecified restriction time, so that every pair is compared over the same window and censoring becomes a nuisance in estimation instead of part of the estimand.\cite{ich2019e9r1,mao2024defining} %

Estimation of censoring-free win and loss probabilities must account for covariate dependent censoring, which randomization alone does not remove. Dong et al.\ \cite{dong2020ipcw,dong2021dependent} developed inverse probability of censoring weighting (IPCW) for win statistics, leading to consistent estimators when the censoring model is correctly specified. The pairwise weight, however, is the inverse of a product of two censoring survival probabilities and can be large (hence leading to inefficient estimators) even when censoring is moderate in each arm. IPCW also discards the partial information in unresolved comparisons. Fang and Li\cite{fang2026ctw} proposed the conditional tie weighting, which recovers part of this information by replacing an unconfirmed higher priority tie with its conditional probability given the observed pairwise data, but its validity also rests on the censoring model. Alternatively, estimators based purely on outcome regression predict the unresolved portion of each comparison but rely on correct specification of the outcome model. Doubly robust estimation combines both models so that consistency holds when either is correctly specified, while allowing covariates to address censoring and improve precision.\cite{tsiatis2006semiparametric,fang2026estimands,fang2026doubly} Datta et al.\ \cite{datta2010ipcw} developed IPCW U-statistics for right censored data and proposed a doubly robust extension based on conditional expectations of the comparison kernel, allowing censoring to depend on observed covariate histories. With independent data, Zhang et al.\ \cite{zhang2019mannwhitney,zhang2020efficient} developed doubly robust estimators of probabilistic index type effects for a single right censored survival outcome. Related methods for the restricted mean time in favor of treatment integrate the treatment advantage over follow-up.\cite{fang2026doubly} The single outcome and restricted mean formulations concern estimands distinct from the marginal win and loss probabilities of a hierarchical composite endpoint at a prespecified restriction time. With a hierarchical composite endpoint, the comparison depends jointly on two multi-component event histories, so that the censoring correction must be applied to each member in turn. Furthermore, the outcome model should predict how an unresolved comparison would be resolved given the partial histories of both members, and these predictions must be mutually coherent; estimating marginal survival curves alone does not guarantee such coherence when the component event times are dependent. Finally. valid inference must also account for both the dependence induced by reusing participants across pairwise comparisons and the estimation of nuisance functions in the resulting two-sample U-statistic.

These challenges are further complicated in cluster-randomized trials (CRTs), in which groups of individuals, such as practices, hospitals, or communities, are randomized to intervention conditions.\cite{murray1998design} CRTs are now common in pragmatic trials embedded in health care systems, where the intervention is delivered at the level of the practice or provider and individual randomization is impractical. Observations within the same cluster tend to be correlated, and the classic literature on the design and analysis of CRTs has emphasized accounting for this correlation in inference. More recently, it has been recognized that the hierarchical data structure also affects the definition of the treatment effect, because different weighting schemes yield different estimands. Kahan et al.\ \cite{kahan2023estimands,kahan2024demystifying} explained that the cluster-average treatment effect, which gives equal weight to each cluster, and the individual-average treatment effect, which gives equal weight to each individual, address different scientific questions and can differ in magnitude in the presence of informative cluster size, that is, when the cluster size is associated with the outcome or with the cluster specific treatment effect.\cite{kahan2023informative} This recognition has led to an estimand driven literature that provides robust and efficient covariate adjusted estimators for continuous and binary outcomes.\cite{benitez2023defining,wang2024model}. For survival outcomes, Fang et al.\ \cite{fang2026estimands} first formalized cluster-average and individual-average survival probabilities and restricted mean survival times and proposed doubly robust estimators under covariate dependent censoring. For win estimands, Lee et al.\ \cite{lee2026winning} and Chen and Li \cite{chenli2026crt} distinguished cluster-pair and individual-pair targets and developed unadjusted and covariate-adjusted estimators for win estimands. Right-censored survival outcomes, however, have yet to be addressed for win estimands under cluster randomization, where it must be handled together with the clustering, and doing so involves more than combining the IRT methods with a cluster robust variance estimator. Specifically, because participants in the same cluster are correlated, all pairs that share a cluster are dependent, so that the large sample theory must be developed with the cluster as the independent unit. In addition, the individual-pair win estimand involves the average cluster size as a normalizing constant, which must be estimated and contributes additional variability to the estimator. Furthermore, censoring may depend on cluster size and the baseline characteristics of other cluster members, and censoring times may be dependent within clusters. These considerations motivate a unified analysis of win probabilities at a fixed restriction time that combines coherent event history models, covariate adjustment under known randomization, and inference accounting for nuisance estimation in both IRTs and CRTs.

To fill these gaps, we develop a unified doubly robust framework for win estimands with right-censored hierarchical outcomes under individual and cluster randomization. Our framework targets win, loss, and tie probabilities at a fixed restriction time that are determined by the clinical hierarchy, the target population, and the restriction time alone, so that the net benefit, win ratio, and win odds inherit a transparent interpretation that does not depend on the follow-up pattern of a particular trial. In CRTs, we further introduce cluster-pair and individual-pair versions of these estimands in parallel with the estimand distinction for average treatment effects. The proposed estimators are built from a sequential censoring transform that corrects the pairwise comparison for the censoring of each member in turn and yields a pair score with complete pair inverse weighting, two single member corrections, and a double member correction. This construction combines the strengths of the two existing approaches. It recovers the partial information carried by unresolved pairs, which inverse weighting discards, and it remains consistent for the target estimand when either the censoring models or the coherent joint outcome models are correctly specified, but not necessarily both. A notable feature is that the double member correction, which has no analogue in augmented estimators for functionals of a single participant, contributes nothing to the first order variance when the nuisance models are correct and yet is indispensable for double robustness. For inference, we treat the randomization unit as the independent unit and derive an analytic sandwich variance estimator that accounts for the estimation of the nuisance functions under either robustness route, which is valid under arbitrary within cluster dependence and avoids the computationally intensive cluster bootstrap for pairwise statistics. When all nuisance functions are correctly specified, the estimators are locally efficient in IRTs and, under the additional information and range conditions in Section \ref{sec:theory}, in CRTs. %
Table \ref{tab:literature-summary} summarizes existing methods for censored survival outcomes and their main features under individual- and cluster-randomized trials.

\begin{table}[!t]
\centering
\caption{Summary of methods for censored outcomes and pairwise treatment comparisons under individual- and cluster-randomized trial.}
\label{tab:literature-summary}
\begingroup
\fontsize{9.5}{11.5}\selectfont
\setlength{\tabcolsep}{3.5pt}
\renewcommand{\arraystretch}{1.24}
\newcommand{\litcheck}{\ding{51}}
\newcommand{\litdash}{\textendash}
\begin{threeparttable}
\begin{tabularx}{\textwidth}{@{}
  >{\raggedright\arraybackslash}p{3.2cm}
  *{3}{>{\centering\arraybackslash}p{0.55cm}}
  *{4}{>{\centering\arraybackslash}p{0.55cm}}
  >{\raggedright\arraybackslash}X@{}}
\toprule
\rowcolor{black!18}
\textbf{Reference / method} &
\multicolumn{3}{c}{\textbf{IRT}} &
\multicolumn{4}{c}{\textbf{CRT}} &
\textbf{Outcome and main feature} \\
\cmidrule(lr){2-4}\cmidrule(lr){5-8}
\rowcolor{black!18}
& \textbf{RC} & \textbf{CD} & \textbf{DR}
& \textbf{RC} & \textbf{CD} & \textbf{DR} & \textbf{C/I} & \\
\midrule
Dong et al.\ \cite{dong2020ipcw,dong2021dependent}
& \litcheck & \litcheck & \litdash
& \litdash & \litdash & \litdash & \litdash
& \emph{Hierarchical win statistics}. IPCW estimation with adjustment for covariate dependent censoring. \\
\rowcolor{black!5}
Fang and Li\ \cite{fang2026ctw}
& \litcheck & \litcheck & \litdash
& \litdash & \litdash & \litdash & \litdash
& \emph{Hierarchical win statistics}. Conditional tie weighting incorporates partially observed comparisons. \\
Datta et al.\ \cite{datta2010ipcw}\newline
Zhang et al.\ \cite{zhang2019mannwhitney}
& \litcheck & \litcheck & \litcheck
& \litdash & \litdash & \litdash & \litdash
& \emph{Survival functionals and Mann--Whitney effects}. DR estimation using censored U-statistics. \\
\rowcolor{black!5}
Huang et al.\ \cite{huang2026future}\tnote{a}
& \litcheck & \litcheck & \litcheck
& \litdash & \litdash & \litdash & \litdash
& \emph{Win-ratio regression effects}. DR estimation from comparisons accumulated over follow-up.  \\
Parner and Overgaard\ \cite{parner2023recurrent,parner2025win}
& \litcheck & \litdash & \litdash
& \litdash & \litdash & \litdash & \litdash
& \emph{Win and loss probabilities and win ratio}. Prioritized event times or recurrent-event counts. \\
\rowcolor{black!5}
Westling et al.\ \cite{westling2024survival}
& \litcheck & \litcheck & \litcheck
& \litdash & \litdash & \litdash & \litdash
& \emph{Survival probabilities and RMST}. Cross-fitted DR estimation for a single survival outcome. \\
Fang et al.\ \cite{fang2026estimands}
& \litdash & \litdash & \litdash
& \litcheck & \litcheck & \litcheck & \litcheck
& \emph{Survival probabilities and RMST}. Distinct cluster-average and individual-average effects. \\
\rowcolor{black!5}
Fang et al.\ \cite{fang2026doubly}
& \litcheck & \litcheck & \litcheck
& \litcheck & \litcheck & \litcheck & \litcheck
& \emph{Restricted mean time in favor}. Progressive multistate outcomes, with jackknife inference. \\
Fang, Zhao and Li\ \cite{fang2026whilealive}\tnote{a}
& \litcheck & \litcheck & \litcheck
& \litcheck & \litcheck & \litcheck & \litcheck
& \emph{While-alive event rates}. Clinically weighted recurrent-event burden per unit of time alive, with closed form variance estimators \\
\rowcolor{black!5}
Lee et al.\ \cite{lee2026winning}\newline
Chen and Li\ \cite{chenli2026crt}
& \litdash & \litdash & \litdash
& \litdash & \litdash & \litdash & \litcheck
& \emph{Cluster-pair and individual-pair win effects}. Covariate adjustment developed by Chen and Li.\cite{chenli2026crt} \\
\midrule
\rowcolor{black!12}
\textbf{This article}
& \litcheck & \litcheck & \litcheck
& \litcheck & \litcheck & \litcheck & \litcheck
& \textbf{\emph{Fixed-horizon hierarchical win, loss, and tie probabilities}. DR estimation for both CRT targets, with closed form variance estimators} \\
\bottomrule
\end{tabularx}
\begin{tablenotes}[flushleft]
\fontsize{8.2}{10}\selectfont
\item[]  IRT, individually randomized trial; CRT, cluster-randomized trial; RC, right-censored outcomes; CD, adjustment for covariate dependent censoring, including observed histories where allowed; DR, consistency under correct specification of the censoring models or the outcome models when treatment probabilities are known, subject to the method's assumptions; C/I, distinct cluster-average and individual-average estimands. IPCW, inverse probability of censoring weighting; RMST, restricted mean survival time.
\end{tablenotes}
\end{threeparttable}
\endgroup
\end{table}

The remainder of this article is organized as follows. In Section \ref{sec:irt}, we introduce the win estimands at a fixed restriction time in IRTs, present the identification results, and describe the doubly robust estimator. In Section \ref{sec:crt}, we extend the framework to CRTs and formalize the cluster-pair and individual-pair win estimands. In Section \ref{sec:theory}, we establish the asymptotic properties and local efficiency of the proposed estimators for both designs and present the proposed variance estimators. In Section \ref{sec:simulation}, we carry out simulation studies to examine the finite-sample performance of the proposed methods under various data-generating processes, considering different degrees of censoring, informative cluster size, within cluster dependence, and model misspecification. Section \ref{sec:application} provides illustrative analyses of the HF-ACTION individual-randomized trial and the STRIDE cluster-randomized trial, and Section \ref{sec:discussion} concludes with a discussion and potential directions for future research.

\section{Doubly robust win statistics in individual-randomized trials}
\label{sec:irt}

\subsection{Notation and estimands}
\label{subsec:irt-estimands}

We consider an IRT with $n$ participants indexed by $i=1,\ldots,n$ and two treatment conditions. Let $A_i\in\{0,1\}$ denote the treatment assignment of participant $i$, with $A_i=1$ indicating the treatment and $A_i=0$ indicating the control, and let $\bm L_i$ denote the vector of baseline covariates. The prioritized composite endpoint consists of a terminal event and $Q$ nonfatal events ordered by clinical importance. For $a\in\{0,1\}$, let $D_i(a)$, \(T_i^{q}(a)\), and \(C_i(a)\) denote the potential death time, nonfatal event for priority $q=1,\ldots,Q$, and censoring time under assignment $a$, respectively. The potential event history $\mathcal H_i(t;a)$ records the occurrence indicators and times of the component events through time $t$, respectively. The nonfatal events are semicompeting with death, which means death prevents a nonfatal event that has not yet occurred, whereas a nonfatal event does not prevent subsequent death. A nonfatal event time $T_i^q(a)$ is therefore defined only while the participant is alive. If death occurs first, we assign $T_i^q(a)=\infty$. The following assumption relates these potential quantities to the observed outcomes.

\begin{assumption}[Stable Unit Treatment Value Assumption (SUTVA)]
\label{ass:irt-sutva}
For each participant $i$, the potential quantities $\{D_i(a),T_i^q(a),C_i(a),\mathcal H_i(t;a):a=0,1,\ q=1,\ldots,Q\}$ do not depend on the assignments of the other participants, and the realized quantities are $D_i=D_i(A_i)$, $T_i^q=T_i^q(A_i)$, $C_i=C_i(A_i)$, and $\mathcal H_i(t)=\mathcal H_i(t;A_i)$.
\end{assumption}

Assumption \ref{ass:irt-sutva} rules out interference between participants and links each realized quantity to its potential value under the assigned arm. It places no restriction on the dependence among the terminal and nonfatal event times of the same participant.

We compare outcomes over a prespecified restriction time $\tau$. Define the restricted potential outcome vector \(\bm Y_i(a)=\{D_i(a)\wedge\tau,T_i^1(a)\wedge\tau,\ldots,T_i^Q(a)\wedge\tau\}^{\mathsf T}\), with $Y_i^0(a)=D_i(a)\wedge\tau$ and $Y_i^q(a)=T_i^q(a)\wedge\tau$ for $q\geq1$. Larger values are preferred on every tier. A nonfatal event precluded by death receives value $\tau$, but its assigned value can affect the comparison only if the higher priority death comparison does not distinguish the two participants. Because death precludes all subsequent nonfatal events and survival through $\tau$ determines every restricted component, $\bm Y_i(a)$ is fully determined at $R_i(a)=D_i(a)\wedge\tau$, which is called the resolution time. Under Assumption \ref{ass:irt-sutva}, the realized restricted vector and resolution time are $\bm Y_i=\bm Y_i(A_i)$ and $R_i=R_i(A_i)=D_i\wedge\tau$. The event history is observed only while participant $i$ remains under follow-up. Let \(U_i=R_i\wedge C_i\) denote the observed follow-up time and \(\Delta_i^R=\mathbf 1(R_i\leq C_i)\) the resolution indicator. The full vector $\bm Y_i$ is completely observed if and only if $\Delta_i^R=1$. Administrative end of follow-up at $\tau$ therefore resolves the restricted outcome and is not treated as censoring. The observed data are $O_i=\{\bm L_i,A_i,U_i,\Delta_i^R,\mathcal H_i(U_i)\}$. When $\Delta_i^R=0$, only the history before censoring is available. The vector $\bm Y_i$ is incomplete, although some pairwise comparisons may already be determined by the observed events. To describe the censoring, define the at-risk processes \(Y_i^C(t)=\mathbf 1(U_i\geq t)\) and the censoring counting process \(N_i^C(t)=\mathbf 1(U_i\leq t,\Delta_i^R=0)\). Both stop at resolution or censoring time. We write $\mathcal F_{i,t}^O$ as the filtration generated by $\bm L_i$, $A_i$, and the observed event and censoring history of participant $i$ through $t$.

The hierarchical rule compares two participants in order of clinical priority, beginning with the terminal event. A tier determines the comparison only when the outcome differs by more than a prespecified clinically meaningful margin. Otherwise, the comparison proceeds to the next tier. Let $\delta_q\geq0$ be a prespecified indifference threshold for tier $q=0,\ldots,Q$, where tier $0$ corresponds to the terminal event. A difference of at most $\delta_q$ on tier $q$ does not distinguish the two participants on that endpoint. For restricted vectors $\bm y=(y^0,\ldots,y^Q)^{\mathsf T}$ and $\bm y'=(y'^0,\ldots,y'^Q)^{\mathsf T}$, define \(e_q(\bm y,\bm y')=\mathbf 1\{|y^q-y'^q|\leq\delta_q\}\) and \(w_q(\bm y,\bm y')=\mathbf 1(y^q-y'^q>\delta_q)\). Thus, $e_q$ indicates that the difference is within the specified threshold, and $w_q$ indicates a clinically meaningful advantage for the first participant. The indicators \(J_0(\bm y,\bm y')=1\) and \(J_q(\bm y,\bm y')=\prod_{\ell=0}^{q-1}e_\ell(\bm y,\bm y')\) for \(q\geq1\) identify pairs whose differences on all higher priority tiers are within their thresholds. For those pairs, the comparison reaches tier \(q\). The indicator \(h_{+,q}(\bm y,\bm y')=J_q(\bm y,\bm y')w_q(\bm y,\bm y')\) then equals one when tier $q$ is the first to distinguish the participants and favors the first participant. Define \(h_+(\bm y,\bm y')=\sum_{q=0}^{Q}h_{+,q}(\bm y,\bm y')\), \(h_-(\bm y,\bm y')=h_+(\bm y',\bm y)\), and \(h_{=}(\bm y,\bm y')=\prod_{q=0}^{Q}e_q(\bm y,\bm y')\) to indicate a win, a loss, and a tie for the first participant, respectively. Exactly one of $h_+$, $h_-$, and $h_{=}$ equals one for each pair of restricted vectors. A tie, $h_{=}=1$, means that none of the outcome differences exceeds its clinical threshold. These rules are defined by the complete outcomes and the prespecified comparison rule. 

For two independent draws $i$ and $k$ from the participant population, we define the estimands, i.e., the win, loss, and tie probabilities
\begin{equation}
\lambda^r=\mathbb E[h_r\{\bm Y_i(1),\bm Y_k(0)\}],\qquad r\in\{+,-,=\},
\label{eq:irt-target}
\end{equation}
so that $\lambda^+$, $\lambda^-$, and $\lambda^{=}=1-\lambda^+-\lambda^-$ are the probabilities that a treated participant wins, loses, or ties when two participants independently selected from the target population are evaluated under treatment and control, respectively. The estimands in \eqref{eq:irt-target} are generalized causal effects because they compare independent draws from the two counterfactual outcome distributions.\cite{chenli2026principal} The win statistics of interest include
\begin{equation}
\mathrm{NB}=\lambda^+-\lambda^-,\qquad
\mathrm{WR}=\frac{\lambda^+}{\lambda^-},\qquad
\mathrm{WO}=\frac{\lambda^++\lambda^{=}/2}{\lambda^-+\lambda^{=}/2}
=\frac{1+\mathrm{NB}}{1-\mathrm{NB}},
\label{eq:irt-summaries}
\end{equation}
where we require $\lambda^->0$ for the win ratio and $|\mathrm{NB}|<1$ for the win odds to be strictly positive and finite. Because all three summaries are functions of the same probabilities, we develop estimators for $\lambda^r$ and obtain the summaries by transformation.

The estimands in \eqref{eq:irt-target} depend on the complete restricted outcomes, which may be incompletely observed because of censoring. The following assumptions allow these estimands to be identified from the observed data $O_i$.
\begin{assumption}[Random sampling]
\label{ass:irt-independence}
The vectors $\{\bm L_i,D_i(a),T_i^q(a),C_i(a):a=0,1,\ q=1,\ldots,Q\}$, $i=1,\ldots,n$, are independent and identically distributed.
\end{assumption}

\begin{assumption}[Individual randomization]
\label{ass:irt-randomization}
$A_i\ \perp\!\!\!\perp\ \{D_i(a),T_i^q(a),C_i(a):a=0,1,\ q=1,\ldots,Q\}\mid\bm L_i$, and the assignment probabilities $\pi_s(\bm L_i)=\mathbb P(A_i=s\mid\bm L_i)$, $s=0,1$, are known and satisfy $\epsilon_\pi\leq\pi_s(\bm L_i)\leq1-\epsilon_\pi$ almost surely for some $\epsilon_\pi>0$. Conditional on all baseline and potential event and censoring records, the assignments are independent across participants with probabilities $\pi_s(\bm L_i)$.
\end{assumption}

Assumption \ref{ass:irt-independence} specifies the population from which the independent draws in \eqref{eq:irt-target} are taken. Independence is imposed across participants only, and not across the outcome components or across the two potential outcomes of the same participant. Under Assumption \ref{ass:irt-randomization}, assignments are conditionally independent across participants. The probability that participant $i$ receives treatment and participant $k$ receives control is therefore $\pi_1(\bm L_i)\pi_0(\bm L_k)$. Randomization with fixed treatment totals changes this pair probability and is considered in Section \ref{subsec:irt-theory}. Randomization addresses treatment assignment but does not by itself remove selection due to censoring. The next assumption specifies how censoring may depend on the observed information.

\begin{assumption}[Covariate dependent censoring]
\label{ass:irt-car}
For each treatment condition $a\in\{0,1\}$, 
\begin{equation}
C_i(a)\ \perp\!\!\!\perp\ \{D_i(a),T_i^1(a),\ldots,T_i^Q(a)\}\mid\bm L_i.
\label{eq:irt-censoring}
\end{equation}
Let $K_{0,a}(t\mid\bm l) =\mathbb P\{C_i(a)\geq t\mid\bm L_i=\bm l\}$ denote the conditional censoring survival function. There exists a constant $\epsilon_C>0$ such that $K_{0,a}(\tau\mid\bm L_i)\geq\epsilon_C$ almost surely for both treatment conditions.
\end{assumption}

Assumption \ref{ass:irt-car} allows censoring to be associated with the event outcomes through baseline covariates, but requires this association to be explained by the measured covariates
$\bm L_i$ within each treatment condition.\cite{zhang2019mannwhitney,fang2026doubly} The positivity condition ensures that participants with the baseline characteristics represented in the target population have a positive probability of remaining uncensored through $\tau$ in either treatment arm. Under consistency and randomization, the censoring survival function for participant $i$ is
$K_{0,i}(t)=K_{0,A_i}(t\mid\bm L_i) =\mathbb P(C_i\geq t\mid A_i,\bm L_i)$. For the estimation, we assume continuous cumulative censoring hazards and write $\Lambda_{0,i}^C(t)=-\log K_{0,i}(t)$. Thus, the win, loss, and tie probabilities are identified from the observed data. Let
\[
\mu_0^r(\bm L_i,\bm L_k)=\mathbb E\!\left[h_r\{\bm Y_i(1),\bm Y_k(0)\}\mid\bm L_i,\bm L_k\right]
\]
denote the conditional probability of comparison result $r$ given the baseline covariates of the two participants. This probability is obtained by averaging the comparison indicator over the joint outcome distribution in each treatment arm. In the weighting representation, the complete-pair contribution is defined as zero unless both members are resolved, so that it can be evaluated entirely from the observed data.

\begin{theorem}[Identification in IRTs]
\label{thm:irt-identification}
Under Assumptions \ref{ass:irt-sutva} to \ref{ass:irt-car}, for $r\in\{+,-,=\}$,
\begin{align}
\lambda^r=\mathbb E\left[\frac{\mathbf 1(A_i=1,A_k=0)}{\pi_1(\bm L_i)\pi_0(\bm L_k)}\frac{\Delta_i^R\Delta_k^R h_r(\bm Y_i,\bm Y_k)}
{K_{0,i}(R_i)K_{0,k}(R_k)}\right]=\mathbb E\{\mu_0^r(\bm L_i,\bm L_k)\}.
\label{eq:irt-identification}
\end{align}
The same representations apply to the tier specific win probabilities $\lambda^{+,q}=\mathbb E[h_{+,q}\{\bm Y_i(1),\bm Y_k(0)\}]$ after replacing $h_r$ by $h_{+,q}$.
\end{theorem}

Theorem \ref{thm:irt-identification} provides inverse probability of censoring weighting (IPCW) and outcome regression representations of the same estimand. The first uses comparisons for which both participants'
restricted outcome vectors are observed, with inverse probability weights accounting for treatment assignment and censoring. The second uses the joint outcome distributions in the two treatment arms to calculate the probability of a win, loss, or tie conditional on the two participants' baseline covariates. These probabilities are then averaged over the baseline covariate distribution in the target population. The proof is provided in Appendix \ref{subsec:app-irt-identification}. 

Because the treatment assignment probabilities are known by design, the first representation requires estimation of the censoring survival functions. Substituting estimates from fitted censoring models yields the IPCW estimator. The second representation requires estimation of the outcome regression, which can be calculated from fitted joint outcome models in the two treatment arms, yielding the outcome regression (OR) estimator. The IPCW estimator is consistent if the censoring models for both arms are correctly specified, whereas the OR estimator is consistent if the joint outcome models for both arms are correctly specified. Each approach therefore relies on correct specification of its respective working models.

\subsection{Doubly robust estimation}
\label{subsec:irt-estimation}

We proposed to combine the two representations in Theorem \ref{thm:identification} to obtain an estimator that is consistent when either set of working models is correctly specified. Here, a pairwise comparison involves two participants, either of whom may be censored. We first define an augmented IPCW contribution for one participant and then apply the same correction to both members of a pair. This produces the two single member corrections and the additional term involving both members that are needed for double robustness. Let $\eta$ collect the nuisance functions from the two censoring models, the arm-specific joint outcome models, and a baseline pair regression $\mu^r(\bm L_i,\bm L_k;\eta)$. The baseline pair regression may be obtained from the outcome models or fitted separately. Write $\widehat\eta$ for the fitted nuisance functions, $\eta^*$ for their probability limit, which may differ from the truth under misspecification, and $\eta_0$ for the true nuisance functions. For a single participant $i$, let $B_i$ be an integrable function of the restricted outcome vector $\bm Y_i$ and baseline covariates $\bm L_i$. Since $\bm Y_i$ is determined at $R_i$, the contribution $B_i$ is available whenever $\Delta_i^R=1$, although a particular contribution may be known earlier. In the pairwise construction, $B_i$ is the comparison indicator $h_r(\bm Y_i,\bm y)$ with the other member's outcome fixed at $\bm y$, or its conditional expectation given the other member's baseline covariates. We construct a contribution from the observed data whose expectation is $\mathbb E(B_i)$ under the censoring assumptions in Assumption \ref{ass:irt-car} and either of the model correctness conditions stated below.

The true cumulative censoring hazard and its associated survival functions are $\Lambda_{0,i}^C$ and $K_{0,i}$. Let $\Lambda_i^C$ denote a working cumulative censoring hazard, with associated survival function \(K_i(t)=\exp\{-\Lambda_i^C(t)\}\). Define the censoring residual process \(M_i^C(t;\Lambda_i^C)=N_i^C(t)-\int_0^tY_i^C(u)\,d\Lambda_i^C(u)\), with \(M_{0,i}^C(t)=M_i^C(t;\Lambda_{0,i}^C)\). If $\Lambda_i^C$ equals $\Lambda_{0,i}^C$ while the participant is at risk of censoring, $M_i^C(t;\Lambda_i^C)$ is a mean-zero martingale with respect to $\mathcal F_{i,t}^O$. Otherwise, its increments need not have mean zero. Let \(q_{0,i}(t)=\mathbb E(B_i\mid\mathcal F_{i,t^-}^{O},R_i\geq t)\) denote the observed history regression. It predicts $B_i$ from the participant's observed history just before $t$, while the participant remains under observation and unresolved. For a comparison indicator, this is the conditional probability of the corresponding win, loss, or tie. Let $q_i(t)$ be a predictable working version of $q_{0,i}(t)$. Define the augmented IPCW contribution by
\begin{equation}
\mathcal A_i(B_i;K_i,q_i)=\frac{\Delta_i^RB_i}{K_i(R_i)}+\int_0^{U_i}\frac{q_i(t)}{K_i(t)}\,dM_i^C(t;\Lambda_i^C),
\label{eq:one-member-score}
\end{equation}
where the integral stops at resolution or censoring, whichever occurs first. The first term uses $B_i$ from resolved participants and applies the inverse censoring weight. The second term augments this contribution using the observed history regression $q_i(t)$ and the censoring residual process. Partial histories therefore contribute information even when a participant is censored before resolution. Two identities explain the resulting censoring adjustment. When both working functions are correct,
\begin{equation}
\mathcal A_i(B_i;K_{0,i},q_{0,i})=B_i+\int_0^{U_i}\frac{q_{0,i}(t)-B_i}{K_{0,i}(t)}\,dM_{0,i}^C(t).
\label{eq:true-lift-representation}
\end{equation}
Equation \eqref{eq:true-lift-representation} expresses the augmented contribution as the complete-data value $B_i$ plus a mean-zero stochastic integral under the censoring assumption. Its expectation is therefore $\mathbb E(B_i)$. For arbitrary working functions, the mean error is
\begin{equation}
\mathbb E\{\mathcal A_i(B_i;K_i,q_i)-B_i\}=\mathbb E\int_0^\tau\frac{Y_i^C(t)}{K_i(t)}\{q_i(t)-q_{0,i}(t)\}\{d\Lambda_{0,i}^C(t)-d\Lambda_i^C(t)\},
\label{eq:one-member-bias}
\end{equation}
where the integrand is the product of the error in the observed history regression and the error in the censoring hazard. The contribution is therefore unbiased for $\mathbb E(B_i)$ if either the censoring model or the observed history regression is correctly specified. (see \ref{subsec:app-irt-one-member}).

Consider a treated participant 1 and a control participant 2. Let $H^r=h_r(\bm Y_1,\bm Y_2)$ and $W_j=\Delta_j^R/K_j(R_j)$ for $j=1,2$. The IPCW contribution $W_1W_2H^r$ is defined as zero unless both participants' restricted outcome vectors are observed. To incorporate information recorded before censoring, we augment this contribution using conditional expectations of the comparison indicator under the working joint outcome models. For fixed restricted outcome vectors $\bm y_1$ and $\bm y_2$, let $Q_1^r(t;\bm y_2)$ denote the conditional expectation of $h_r(\bm Y_1,\bm y_2)$ given participant 1's observed history just before $t$ and $R_1\geq t$. Similarly, let $Q_2^r(\bm y_1;s)$ denote the conditional expectation of $h_r(\bm y_1,\bm Y_2)$ given participant 2's observed history just before $s$ and $R_2\geq s$. Let $Q_{12}^r(t,s)$ denote the conditional expectation of $H^r$ given participant 1's observed history just before $t$, participant 2's observed history just before $s$, and $R_1\geq t$, $R_2\geq s$. The first two expectations predict the comparison result when the other participant's outcome is fixed, whereas the third uses the observed histories of both participants. All three conditional expectations are calculated from the same joint outcome models for the treatment and control arms. Thus, averaging $Q_1^r(t;\bm y_2)$ over the conditional distribution of participant 2's outcome gives $Q_{12}^r(t,s)$, and the corresponding relationship holds for $Q_2^r(\bm y_1;s)$. The joint models should also account for dependence among the component event times.

Apply \eqref{eq:one-member-score} first to member 1 with $B_1=h_r(\bm Y_1,\bm y_2)$ and $q_1=Q_1^r(\cdot;\bm y_2)$, and then to member 2 with $B_2$ equal to the resulting quantity evaluated at $\bm y_2=\bm Y_2$. This gives the augmented pair contribution

\begin{align}
 \Gamma_{12}^{r}&=W_1W_2H^r+W_2\int_0^{U_1}\frac{Q_1^r(t;\bm Y_2)}{K_1(t)}dM_1^C(t) \nonumber\\
 &\quad +W_1\int_0^{U_2}\frac{Q_2^r(\bm Y_1;s)}{K_2(s)}dM_2^C(s) +\int_0^{U_1}\int_0^{U_2}\frac{Q_{12}^r(t,s)}{K_1(t)K_2(s)}dM_2^C(s)dM_1^C(t) \nonumber\\
 & = \mathcal T_0^r + \mathcal T_1^r + \mathcal T_2^r + \mathcal T_{12}^r \label{eq:irt-gamma}
\end{align}
The first term \(\mathcal T_{0}^r\) is the IPCW contribution from a pair with both members resolved. The terms \(\mathcal T_{1}^r\), \(\mathcal T_{2}^r\) use the observed history of one member and the complete outcome of the other, and \(\mathcal T_{12}^r\) uses the observed histories of both members. Terms that require an unavailable outcome are defined as zero when the corresponding resolution indicator is zero. Thus, the contribution can be evaluated from the observed data. The residual integrals need not be zero for an uncensored participant, because the censoring hazard enters the residual process even when censoring does not occur. In the absence of censoring and with a working censoring hazard of zero, the pair contribution reduces to $H^r$. The double integral is generally needed when the censoring models are misspecified, because both members require censoring adjustment.

For a symmetric kernel $f$, write $U_nf=\binom n2^{-1}\sum_{i<k}f(O_i,O_k)$ for the U-statistic with kernel $f$. Let $\Gamma_{ik}^r$ denote the augmented pair contribution in \eqref{eq:irt-gamma}, with participant $i$ as member 1 and participant $k$ as member 2. The proposed estimator is
\begin{equation}
    \widehat\lambda_{\mathrm{DR}}^r=U_ng^r_{\widehat\eta}, \label{eq:irt-estimator}
\end{equation}
where
\[ G_{ik}^r(\eta)=\mu^r(\bm L_i,\bm L_k)+\frac{\mathbf 1(A_i=1,A_k=0)}{\pi_1(\bm L_i)\pi_0(\bm L_k)} \{\Gamma_{ik}^r(\eta)-\mu^r(\bm L_i,\bm L_k)\},\]
and \(g^r(O_i,O_k;\eta)=\{G_{ik}^r(\eta)+G_{ki}^r(\eta)\}/2\). In $G_{ik}^r$, the outcome regression $\mu^r(\bm L_i,\bm L_k)$ is adjusted by the weighted difference $\Gamma_{ik}^r-\mu^r(\bm L_i,\bm L_k)$. This adjustment contributes when participant $i$ receives treatment and participant $k$ receives control, and is weighted by the inverse probability of that treatment assignment. Averaging $G_{ik}^r$ and $G_{ki}^r$ gives a kernel that is symmetric in the two participants. Because the treatment assignment probabilities are known, the terms involving $\mu^r$ cancel in expectation conditional on the baseline covariates. Thus, $\mu^r$ need not be correctly specified. Its inclusion provides covariate adjustment that can improve precision.

For comparison, the outcome regression (OR) estimator averages $\{\widehat\mu^r(\bm L_i,\bm L_k)+\widehat\mu^r(\bm L_k,\bm L_i)\}/2$ over all pairs $i<k$. The IPCW estimator averages $\mathbf 1(A_i=1,A_k=0)W_iW_kh_r(\bm Y_i,\bm Y_k)/\{\pi_1(\bm L_i)\pi_0(\bm L_k)\}$ over all ordered pairs $i\ne k$, thereby including both orderings of each pair. The OR estimator is consistent when the joint outcome models for both treatment arms are correctly specified, whereas the IPCW estimator is consistent when the censoring models for both arms are correctly specified. The proposed estimator is consistent when either set of models is correctly specified, as established in the following theorem.

\begin{theorem}[Double robustness in IRTs]
\label{thm:irt-dr}
Under Assumptions \ref{ass:irt-sutva} to \ref{ass:irt-car} and the regularity conditions in Appendix \ref{subsec:app-irt-composition}, $\widehat\lambda_{\mathrm{DR}}^r$ is a consistent estimator
of $\lambda^r$ if either (i) the censoring models are correctly specified for both treatment arms, or (ii) the joint outcome models are correctly specified for both treatment arms.
\end{theorem}
Theorem \ref{thm:irt-dr} establishes the double robustness of the proposed estimator, providing additional protection against model misspecification. The proof is given in Appendix \ref{subsec:app-irt-composition}.

\subsection{Specification of working models}
\label{subsec:irt-working-models}

The proposed estimator requires models for censoring and for the joint distribution of the component event times. As in the work on survival and RMT-IF estimands, these quantities can be estimated using proportional hazards models.\cite{fang2026estimands,fang2026doubly} The models can be fitted separately in each treatment arm, allowing the baseline hazards and covariate associations to differ between treatment groups. For participant $i$ with $A_i=a$, a Cox censoring model specifies $\Lambda_i^C(t)=\Lambda_{0a}^C(t)\exp(\bm\gamma_a^{\mathsf T}\bm L_i)$ and $K_i(t)=\exp\{-\Lambda_i^C(t)\}$, where $\Lambda_{0a}^C$ is the unspecified baseline cumulative hazard in arm $a$ and $\bm\gamma_a$ is a vector of regression coefficients. The response consists of $U_i$ and the event indicator $1-\Delta_i^R$. Censoring before resolution is the event of interest, whereas the censoring time is right-censored at death or at $\tau$. The coefficient estimate $\widehat{\bm\gamma}_a$ maximizes the Cox partial likelihood, and the baseline cumulative hazard is estimated by the Breslow estimator,
\begin{align*}
L_a^C(\bm\gamma_a)&=\prod_{i:A_i=a,\,\Delta_i^R=0}\frac{\exp(\bm\gamma_a^{\mathsf T}\bm L_i)}{\sum_{k:A_k=a}Y_k^C(U_i)\exp(\bm\gamma_a^{\mathsf T}\bm L_k)},\\\widehat\Lambda_{0a}^C(t)
&=\sum_{i:A_i=a,\,U_i\leq t}\frac{1-\Delta_i^R}{\sum_{k:A_k=a}Y_k^C(U_i)
\exp(\widehat{\bm\gamma}_a^{\mathsf T}\bm L_k)}.
\end{align*}
Substituting these estimates gives $\widehat K_i(t)=\exp[-\widehat\Lambda_{0a}^C(t)\exp(\widehat{\bm\gamma}_a^{\mathsf T}\bm L_i)]$ and the estimated cumulative hazard used in the censoring residual process in \eqref{eq:irt-gamma}. If censoring is independent of the event outcomes and baseline covariates within each arm, the arm-specific Kaplan--Meier estimator provides a nonparametric alternative. Writing $d_a^C(u)$ for the number censored at time $u$ and $r_a^C(u)=\sum_{i:A_i=a}Y_i^C(u)$ for the number at risk, this estimator is $\widehat K_a(t)=\prod_{u<t}\{1-d_a^C(u)/r_a^C(u)\}$, with the product taken over observed censoring times before $t$.

For the outcome model, we illustrate the specification with death and one nonfatal component $q$, such as hospitalization. Let $T_D=D$ denote the death time and $T_q$ the latent nonfatal event time in the working joint model. The clinical nonfatal event time equals $T_q$ when $T_q<T_D$, following Section \ref{subsec:irt-estimands}. The proportional marginal hazard models are $S_{D,a}(t\mid\bm L)=\exp[-\Lambda_{0a}^{D}(t)\exp(\bm\beta_{D,a}^{\mathsf T}\bm L)]$ and $S_{q,a}(t\mid\bm L)=\exp[-\Lambda_{0a}^{q}(t)\exp(\bm\beta_{q,a}^{\mathsf T}\bm L)]$, where $\Lambda_{0a}^{D}$ and $\Lambda_{0a}^{q}$ are the outcome specific baseline cumulative hazards in arm $a$, with corresponding regression coefficients $\bm\beta_{D,a}$ and $\bm\beta_{q,a}$. A survival copula combines these margins as
\begin{equation}
S_{Dq,a}(d,u\mid\bm L)=C_{\theta_{W,a}}\{S_{D,a}(d\mid\bm L),S_{q,a}(u\mid\bm L)\},
\label{eq:irt-working-copula}
\end{equation}
where $\theta_{W,a}$ describes the association between the component event times of the same participant in treatment arm $a$. In an IRT, this is the outcome association parameter. The copula families considered for this within-participant association are summarized in Table \ref{tab:main-copula-choices}. The regression coefficients, baseline cumulative hazards, and within-participant association parameter are estimated jointly by maximizing the likelihood of the observed event histories. The likelihood contributions are given in Appendix \ref{subsec:app-continuation}.

The fitted model in \eqref{eq:irt-working-copula} supplies the outcome regressions in \eqref{eq:irt-gamma}. Let $\widehat F_a(\bm y\mid\bm L)$ denote the distribution of the restricted outcome vector obtained from the fitted joint model in arm $a$. The fitted baseline outcome regression is
\begin{equation*}
\widehat\mu^r(\bm L_i,\bm L_k)=\int\!\int h_r(\bm y_1,\bm y_2)\,d\widehat F_1(\bm y_1\mid\bm L_i)\,d\widehat F_0(\bm y_2\mid\bm L_k).
\end{equation*}
The same model provides the fitted conditional distribution $\widehat F_a(\bm y\mid\mathcal F_{i,t^-}^{O},R_i\geq t)$ given the participant's observed history while the restricted outcome remains unresolved. Before either event, predictions condition on $T_D>t$ and $T_q>t$. For a participant who experiences the nonfatal event at $u<t$ and remains alive at $t$, the conditional probability of survival beyond $d\geq t$ is $\{-\partial_u S_{Dq,a}(d,u\mid\bm L)\}/ \{-\partial_u S_{Dq,a}(t,u\mid\bm L)\}$, evaluated at the fitted parameters. The fitted observed history regressions are therefore
\begin{align*}
\widehat Q_1^r(t;\bm y_2)&=\int h_r(\bm y_1,\bm y_2)\, d\widehat F_1(\bm y_1\mid\mathcal F_{1,t^-}^{O},R_1\geq t),\\
\widehat Q_2^r(\bm y_1;s)&=\int h_r(\bm y_1,\bm y_2)\,d\widehat F_0(\bm y_2\mid\mathcal F_{2,s^-}^{O},R_2\geq s),\\
\widehat Q_{12}^r(t,s)&=\int\!\int h_r(\bm y_1,\bm y_2)\,d\widehat F_1(\bm y_1\mid\mathcal F_{1,t^-}^{O},R_1\geq t)\,d\widehat F_0(\bm y_2\mid\mathcal F_{2,s^-}^{O},R_2\geq s).
\end{align*}
In these conditional distributions, observed components are held fixed and integration averages over the components that remain unobserved. All three regressions are thus calculated from the same two fitted joint outcome models. For $Q$ nonfatal components, the working joint survival model extends to
\begin{align*}
\mathbb P_\eta(T_D>d,T_1>t_1,\ldots,T_Q>t_Q \mid A=a,\bm L) =C_{\theta_{W,a}}\{S_{D,a}(d\mid\bm L),S_{1,a}(t_1\mid\bm L),\ldots, S_{Q,a}(t_Q\mid\bm L)\},
\end{align*}
where $T_1,\ldots,T_Q$ are the latent nonfatal event times in the working joint model. The parameters are estimated from the corresponding observed-data likelihood, and the outcome regressions are obtained by the same conditional integration. Further details are given in Appendix \ref{subsec:app-continuation}.

\begin{table}[!t]
\centering
\caption{Copula working models for component event times in individual- and cluster-randomized trials.}
\label{tab:main-copula-choices}
\newcommand{\thW}{\theta_{W,a}}
\newcommand{\thB}{\theta_{B,a}}
\renewcommand{\arraystretch}{1.3}
\begin{adjustbox}{max width=\textwidth}
\begin{tabular}{@{}lllll@{}}
\toprule
 & \multicolumn{2}{c}{\textbf{Individual-randomized trial}} & \multicolumn{2}{c}{\textbf{Cluster-randomized trial}} \\
\cmidrule(lr){2-3} \cmidrule(l){4-5}
\textbf{Family} & \textbf{Copula} $C(u,v)$ & \textbf{Parameter} & \textbf{Joint survival} $S_{i,a}$ & \textbf{Parameter} \\
\midrule
Independence & $uv$ & -- & $\prod_{j=1}^{N_i} w_{ij,a}$ & -- \\
Gumbel & $\exp\left[-\left\{(-\log u)^{\thW}+(-\log v)^{\thW}\right\}^{1/\thW}\right]$ & $\thW\geq 1$ & $C_{\thB}(w_{i1,a},\ldots,w_{iN_i,a})$ & $1\leq\thB\leq\thW$ \\
Clayton & $\left(u^{-\thW}+v^{-\thW}-1\right)^{-1/\thW}$ & $\thW>0$ & $C_{\thB}(w_{i1,a},\ldots,w_{iN_i,a})$ & $0<\thB\leq\thW$ \\
Frank & $-\thW^{-1}\log\left[1+\dfrac{(e^{-\thW u}-1)(e^{-\thW v}-1)}{e^{-\thW}-1}\right]$ & $\thW>0$ & $C_{\thB}(w_{i1,a},\ldots,w_{iN_i,a})$ & $0<\thB\leq\thW$ \\
Joe & $1-\left\{(1-u)^{\thW}+(1-v)^{\thW}-(1-u)^{\thW}(1-v)^{\thW}\right\}^{1/\thW}$ & $\thW\geq 1$ & $C_{\thB}(w_{i1,a},\ldots,w_{iN_i,a})$ & $1\leq\thB\leq\thW$ \\
\bottomrule
\end{tabular}
\end{adjustbox}
\par\smallskip
\begin{minipage}{\textwidth}
\footnotesize
In the individual-randomized trial columns, $u$ and $v$ are marginal survival probabilities for two component event times of the same participant, and $\thW$ describes their association. In the cluster-randomized trial columns, $w_{ij,a}=C_{\thW}\{S_{D,a}(d_j\mid\bm Z_{ij}),S_{q,a}(u_j\mid\bm Z_{ij})\}$ is the within-participant joint survival probability for participant $j$. The outer copula uses the same family and has parameter $\thB$, which describes association among participants in the same cluster.\par
For each dependent family, the outer parameter cannot exceed the inner parameter under the sufficient nesting condition. For the Gumbel family, $\tau_{W,a}=1-\thW^{-1}$ and $\tau_{B,a}=1-\thB^{-1}$ are the corresponding Kendall association measures.
\end{minipage}
\end{table}

\section{Doubly robust win statistics in cluster-randomized trials}
\label{sec:crt}

Building on the win estimands and doubly robust estimator developed for IRTs in Section \ref{sec:irt}, in which each participant is randomized and sampled independently, we consider CRTs, in which practices, hospitals, or communities are randomized and participants within a cluster share the same treatment assignment.%
Their outcomes and censoring tend to be correlated. %
We use the same clinical comparison $h_r$ to define cluster-average and individual-average win estimands, which give equal weight to clusters and participants, respectively. Estimation accounts for dependence within clusters, and censoring model may depend on cluster size and the baseline covariates of all participants in the cluster. Section \ref{subsec:crt-estimands} defines the estimands and establishes identification, and Section \ref{subsec:crt-estimation} presents the corresponding estimators.

\subsection{Notation and estimands}
\label{subsec:crt-estimands}

Consider a parallel arm CRT with $m$ clusters indexed by $i=1,\ldots,m$. Cluster $i$ contains $N_i$ participants indexed by $j=1,\ldots,N_i$ and receives treatment assignment $A_i\in\{0,1\}$. Let $\bm V_i$ and $\bm X_{ij}$ denote the cluster-level and participant-level baseline covariates, respectively, and write \(\bm L_i=(N_i,\bm V_i^{\mathsf T},\bm X_{i1}^{\mathsf T},\ldots,\bm X_{iN_i}^{\mathsf T})^{\mathsf T}\) for the complete baseline information. We include $N_i$ in $\bm L_i$ because cluster size may be associated with prognosis or treatment response. For $a\in\{0,1\}$, define $D_{ij}(a)$, $T_{ij}^q(a)$ for $q=1,\ldots,Q$, $C_{ij}(a)$, and $\mathcal H_{ij}(t;a)$ as in Section \ref{sec:irt}, now for participant $j$ in cluster $i$ assigned to treatment $a$. The restricted outcome vector $\bm Y_{ij}(a)$ and resolution time $R_{ij}(a)=D_{ij}(a)\wedge\tau$ are defined similarly. Write \(\mathcal Y_i(a)=\{\bm Y_{ij}(a):j=1,\ldots,N_i\}\) for the restricted outcomes of cluster $i$ and \(\mathcal W_i(a)=\{D_{ij}(a),T_{ij}^1(a),\ldots,T_{ij}^Q(a),C_{ij}(a):j=1,\ldots,N_i\}\) for its complete event and censoring data. The restriction time $\tau$, priority order, thresholds, and indicators $h_+$, $h_-$, $h_{=}$, and $h_{+,q}$ retain their definitions in Section \ref{sec:irt}. The following assumption relates the potential and observed outcomes.

\begin{assumption}[Cluster-level SUTVA]
\label{ass:sutva}
For each cluster $i$, the potential quantities $\{\mathcal W_i(a),\mathcal H_{ij}(t;a):a=0,1,\ j=1,\ldots,N_i\}$ do not depend on the assignments of the other clusters, and the realized quantities of every member of cluster $i$ are $D_{ij}=D_{ij}(A_i)$, $T_{ij}^q=T_{ij}^q(A_i)$, $C_{ij}=C_{ij}(A_i)$, and $\mathcal H_{ij}(t)=\mathcal H_{ij}(t;A_i)$.
\end{assumption}

The potential outcome $\bm Y_{ij}(a)$ represents the response to assigning the entire cluster to treatment $a$. Under Assumption \ref{ass:sutva}, the observed restricted vector and resolution time are $\bm Y_{ij}=\bm Y_{ij}(A_i)$ and $R_{ij}=D_{ij}\wedge\tau$. Define \(U_{ij}=R_{ij}\wedge C_{ij}\) and \(\Delta_{ij}^R=\mathbf 1(R_{ij}\leq C_{ij})\), so that $\bm Y_{ij}$ is completely observed when $\Delta_{ij}^R=1$. The observed cluster data are \(O_i=\{\bm L_i,A_i,(U_{ij},\Delta_{ij}^R,\mathcal H_{ij}(U_{ij})):j=1,\ldots,N_i\}\). The censoring processes are $Y_{ij}^C(t)=\mathbf 1(U_{ij}\geq t)$ and $N_{ij}^C(t)=\mathbf 1(U_{ij}\leq t,\Delta_{ij}^R=0)$. Let $\mathcal F_{i,t}^O$ denote the information generated by $\bm L_i$, $A_i$, and the observed event and censoring histories of all cluster members through time $t$. This cluster filtration extends the individual information defined in Section \ref{subsec:irt-estimands}.

For a prespecified clinical comparison, the estimand definition in CRT also depends on how participants are sampled from the target population. We consider equal weighting of clusters and equal weighting of participants. Let $i$ and $k$ index two independent clusters and, for $a\in\{0,1\}$ and $r\in\{+,-,=\}$, define
\begin{equation}
H_{ik,a}^{r}=\frac{1}{N_iN_k}\sum_{j=1}^{N_i}\sum_{\ell=1}^{N_k} h_r\{\bm Y_{ij}(a),\bm Y_{k\ell}(1-a)\},
\label{eq:H-cluster}
\end{equation}
which averages the comparisons over all participant pairs from the two clusters, with cluster $i$ assigned to $a$ and cluster $k$ to $1-a$. The cluster-average win, loss, and tie probabilities are \(\lambda_{C,a}^{r}=\mathbb E(H_{ik,a}^{r})\).  The expectation in \(\lambda_{C,a}^{r}\) first samples two clusters with equal probability regardless of their sizes and then averages equally over all participant pairs within the sampled clusters. Each cluster therefore carries equal weight in the target population, and $\lambda_{C,a}^{r}$ describes the comparison between two arbitrary clusters. For the individual-average target, we assume $1\leq\nu=\mathbb E(N_i)<\infty$ and define \(\theta_{I,a}^{r}=\mathbb E(N_iN_kH_{ik,a}^{r})\), \(\lambda_{I,a}^{r}=\frac{\theta_{I,a}^{r}}{\nu^2}\).  The factors $N_iN_k$ weight each cluster in proportion to its size, because a larger cluster contributes more members to the population of participants, and the normalization by $\nu^2$ converts $\theta_{I,a}^{r}$ into an individual-level estimand. Each participant therefore carries equal weight in the target population, and $\lambda_{I,a}^{r}$ describes the comparison between two arbitrary participants, which is equivalent to the target estimand in IRT. Because $\nu$ is unknown, $\lambda_{I,a}^{r}$ is a ratio of two population functionals, which is a feature that may affect estimation and inference. Figure \ref{fig:target-populations} illustrates the two target population weights. The tier specific win probabilities $\lambda_{C,a}^{+,q}$ and $\lambda_{I,a}^{+,q}$ are defined in the same way after replacing $h_r$ by $h_{+,q}$ in \eqref{eq:H-cluster}. For $v\in\{C,I\}$, $\lambda_{v,1}^{+}$ and $\lambda_{v,1}^{-}$ are the win and loss probabilities for intervention versus control, and $\lambda_{v,1}^{=}=1-\lambda_{v,1}^{+}-\lambda_{v,1}^{-}$ is the tie probability. As in Section \ref{subsec:irt-estimands}, these probabilities compare independent participants drawn from the corresponding intervention and control populations.\cite{mao2018causal}

The distinction between $v=C$ and $v=I$ parallels the cluster-average and individual-average treatment effects for a single outcome.\cite{kahan2023estimands,wang2024model} The two targets coincide when the expected comparison is constant across cluster sizes. They can differ under informative cluster size, when larger clusters have different prognoses or treatment responses.\cite{kahan2023informative,lee2026winning} The choice of target should therefore reflect whether the clinical question gives equal importance to clusters or to participants. When $N_i\equiv1$, $m=n$, and $\bm Y_{i1}(a)=\bm Y_i(a)$, we have $\nu=1$ and $\lambda_{C,1}^r=\lambda_{I,1}^r=\lambda^r$, recovering the IRT estimands. Applying \eqref{eq:irt-summaries} to either target gives
\begin{equation}
\mathrm{NB}_v=\lambda_{v,1}^{+}-\lambda_{v,1}^{-}, \quad \mathrm{WR}_v=\frac{\lambda_{v,1}^{+}}{\lambda_{v,1}^{-}}, \quad \mathrm{WO}_v=
\frac{\lambda_{v,1}^{+}+\lambda_{v,1}^{=}/2}{\lambda_{v,1}^{-}+\lambda_{v,1}^{=}/2}= \frac{1+\mathrm{NB}_v}{1-\mathrm{NB}_v} \label{eq:WO}.
\end{equation}
The net benefit is the difference between win and loss probabilities and ranges over $[-1,1]$. The win ratio compares these probabilities on a relative scale, and the win odds allocate half of the tie probability to each condition. By \eqref{eq:WO}, the win odds are a monotone transformation of the net benefit. As in Section \ref{sec:irt}, we estimate $(\lambda_{v,1}^{+},\lambda_{v,1}^{-})$ jointly and obtain all three summaries by transformation. Both targets depend on complete restricted outcomes, whereas censoring may prevent some pairwise comparisons from being observed. The following assumptions specify cluster sampling, randomization, and censoring for identification from the observed data $O_i$.

\begin{assumption}[Independent cluster sampling]
\label{ass:independence}
The cluster-level vectors $\{\bm L_i,\mathcal W_i(0),\mathcal W_i(1)\}$, $i=1,\ldots,m$, are independent and identically distributed with common joint distribution $\mathbb P_0$.
\end{assumption}

\begin{assumption}[Cluster randomization]
\label{ass:randomization}
$A_i\ \perp\!\!\!\perp\ \{\mathcal W_i(0),\mathcal W_i(1)\}\mid\bm L_i$, and the assignment probabilities $\pi_s(\bm L_i)=\mathbb P(A_i=s\mid\bm L_i)$, $s=0,1$, are known and satisfy $\epsilon_\pi\leq\pi_s(\bm L_i)\leq1-\epsilon_\pi$ almost surely for some $\epsilon_\pi>0$. Conditional on all baseline and potential event and censoring records, the assignments are independent across clusters with probabilities $\pi_s(\bm L_i)$.
\end{assumption}

Assumption \ref{ass:independence} replaces independent sampling of participants by independent sampling of whole clusters. Dependence among the event and censoring processes of members within a cluster is left unrestricted, and the independence between clusters is what allows the censoring corrections of the two members of a cross cluster pair to be integrated separately. Under Assumption \ref{ass:randomization}, the probability that cluster $i$ is assigned to $a$ and cluster $k$ to $1-a$ is $\pi_a(\bm L_i)\pi_{1-a}(\bm L_k)$. Because all members of a cluster receive the same assignment, an intervention control comparison involves participants from different clusters.
\begin{assumption}[Covariate dependent censoring]
\label{ass:car}
For each cluster $i$ and treatment arm $a\in\{0,1\}$,
\[
\{C_{ij}(a):j=1,\ldots,N_i\}\ \perp\!\!\!\perp\ \{D_{ij}(a),T_{ij}^1(a),\ldots,T_{ij}^Q(a): j=1,\ldots,N_i\} \mid\bm L_i.
\]
Furthermore, there exists a constant $\epsilon_C>0$ such that \(\mathbb P\{C_{ij}(a)\geq\tau\mid\bm L_i\} \geq\epsilon_C\) almost surely for every $j=1,\ldots,N_i$ and $a\in\{0,1\}$.
\end{assumption}
Assumption \ref{ass:car} extends Assumption \ref{ass:irt-car} to the cluster setting by conditioning on the baseline information of the entire cluster. Censoring may depend on cluster size and the baseline covariates of all cluster members. Dependence among participants' censoring times and among their event times is allowed. We write $K_{0,ij}(t)=\mathbb P(C_{ij}\geq t\mid\bm L_i,A_i)$ for the conditional censoring survival function of participant $j$ in cluster $i$ under the assigned treatment.

For the treatment comparison in \eqref{eq:H-cluster}, let $\mu_{C,a,0}^r(\bm L_i,\bm L_k)=\mathbb E(H_{ik,a}^r\mid\bm L_i,\bm L_k)$ and $\mu_{I,a,0}^r(\bm L_i,\bm L_k)=\mathbb E(N_iN_kH_{ik,a}^r\mid\bm L_i,\bm L_k)$ denote the outcome regression functions for the cluster-average probability and the numerator of the individual-average probability, respectively. Since cluster sizes are included in $\bm L_i$ and $\bm L_k$, $\mu_{I,a,0}^r=N_iN_k\mu_{C,a,0}^r$. As in Theorem \ref{thm:irt-identification}, the IPCW contribution of a participant pair is set to zero unless the restricted outcome vectors of both participants are completely observed.

\begin{theorem}[Identification in CRTs]
\label{thm:identification}
Under Assumptions \ref{ass:sutva} to \ref{ass:car} and $\nu<\infty$, for $a\in\{0,1\}$ and $r\in\{+,-,=\}$,
\begin{align}
\lambda_{C,a}^{r}
&=\mathbb E\Bigg[\frac{\mathbf 1(A_i=a,A_k=1-a)}{\pi_a(\bm L_i)\pi_{1-a}(\bm L_k)}\frac{1}{N_iN_k}\sum_{j=1}^{N_i}\sum_{\ell=1}^{N_k}\frac{\Delta_{ij}^{R}\Delta_{k\ell}^{R}h_r(\bm Y_{ij},\bm Y_{k\ell})}{K_{0,ij}(R_{ij})K_{0,k\ell}(R_{k\ell})}\Bigg]=\mathbb E\{\mu_{C,a,0}^r(\bm L_i,\bm L_k)\},
\label{eq:crt-identification-C}\\
\lambda_{I,a}^{r}
&=\frac{1}{\nu^2}\mathbb E\Bigg[\frac{\mathbf 1(A_i=a,A_k=1-a)}{\pi_a(\bm L_i)\pi_{1-a}(\bm L_k)}\sum_{j=1}^{N_i}\sum_{\ell=1}^{N_k}\frac{\Delta_{ij}^{R}\Delta_{k\ell}^{R}h_r(\bm Y_{ij},\bm Y_{k\ell})}{K_{0,ij}(R_{ij})K_{0,k\ell}(R_{k\ell})}\Bigg]=\frac{\mathbb E\{\mu_{I,a,0}^r(\bm L_i,\bm L_k)\}}{\nu^2}.
\label{eq:crt-identification-I}
\end{align}
The unnormalized expectation in \eqref{eq:crt-identification-I} identifies $\theta_{I,a}^r$, and the same representations apply to the tier specific win probabilities after replacing $h_r$ by $h_{+,q}$.
\end{theorem}
Theorem \ref{thm:identification} extends Theorem \ref{thm:irt-identification} by averaging participant comparisons with the weights appropriate to each CRT target. The IPCW representation uses pairs with both restricted outcome vectors observed, and the outcome regression representation averages predictions conditional on the two baseline cluster records. Division by $N_iN_k$ gives equal weight to cluster pairs. For the individual-average target, the unnormalized sum is divided by $\nu^2$, with $\mathbb E(N_iN_k)=\nu^2$ under independent cluster sampling. The proof and the reduction to \eqref{eq:irt-identification} when $N_i\equiv1$ and $a=1$ are given in Appendix \ref{subsec:app-identification}. Replacing the censoring models or the outcome regression functions by fitted models gives the corresponding IPCW or outcome regression estimator. These estimators target $\lambda_{C,a}^r$ or $\lambda_{I,a}^r$ according to the chosen population weights.

\subsection{Doubly robust estimation}
\label{subsec:crt-estimation}

The CRT estimator uses the pairwise contribution from Section \ref{subsec:irt-estimation} and averages it according to the target population. Let $\mathcal F_{ij,t}^O$ denote the information generated by the complete baseline cluster record $\bm L_i$, treatment $A_i$, and participant $j$'s observed event and censoring history through time $t$. Define $\mathcal F_{k\ell,t}^O$ analogously. For participant $i,j$, let $\Lambda_{ij}^C$ denote a working cumulative censoring hazard predictable with respect to $\mathcal F_{ij,t}^O$, and write $K_{ij}(t)=\exp\{-\Lambda_{ij}^C(t)\}$, \(M_{ij}^C(t)=N_{ij}^C(t)-\int_0^tY_{ij}^C(u)\,d\Lambda_{ij}^C(u)\), and $W_{ij}=\Delta_{ij}^R/K_{ij}(R_{ij})$. Define $\Lambda_{k\ell}^C$, $K_{k\ell}$, $M_{k\ell}^C$, and $W_{k\ell}$ analogously. The required conditional expectations are \(Q_{ij,a}^{r}(t;\bm y)=\mathbb E_\eta\bigl[h_r\{\bm Y_{ij}(a),\bm y\}\mid\mathcal F_{ij,t^-}^O,R_{ij}\geq t\bigr]\), \(Q_{k\ell,b}^{r}(\bm y;s)=\mathbb E_\eta\bigl[h_r\{\bm y,\bm Y_{k\ell}(b)\}\mid\mathcal F_{k\ell,s^-}^O,R_{k\ell}\geq s\bigr]\), and \(Q_{ij,k\ell,a}^{r}(t,s)=\mathbb E_\eta\bigl[h_r\{\bm Y_{ij}(a),\bm Y_{k\ell}(b)\}\mid\mathcal F_{ij,t^-}^O,\mathcal F_{k\ell,s^-}^O,R_{ij}\geq t,R_{k\ell}\geq s\bigr]\), where $\mathbb E_\eta$ denotes expectation under the working joint outcome models and $\bm y$ is a fixed restricted outcome vector. The first two expectations hold the other participant's outcome fixed, and the third averages over the remaining outcomes of both participants. All three are calculated from the same two joint outcome models. In particular, averaging $Q_{ij,a}^{r}(t;\bm y)$ over the conditional distribution of $\bm Y_{k\ell}(b)$ given $\mathcal F_{k\ell,s^-}^O$ and $R_{k\ell}\geq s$ gives $Q_{ij,k\ell,a}^{r}(t,s)$.

Using \eqref{eq:one-member-score} for both participants gives
\begin{equation}
\begin{split}
\Gamma_{ij,k\ell,a}^{r}(\eta)
={}&W_{ij}W_{k\ell}h_r(\bm Y_{ij},\bm Y_{k\ell}) +W_{k\ell}\int_0^{U_{ij}}\frac{Q_{ij,a}^{r}(t;\bm Y_{k\ell})}{K_{ij}(t)}\,dM_{ij}^C(t)\\
&+W_{ij}\int_0^{U_{k\ell}}\frac{Q_{k\ell,b}^{r}(\bm Y_{ij};s)}{K_{k\ell}(s)}\,dM_{k\ell}^C(s)+\int_0^{U_{ij}}\int_0^{U_{k\ell}}\frac{Q_{ij,k\ell,a}^{r}(t,s)}{K_{ij}(t)K_{k\ell}(s)}\,dM_{k\ell}^C(s)\,dM_{ij}^C(t),
\end{split}
\label{eq:crt-gamma}
\end{equation}
which extends \eqref{eq:irt-gamma} to participants from independent clusters. Its conditional expectation is characterized by the following Theorem.

\begin{theorem}[Double robustness of the pairwise contribution in CRTs]
\label{thm:pair-dr}
Under Assumptions \ref{ass:sutva} to \ref{ass:car}, suppose that the integrals in \eqref{eq:crt-gamma} have finite absolute expectations. For $a\in\{0,1\}$, $b=1-a$, and $r\in\{+,-,=\}$, the pairwise contribution satisfies
\[
\mathbb E\bigl\{\Gamma_{ij,k\ell,a}^{r}(\eta^*) \mid A_i=a,A_k=b,\bm L_i,\bm L_k\bigr\} = \mathbb E\bigl[h_r\{\bm Y_{ij}(a),\bm Y_{k\ell}(b)\} \mid\bm L_i,\bm L_k \bigr]
\]
provided that, at $\eta^*$, either (i) the censoring models are correctly specified for both treatment arms, or (ii) the joint outcome models are correctly specified for both treatment arms.
\end{theorem}

The proof of Theorem \ref{thm:pair-dr} is given in Appendix \ref{sec:pair-score-proof}. For each pair of distinct clusters $i$ and $k$, we average the participant contributions for the cluster-average estimand and sum them for the numerator of the individual-average estimand, which can be expressed as
\begin{equation*}
 \overline\Gamma_{C,ik,a}^{r}(\eta)=\frac{1}{N_iN_k}\sum_{j=1}^{N_i}\sum_{\ell=1}^{N_k}\Gamma_{ij,k\ell,a}^{r}(\eta), \quad \overline\Gamma_{I,ik,a}^{r}(\eta) =\sum_{j=1}^{N_i}\sum_{\ell=1}^{N_k}\Gamma_{ij,k\ell,a}^{r}(\eta)
\end{equation*}

These quantities correspond to $H_{ik,a}^{r}$ in \eqref{eq:H-cluster} and $N_iN_kH_{ik,a}^{r}$, respectively. For $v\in\{C,I\}$, let $\mu_{v,a}^{r}(\bm L_i,\bm L_k;\eta)$ denote a working outcome
regression corresponding to $\mu_{v,a,0}^{r}$. Using the known randomization probabilities in Assumption \ref{ass:randomization}, we define the symmetric kernel as
\begin{align} 
g_{v,a}^{r}(O_i,O_k;\eta)=\frac12\Bigg[&
\mu_{v,a}^{r}(\bm L_i,\bm L_k;\eta)+\mu_{v,a}^{r}(\bm L_k,\bm L_i;\eta)
+\frac{\mathbf 1(A_i=a,A_k=b)}{\pi_a(\bm L_i)\pi_b(\bm L_k)}\Big\{\overline\Gamma_{v,ik,a}^{r}(\eta)-\mu_{v,a}^{r}(\bm L_i,\bm L_k;\eta)\Big\}\nonumber\\
&+\frac{\mathbf 1(A_k=a,A_i=b)}{\pi_a(\bm L_k)\pi_b(\bm L_i)}\Big\{\overline\Gamma_{v,ki,a}^{r}(\eta)-\mu_{v,a}^{r}(\bm L_k,\bm L_i;\eta)\Big\}\Bigg].
\label{eq:augmented-U-kernel}
\end{align}
The two weighted terms account for either cluster receiving treatment $a$, with the other receiving treatment $b$. As in the IRT estimator, the known randomization probabilities ensure that the outcome regression $\mu_{v,a}^{r}$ cancels from the conditional expectation of the kernel. This baseline covariate adjustment can improve precision. For a symmetric function $f$, write
$U_mf=\binom{m}{2}^{-1}\sum_{1\leq i<k\leq m}f(O_i,O_k)$ for the U-statistic over cluster pairs. The proposed estimators are
\begin{equation}
\widehat\lambda_{C,a}^{r}=U_mg_{C,a}^{r}(\widehat\eta),
\qquad
\widehat\theta_{I,a}^{r} =U_mg_{I,a}^{r}(\widehat\eta),
\label{eq:theta-estimators}
\end{equation}
and
\begin{equation}
\widehat\lambda_{I,a}^{r}=\frac{\widehat\theta_{I,a}^{r}}{\widehat\nu^{\,2}},
\qquad
\widehat\nu=\frac1m\sum_{i=1}^{m}N_i.
\label{eq:lambda-I-estimator}
\end{equation}
The win and loss probabilities are estimated jointly, and the tie probability is estimated by one minus their sum. The summary win statistics are obtained by substituting these estimates into \eqref{eq:WO}.

\begin{theorem}[Double robustness in CRTs]
\label{thm:dr-efficiency}
Under Assumptions \ref{ass:sutva} to \ref{ass:car} and the positivity and regularity conditions in Appendix \ref{sec:parametric-asymptotic-proof}, the estimators in \eqref{eq:theta-estimators} and \eqref{eq:lambda-I-estimator} are consistent for their respective parameters if either (i) the censoring models are correctly specified for both treatment arms, or (ii) the joint outcome models are correctly specified for both treatment arms.
\end{theorem}
The proof is given in Appendix \ref{sec:parametric-asymptotic-proof}. When $N_i\equiv1$, $\widehat\nu=1$, and, for $a=1$, the cluster-average and individual-average estimators both reduce to the IRT estimator in \eqref{eq:irt-estimator}.

\subsection{Specification of working models}
\label{subsec:crt-working-models}
The working models in Section \ref{subsec:irt-working-models} extend to CRTs through marginal Cox models fitted separately in each treatment arm.\cite{lin1994cox,fang2026estimands} Let $\bm Z_{ij}$ contain selected functions of the complete baseline cluster record $\bm L_i$, such as $\log N_i$, $\bm V_i$, $\bm X_{ij}$, and the cluster mean $\overline{\bm X}_i=N_i^{-1}\sum_{j=1}^{N_i}\bm X_{ij}$. The models specify the conditional distributions given $\bm L_i$ as functions of $\bm Z_{ij}$. Correct specification requires these selected covariates to capture the relevant dependence on the complete baseline record. For participant $j$ in a cluster assigned to $a$, the censoring model specifies
\begin{equation}
   \Lambda_{ij}^C(t)=\Lambda_{0a}^C(t)\exp(\bm\gamma_a^{\mathsf T}\bm Z_{ij}), \quad K_{ij}(t)=\exp\{-\Lambda_{0a}^C(t)\exp(\bm\gamma_a^{\mathsf T}\bm Z_{ij})\}, \label{eq:crt-working-censoring}
\end{equation}
where $\Lambda_{0a}^C$ is the unspecified baseline cumulative censoring hazard in arm $a$. The coefficient vector $\bm\gamma_a$ is estimated by Cox partial likelihood under working independence across participants, using $U_{ij}$ and $1-\Delta_{ij}^R$ as the follow-up time and event indicator. The Breslow estimator is
\begin{equation*}
\widehat\Lambda_{0a}^C(t)=\int_0^t\frac{\sum_{i:A_i=a}\sum_{j=1}^{N_i}dN_{ij}^C(u)}{\sum_{i:A_i=a}\sum_{j=1}^{N_i}Y_{ij}^C(u)\exp(\widehat{\bm\gamma}_a^{\mathsf T}\bm Z_{ij})}.
\end{equation*}
Substitution into \eqref{eq:crt-working-censoring} gives $\widehat K_{ij}(t)$ and the fitted cumulative hazard used in the censoring residual process. If the conditional censoring distribution is the same for all baseline cluster records within an arm, the arm-specific Kaplan--Meier estimator provides a nonparametric alternative.\cite{fang2026doubly}

For death and one nonfatal component $q$, the participant-specific marginal survival models are
\begin{equation*}
    S_{D,a}(t\mid\bm Z_{ij})=\exp\left[-\Lambda_{0a}^{D}(t)\exp\{\bm\beta_{D,a}^{\mathsf T}\bm Z_{ij}\}\right], \quad
    S_{q,a}(t\mid\bm Z_{ij})=\exp\left[-\Lambda_{0a}^{q}(t)\exp\{\bm\beta_{q,a}^{\mathsf T}\bm Z_{ij}\}\right].
\end{equation*}
The joint survival function for the component event times of participant $j$ is
\begin{equation}
w_{ij,a}(d,u\mid\bm L_i)=C_{\theta_{W,a}}\{S_{D,a}(d\mid\bm Z_{ij}),S_{q,a}(u\mid\bm Z_{ij})\},
\label{eq:crt-working-copula}
\end{equation}
where $\theta_{W,a}$ describes association between the component event times of the same participant, as in \eqref{eq:irt-working-copula}.

A CRT may additionally exhibit dependence between different participants in the same cluster. For $\bm d_i=(d_1,\ldots,d_{N_i})^{\mathsf T}$ and $\bm u_i=(u_1,\ldots,u_{N_i})^{\mathsf T}$, we specify the cluster-level joint survival function as
\begin{equation}
S_{i,a}(\bm d_i,\bm u_i\mid\bm L_i)=C_{\theta_{B,a}}\left\{w_{i1,a}(d_1,u_1\mid\bm L_i),\ldots,w_{iN_i,a}(d_{N_i},u_{N_i}\mid\bm L_i)\right\},
\label{eq:crt-cluster-copula}
\end{equation}
where $\theta_{B,a}$ describes association between participants within the same cluster. Thus, the CRT joint outcome model has two association parameters in each treatment arm: $\theta_{W,a}$ for component event times within a participant and $\theta_{B,a}$ for event times from different participants in the same cluster. Both parameters may differ between treatment arms. We use the same Archimedean family at the participant and cluster levels. The sufficient nesting condition requires the outer association parameter $\theta_{B,a}$ to be no larger than the inner association parameter $\theta_{W,a}$. Equivalently, association between participants in a cluster cannot exceed association among the component event times of the same participant under this working model.\cite{hofert2011nested}

Let $\widehat F_a(\bm y\mid\bm Z_{ij})$ denote the fitted distribution of participant $j$'s restricted outcome vector under treatment $a$. Under the working model, this is the conditional distribution given $\bm L_i$, expressed through $\bm Z_{ij}$. For $b=1-a$, the baseline outcome regressions obtained from these models are
\begin{align*}
\widehat\mu_{C,a}^r(\bm L_i,\bm L_k)&=\frac{1}{N_iN_k}\sum_{j=1}^{N_i}\sum_{\ell=1}^{N_k}\int\!\int h_r(\bm y_1,\bm y_2)\,d\widehat F_a(\bm y_1\mid\bm Z_{ij})\,d\widehat F_b(\bm y_2\mid\bm Z_{k\ell}),\\
\widehat\mu_{I,a}^r(\bm L_i,\bm L_k)&=N_iN_k\widehat\mu_{C,a}^r(\bm L_i,\bm L_k).
\end{align*}

For implementation, let $\mathcal F_{ij,t}^{O}$ denote the information generated by $\bm L_i$, $A_i$, and participant $j$'s observed event and censoring history through $t$. Under Assumption \ref{ass:car}, the contribution in \eqref{eq:crt-gamma} can be constructed using this information for both the censoring residual process and the outcome regressions. Using the fitted conditional distributions given these histories,
\begin{align*}
\widehat Q_{ij,a}^{r}(t;\bm y_2)&=\int h_r(\bm y_1,\bm y_2)\,d\widehat F_a(\bm y_1\mid\mathcal F_{ij,t^-}^{O},R_{ij}\geq t),\\\widehat Q_{k\ell,b}^{r}(\bm y_1;s)
&=\int h_r(\bm y_1,\bm y_2)\,d\widehat F_b(\bm y_2\mid\mathcal F_{k\ell,s^-}^{O},R_{k\ell}\geq s),\\\widehat Q_{ij,k\ell,a}^{r}(t,s)
&=\int\!\int h_r(\bm y_1,\bm y_2)\,d\widehat F_a(\bm y_1\mid\mathcal F_{ij,t^-}^{O},R_{ij}\geq t)\,d\widehat F_b(\bm y_2\mid\mathcal F_{k\ell,s^-}^{O},R_{k\ell}\geq s).
\end{align*}
Thus, the same fitted participant models provide the baseline and observed history regressions for both CRT estimands. The averaging rules in \eqref{eq:theta-estimators} and the normalization in \eqref{eq:lambda-I-estimator} determine the target population. Dependence within clusters, including dependence among contributions to estimation of the working models, is accounted for in the cluster-level variance calculation in Section \ref{sec:theory} and Appendix \ref{subsec:app-nuisance-equations}.

\section{Asymptotic properties}
\label{sec:theory}

The estimators average pairwise comparisons that share participants or clusters. Their asymptotic distributions are therefore determined by the number of independent participants in an IRT and the number of independent clusters in a CRT. The variance estimators below account for this dependence and for estimation of the working models. Regularity conditions, derivations of the influence functions, and proofs are given in \ref{sec:app-theory}.

\subsection{Individual-randomized trials}
\label{subsec:irt-theory}

Write $\bm\lambda=(\lambda^+,\lambda^-)^{\mathsf T}$ and $\widehat{\bm\lambda}=(\widehat\lambda_{\mathrm{DR}}^+,\widehat\lambda_{\mathrm{DR}}^-)^{\mathsf T}$, and let $\bm g=(g^+,g^-)^{\mathsf T}$ collect the kernels in \eqref{eq:irt-estimator}. Let $\bm\Psi(\eta)=\mathbb E\{\bm g(O_i,O_k;\eta)\}$. Denote the estimated influence contribution from fitting the working models by $\widehat{\bm\psi}_i$, and estimate the derivative of $\bm\Psi$ by differentiating $U_n\bm g(\eta)$ at $\widehat\eta$. We write this estimated derivative as $\widehat{\dot{\bm\Psi}}$. The calculation includes the regression coefficients, baseline hazards, and association parameters, as described in Appendix \ref{subsec:app-nuisance-equations}.

Under independent treatment assignment, the estimated participant influence contributions and the covariance estimator for $\widehat{\bm\lambda}$ are
\begin{align}
\widehat{\bm\chi}_i&=\frac{2}{n-1}\sum_{k\ne i}\bm g(O_i,O_k;\widehat\eta)-2\widehat{\bm\lambda}+\widehat{\dot{\bm\Psi}}[\widehat{\bm\psi}_i],\nonumber\\
\widehat{\bm V}&=\frac{1}{n(n-1)}\sum_{i=1}^n(\widehat{\bm\chi}_i-\overline{\bm\chi})(\widehat{\bm\chi}_i-\overline{\bm\chi})^{\mathsf T},
\label{eq:irt-variance}
\end{align}
where $\overline{\bm\chi}=n^{-1}\sum_i\widehat{\bm\chi}_i$. The empirical kernel average uses the pairwise comparisons evaluated for the point estimator. The factor two accounts for the occurrence of each participant in $n-1$ pairs, and the final term accounts for fitting the working models. The standard errors of the estimated win and loss probabilities are the square roots of the corresponding diagonal elements of $\widehat{\bm V}$. Let $\bm\Sigma_{\mathrm{IRT}}$ denote the covariance matrix of the participant influence functions in Appendix \ref{subsec:app-irt-asymptotics}.
\begin{theorem}[Asymptotic properties in IRTs]
\label{thm:irt-inference}
Under Assumptions \ref{ass:irt-sutva} to \ref{ass:irt-car}, either model condition in Theorem \ref{thm:irt-dr}, and the regularity conditions in Appendix \ref{subsec:app-irt-asymptotics}, $\widehat{\bm\lambda}\to\bm\lambda$ in probability and 
$\sqrt n(\widehat{\bm\lambda}-\bm\lambda)\xrightarrow{d}\mathrm N(\bm0,\bm\Sigma_{\mathrm{IRT}})$ as $n\to\infty$. Under the additional empirical mean square consistency conditions in Appendix \ref{subsec:app-irt-variance}, $n\widehat{\bm V}\to\bm\Sigma_{\mathrm{IRT}}$ in probability.
\end{theorem}

Double robustness has a direct implication for variance estimation. When only one set of working models is correctly specified, estimation of those models can contribute to the first-order variance. Retaining the nuisance estimation term in \eqref{eq:irt-variance} therefore provides a common variance calculation under either model condition. When both sets are correctly specified, this contribution vanishes to first order. If the baseline outcome regression is also correctly specified, the estimator's influence function is the efficient influence function (EIF). The outcome regression uses the participant's observed history to predict the remaining comparison, while censoring weighting accounts for observing the complete restricted outcome. This augmentation has the same single participant structure as the survival influence function used in previous work on restricted mean time in favor of treatment.\cite{fang2026doubly} The efficiency calculation is given in Appendix \ref{subsec:app-irt-theory}.

For simple complete randomization with fixed treatment totals $n_s=\sum_i\mathbf 1(A_i=s)$, let $\overline{\bm\chi}_s=n_s^{-1}\sum_{i:A_i=s}\widehat{\bm\chi}_i$. The covariance estimator centers the contributions within treatment arms,
\begin{equation}
\widehat{\bm V}_{\mathrm{fix}}=\frac1{n^2}\sum_{s=0}^1\frac{n_s}{n_s-1}\sum_{i:A_i=s}(\widehat{\bm\chi}_i-\overline{\bm\chi}_s) (\widehat{\bm\chi}_i-\overline{\bm\chi}_s)^{\mathsf T}.
\label{eq:app-irt-fixed-cov}
\end{equation}
The implementation in Section \ref{subsec:sim-irt} additionally multiplies the empirical kernel projection in \eqref{eq:irt-variance} by $(n-1)/(n-2)$ before adding the nuisance estimation term. This finite sample adjustment tends to one. The justification for \eqref{eq:app-irt-fixed-cov} and the assignment probabilities under fixed treatment totals is given in Appendix \ref{subsec:app-irt-fixed}. For a differentiable summary $f(\bm\lambda)$, the delta-method variance estimator is
\begin{equation}
\widehat{\operatorname{Var}}\{f(\widehat{\bm\lambda})\}=\nabla f(\widehat{\bm\lambda})^{\mathsf T} \widehat{\bm V}\, \nabla f(\widehat{\bm\lambda}),
\label{eq:win-summary-variance}
\end{equation}
with $\widehat{\bm V}_{\mathrm{fix}}$ used under fixed treatment totals. The gradients are $(-1,-1)^{\mathsf T}$ for the tie probability, $(1,-1)^{\mathsf T}$ for the net benefit, $(1/\lambda^+,-1/\lambda^-)^{\mathsf T}$ for the log win ratio, and $2(1,-1)^{\mathsf T}/(1-\mathrm{NB}^2)$ for the log win odds. For positive win and loss probabilities and $|\mathrm{NB}|<1$, a $100(1-\alpha)\%$ confidence interval for either ratio is $\exp\{\widehat\ell\pm z_{1-\alpha/2}\widehat{\operatorname{se}}(\widehat\ell)\}$, where $\widehat\ell$ is its estimated logarithm. Further derivations are given in Appendix \ref{sec:app-derived-inference}.

\subsection{Cluster-randomized trials}
\label{subsec:crt-theory}

For the treatment versus control comparison, denote the different levels of parameters as $\bm\vartheta=(\lambda_{C,1}^{+},\lambda_{C,1}^{-},\theta_{I,1}^{+},\theta_{I,1}^{-},\nu)^{\mathsf T}$, and let $\widehat{\bm\vartheta}$ contain their estimators in \eqref{eq:theta-estimators} and \eqref{eq:lambda-I-estimator}. Let $\bm\Sigma_{\mathrm{CRT}}$ denote the covariance matrix of the corresponding cluster influence functions in Appendix \ref{sec:parametric-asymptotic-proof}.
\begin{theorem}[Consistency and asymptotic normality in CRTs]
\label{thm:asymptotic-normality}
Under Assumptions \ref{ass:sutva} to \ref{ass:car}, either model condition in Theorem \ref{thm:dr-efficiency}, and the regularity conditions in Appendix \ref{sec:parametric-asymptotic-proof}, $\widehat{\bm\vartheta}\to\bm\vartheta$ in probability and 
$\sqrt m(\widehat{\bm\vartheta}-\bm\vartheta)\xrightarrow{d}\mathrm N(\bm0,\bm\Sigma_{\mathrm{CRT}})$ as $m\to\infty$. The individual-average probability estimators are consistent and jointly asymptotically normal under the normalization in \eqref{eq:lambda-I-estimator}.
\end{theorem}

Variance estimation uses clusters as the independent sampling units. For $v\in\{C,I\}$, estimate the contribution of cluster $i$ to the kernel average by $\widehat g_{v,a,i}^r=(m-1)^{-1}\sum_{k\ne i}g_{v,a}^r(O_i,O_k;\widehat\eta)$. Let $\widehat{\bm D}_{v,a}^r=\partial U_mg_{v,a}^r(\eta)/\partial\eta\vert_{\eta=\widehat\eta}$, and let $\widehat{\bm\psi}_i$ denote the estimated cluster influence contribution from fitting the working models. The estimated influence contributions are
\begin{align}
\widehat\chi_{\Psi,v,a,i}^r&=2\{\widehat g_{v,a,i}^r-U_mg_{v,a}^r(\widehat\eta)\} +(\widehat{\bm D}_{v,a}^r)^{\mathsf T}\widehat{\bm\psi}_i,\nonumber\\
\widehat\chi_{\nu,i}&=N_i-\widehat\nu.
\label{eq:hi-hat}
\end{align}
The derivative includes all fitted nuisance components. Its interpretation for estimated baseline hazard functions and the calculation of $\widehat{\bm\psi}_i$ are given in Appendices \ref{subsec:app-functional-nuisance} and \ref{subsec:app-nuisance-equations}.

For the five components of $\bm\vartheta$, set $\widehat{\bm\chi}_i=(\widehat\chi_{\Psi,C,1,i}^+,\widehat\chi_{\Psi,C,1,i}^-,\widehat\chi_{\Psi,I,1,i}^+,\widehat\chi_{\Psi,I,1,i}^-,\widehat\chi_{\nu,i})^{\mathsf T}$. The analytic sandwich estimator is
\begin{equation}
\widehat{\bm\Sigma}_{\mathrm{CRT}}=\frac1m\sum_{i=1}^m\widehat{\bm\chi}_i\widehat{\bm\chi}_i^{\mathsf T}.
\label{eq:sandwich}
\end{equation}
The estimated covariance of $\widehat{\bm\vartheta}$ is $\widehat{\bm\Sigma}_{\mathrm{CRT}}/m$. For the individual-average probability, the delta method gives
\begin{equation}
\widehat\chi_{I,a,i}^r=\frac{\widehat\chi_{\Psi,I,a,i}^r}{\widehat\nu^{\,2}}-\frac{2\widehat\lambda_{I,a}^r}{\widehat\nu}(N_i-\widehat\nu).
\label{eq:crt-individual-influence}
\end{equation}
The second term accounts for estimation of mean cluster size. Computing both terms from the same cluster retains their covariance, which is relevant when cluster size is associated with participant outcomes. With $\widehat{\bm\chi}_{v,i}=(\widehat\chi_{v,1,i}^+,\widehat\chi_{v,1,i}^-)^{\mathsf T}$, the covariance of the estimated win and loss probabilities is estimated by $\widehat{\bm V}_v=m^{-2}\sum_i\widehat{\bm\chi}_{v,i}\widehat{\bm\chi}_{v,i}^{\mathsf T}$. Their standard errors are the square roots of its diagonal elements. The variances and confidence intervals for the win statistics follow from \eqref{eq:win-summary-variance}, using the target-specific probability estimates and $\widehat{\bm V}_v$. For the individual-average win ratio, the common term $\widehat\nu^{\,2}$ cancels from the ratio, and its variance can equivalently be calculated directly from the two numerator estimates. Under the conditions of Theorem \ref{thm:asymptotic-normality} and the empirical mean square consistency conditions in Appendix \ref{sec:variance-proof}, $\widehat{\bm\Sigma}_{\mathrm{CRT}}\to\bm\Sigma_{\mathrm{CRT}}$ in probability. The resulting Wald intervals have nominal asymptotic coverage for differentiable summaries with positive limiting variance.

For simple complete randomization with fixed numbers of clusters per arm, \eqref{eq:app-irt-fixed-cov} applies with $n$ and $n_s$ replaced by $m$ and $m_s$, and with the full cluster influence vectors in place of the participant contributions. The within-arm centering includes the contribution from estimating $\nu$. In Section \ref{subsec:sim-crt}, the empirical kernel projection in \eqref{eq:hi-hat} is multiplied by $(m-1)/(m-2)$ before the nuisance estimation contribution is added. The corresponding derivation is given in Appendix \ref{sec:fixed-allocation-proof}.

The implementation uses participant-specific outcome and censoring models fitted under working independence, while the cluster influence functions and their covariance account for dependence within clusters. The full pairwise kernel and the contributions from fitting the working models are retained in \eqref{eq:hi-hat}, providing the variance calculation under either model condition in Theorem \ref{thm:dr-efficiency}. This use of cluster-level influence functions follows the same procedure on while-alive estimands.\cite{fang2026whilealive} In the joint cluster model, outcomes recorded for one participant can also provide information about another participant's censored outcomes. The efficient influence function can therefore involve additional terms depending on the observed records of several cluster members. Evaluating these terms requires conditional expectations under their joint distribution and becomes more demanding as cluster size increases, as also discussed for single survival outcomes in CRTs.\cite{fang2026estimands} %
Appendices \ref{subsec:app-canonical} and \ref{subsec:app-eif-computation} examine higher-order terms within a cluster and establish conditions under which an estimator incorporating the full joint cluster projection attains the semiparametric efficiency bound. Computing this projection requires additional modeling of joint cluster histories and can be computationally demanding, with potentially limited efficiency gains, motivating the practical estimator under working independence.

\section{Simulation studies}
\label{sec:simulation}

We evaluate the finite sample performance of the proposed estimators under individual- and cluster-randomized trial. Both studies prioritize death over first hospitalization, use indifference thresholds $\delta_0=\delta_1=0$, and evaluate restriction times $\tau\in\{1,1.5,2\}$. Each dataset is analyzed under four combinations of censoring and outcome models, denoted by $\mathrm{CC}$, $\mathrm{CM}$, $\mathrm{MC}$, and $\mathrm{MM}$. The first letter refers to the censoring model and the second to the outcome model. Here, $\mathrm C$ denotes a model containing the generating specification, and $\mathrm M$ denotes a reduced model described below. This comparison evaluates the effect of specifying both models correctly, specifying either model correctly, and reducing both models.

For an estimand with true value $\psi_0$, let $\widehat\psi_b$ be its estimate in replication $b$ and let $\overline\psi$ be the average over the available replications. We report relative bias $100(\overline\psi-\psi_0)/\psi_0$ for the win ratio and win odds. For the net benefit, we report absolute bias $|\overline\psi-\psi_0|$ in both IRT and CRT. We also report the Monte Carlo standard deviation (MCSD), the average analytic standard error (ASE), and coverage (COV) of nominal $95\%$ normal Wald intervals. The ASE is the arithmetic mean of the estimated standard errors. For the win ratio and win odds, standard deviations, standard errors, and intervals are calculated on the logarithmic scale. The net benefit is evaluated on its natural scale. Both studies use exact equal allocation, and variance estimation centers the estimated influence contributions within treatment arms as described in Section \ref{sec:theory}.

\subsection{Individual-randomized trials}
\label{subsec:sim-irt}

We generate $n=500$ participants and randomly assign exactly $250$ to each treatment arm. The baseline covariates $V_i$ and $X_{2i}$ are independent $\operatorname{Unif}(-1,1)$ variables, and $X_{1i}$ independently takes values $-1$ and $1$ with equal probability. Conditional on these covariates, the latent hospitalization and death times under treatment $a$, denoted by $T_{H,i}^{a}$ and $T_{D,i}^{a}$, have Weibull proportional hazards margins, for $q\in\{H,D\}$,
\begin{equation}
S_{q,i}^{a}(t)=\exp\{-\kappa_qt^{\alpha_q}\exp(\ell_{q,i}^{a})\},
\label{eq:sim-irt-margins}
\end{equation}
where $(\kappa_H,\alpha_H)=(0.320,1.200)$ and $(\kappa_D,\alpha_D)=(0.110,1.350)$. The linear predictors are
\begin{align}
\ell_{H,i}^{a}
&=0.15V_i+1.25X_{1i}+0.55X_{2i}+0.80(X_{2i}^2-1) +a(-0.35+0.30X_{2i}),\nonumber\\
\ell_{D,i}^{a}
&=0.20V_i+1.25X_{1i}+0.60X_{2i}+0.80(X_{2i}^2-1)+a(-0.25+0.25X_{2i}).
\label{eq:sim-irt-event-predictors}
\end{align}
Thus, both event times depend strongly on the binary covariate, and the treatment effect varies with $X_{2i}$. Their conditional joint survival function is specified by the Gumbel survival copula,
\begin{equation}
\mathbb P(T_{H,i}^{a}>h,T_{D,i}^{a}>d\mid\bm L_i)=\exp\left[-\left\{[-\log S_{H,i}^{a}(h)]^{\theta_W}+[-\log S_{D,i}^{a}(d)]^{\theta_W}\right\}^{1/\theta_W}\right],
\label{eq:sim-irt-copula}
\end{equation}
where $\theta_W=(1-\tau_W)^{-1}$ and the conditional Kendall coefficient is $\tau_W\in\{0.450,0.700\}$. We set $D_i(a)=T_{D,i}^{a}$ and set $T_i^1(a)=T_{H,i}^{a}$ when $T_{H,i}^{a}<D_i(a)$ and $T_i^1(a)=\infty$ otherwise. This allows hospitalization followed by death and death without a preceding hospitalization. The event times are generated using positive stable variables and inverse Weibull margins, as detailed in Appendix \ref{subsec:app-irt-generation}. Censoring is independent of the event times conditional on treatment and baseline covariates, with survival function
\begin{align}
K_i^{a}(t)
&=\exp\{-\kappa_Ct^{0.650}\exp(\ell_{C,i}^{a})\},\nonumber\\
\ell_{C,i}^{a}&=0.05V_i+0.05X_{2i}+0.70(2a-1)X_{1i}+0.80(2a-1)(X_{2i}^2-1/3).
\label{eq:sim-irt-censoring}
\end{align}
Administrative follow-up ends at time $2$. For each dependence and censoring setting, $\kappa_C$ is calibrated from Monte Carlo such that $\mathbb P\{C_i<D_i\wedge2\}=c$, with $c\in\{0.300,0.600\}$. The realized censoring proportions range from $0.299$ to $0.601$. The details are reported in Table \ref{tab:irt-calibration}.  The correctly specified censoring model is a Cox regression with covariates $V_i$, $X_{1i}$, $X_{2i}$, and $X_{2i}^2-1$ and a Breslow estimate of the baseline cumulative hazard, fitted separately in each treatment arm. The correctly specified outcome model joins proportional hazards margins with the same covariates through a Gumbel copula. Its regression coefficients, nonnegative baseline hazard function, and association parameter are estimated jointly by maximizing the observed data likelihood in Appendix \ref{subsec:app-continuation}. The reduced censoring and outcome models retain $V_i$ alone. For each of the four combinations of $(\tau_W,c)$, we generate $1{,}000$ datasets and fit all four model specifications to each dataset. Every fitted estimator is evaluated at all three restriction times. For each dependence level, true values are calculated from four independent uncensored populations of $10^6$ participants. Table \ref{tab:irt-reference} reports these true values.

\begin{table}[htbp!]
\centering
\caption{Performance of the doubly robust win ratio estimator in individually randomized trials with $n=500$ and $\tau_W=0.450$, under $30\%$ and $60\%$ random censoring. Each trial assigns 250 participants to each arm. Death is prioritized over first hospitalization, with both indifference thresholds equal to zero. The conditional Kendall coefficient $\tau_W$ describes dependence between the event times within a participant, and $c$ is the probability of random censoring before death or time $2$. RB is signed relative bias. COV is coverage of nominal $95\%$ normal Wald intervals conditional on an available interval. The first model letter refers to censoring and the second to the outcome model; C denotes the aligned specification and M the reduced specification. The reduced models omit prognostic covariates, and the reduced outcome model also imposes independence between the event times.}
\label{tab:irt-main-wr}
\begingroup
\begin{tabular*}{\textwidth}{@{\extracolsep{\fill}}lrrrr@{\hspace{16pt}}rrrr@{}}
\toprule
& \multicolumn{4}{c}{$c=0.300$} & \multicolumn{4}{c}{$c=0.600$} \\
\cmidrule(lr){2-5}\cmidrule(l){6-9}
Model & RB (\%) & MCSD & ASE & COV & RB (\%) & MCSD & ASE & COV \\
\midrule
\multicolumn{9}{@{}l}{$\tau=1.0$, $\mathrm{WR}_0=1.278$} \\
$\mathrm{CC}$ & 1.213 & 0.170 & 0.167 & 0.940 & 2.850 & 0.212 & 0.211 & 0.954 \\
$\mathrm{CM}$ & 1.903 & 0.195 & 0.194 & 0.942 & 3.648 & 0.238 & 0.237 & 0.949 \\
$\mathrm{MC}$ & 2.017 & 0.170 & 0.167 & 0.945 & 6.010 & 0.206 & 0.207 & 0.948 \\
$\mathrm{MM}$ & 24.709 & 0.195 & 0.196 & 0.829 & 85.841 & 0.237 & 0.235 & 0.280 \\
\addlinespace[5pt]
\multicolumn{9}{@{}l}{$\tau=1.5$, $\mathrm{WR}_0=1.263$} \\
$\mathrm{CC}$ & 0.324 & 0.140 & 0.137 & 0.936 & 2.644 & 0.192 & 0.185 & 0.950 \\
$\mathrm{CM}$ & 1.063 & 0.166 & 0.167 & 0.945 & 3.764 & 0.220 & 0.215 & 0.951 \\
$\mathrm{MC}$ & 1.147 & 0.140 & 0.137 & 0.935 & 5.619 & 0.178 & 0.179 & 0.959 \\
$\mathrm{MM}$ & 30.296 & 0.169 & 0.171 & 0.713 & 111.736 & 0.212 & 0.212 & 0.059 \\
\addlinespace[5pt]
\multicolumn{9}{@{}l}{$\tau=2.0$, $\mathrm{WR}_0=1.245$} \\
$\mathrm{CC}$ & 0.279 & 0.123 & 0.121 & 0.943 & 2.218 & 0.187 & 0.174 & 0.928 \\
$\mathrm{CM}$ & 1.151 & 0.155 & 0.154 & 0.950 & 3.814 & 0.224 & 0.208 & 0.942 \\
$\mathrm{MC}$ & 1.063 & 0.123 & 0.121 & 0.939 & 5.254 & 0.168 & 0.166 & 0.942 \\
$\mathrm{MM}$ & 35.942 & 0.160 & 0.158 & 0.549 & 134.198 & 0.206 & 0.201 & 0.017 \\
\bottomrule
\end{tabular*}
\endgroup
\end{table}

Table \ref{tab:irt-main-wr} shows the win ratio results for $\tau_W=0.450$. With $30\%$ censoring, relative bias under $\mathrm{CC}$, $\mathrm{CM}$, and $\mathrm{MC}$ ranges from $0.279\%$ to $2.017\%$ over the three restriction times, the ASE/MCSD ratio ranges from $0.976$ to $1.008$, and coverage ranges from $0.935$ to $0.950$. At $\tau=2$, the MCSD is $0.123$ under both $\mathrm{CC}$ and $\mathrm{MC}$ and $0.155$ under $\mathrm{CM}$, showing the precision gained from the correctly specified outcome model. Under $\mathrm{MM}$, relative bias ranges from $24.709\%$ at $\tau=1$ to $35.942\%$ at $\tau=2$, with corresponding coverage of $0.829$ and $0.549$. With $60\%$ censoring, sampling variability increases. Relative bias remains below $4\%$ under $\mathrm{CC}$ and $\mathrm{CM}$ and ranges from $5.254\%$ to $6.010\%$ under $\mathrm{MC}$. Coverage across these three specifications ranges from $0.928$ to $0.959$. At $\tau=2$ under $\mathrm{CC}$, the ASE is $0.174$ compared with an MCSD of $0.187$. Under $\mathrm{MM}$, relative bias increases from $85.841\%$ to $134.198\%$ as the restriction time increases from $1$ to $2$, while coverage decreases from $0.280$ to $0.017$. %

Results for $\tau_W=0.700$, the win odds, and the net benefit are given in Appendix \ref{subsec:app-irt-supp-results}. Across both dependence levels, the absolute bias of the net benefit under $\mathrm{CC}$, $\mathrm{CM}$, and $\mathrm{MC}$ is at most $0.0034$ with $30\%$ censoring and $0.0098$ with $60\%$ censoring. Stronger event-time dependence has a comparatively modest effect on sampling variability. With $60\%$ censoring and $\tau_W=0.700$, coverage under $\mathrm{CC}$ at $\tau=2$ is $0.925$, with ASE $0.175$ and MCSD $0.192$. Figures \ref{fig:irt-wr-all}, \ref{fig:irt-wo-all}, and \ref{fig:irt-nb-all} summarize all IRT configurations.

\subsection{Cluster-randomized trials}
\label{subsec:sim-crt}

We generate parallel-arm CRTs with $m\in\{50,100\}$ independent clusters and assign exactly half to each treatment arm. Cluster size $N_i$ is discrete uniform on $\{5,\ldots,35\}$, with mean $20$, and $\ell_i=\log N_i-\mathbb E(\log N_i)$. We generate a cluster-level covariate $V_i=2\Phi(Z_{Vi})-1$, a binary participant covariate $X_{1ij}=2\mathbf 1(Z_{1ij}>0)-1$, and a continuous participant covariate $X_{2ij}=2\Phi(Z_{2ij})-1$, where the underlying $Z$ variables are independent standard normal variables and $\Phi$ is the standard normal distribution function. When cluster size is noninformative, indexed by $I_N=0$, it is independent of these covariates. When $I_N=1$, we set $Z_{Ni}=0.800Z_{Vi}+\sqrt{1-0.800^2}\,E_i$, with $E_i\sim N(0,1)$ independent of $Z_{Vi}$, and take $N_i=F_N^{-1}\{\Phi(Z_{Ni})\}$, where $F_N$ is the discrete uniform distribution function. This construction preserves the cluster-size distribution while associating larger clusters with larger values of $V_i$.

The latent hospitalization and death times have Weibull proportional hazards margins 
\begin{equation*}
S_{q,ij}^{a}(t)=\exp\{-\kappa_qt^{\alpha_q}\exp(\ell_{q,ij}^{a})\}
\end{equation*}
 for $q\in\{H,D\}$, with the same $(\kappa_q,\alpha_q)$ as in the IRT study. The linear predictors are
\begin{align*}
\ell_{H,ij}^{a}
&=0.15V_i+1.25X_{1ij}+0.55X_{2ij}+0.22(X_{2ij}^2-1)+0.50I_N\ell_i+a(-0.35+1.20I_N\ell_i+0.30X_{2ij}),\\
\ell_{D,ij}^{a}
&=0.20V_i+1.25X_{1ij}+0.60X_{2ij}+0.18(X_{2ij}^2-1)+0.55I_N\ell_i+a(-0.25+1.05I_N\ell_i+0.25X_{2ij}).
\end{align*}
Treatment effects vary with $X_{2ij}$ in every configuration and with cluster size when $I_N=1$. Dependence among the $2N_i$ latent event times is specified by a nested Gumbel survival copula. For $\theta\geq1$, write \mbox{$\mathcal G_\theta(u_1,\ldots,u_d)=\exp[-\{\sum_{q=1}^d(-\log u_q)^\theta\}^{1/\theta}]$}. With $\theta_B=(1-\tau_B)^{-1}$ and $\theta_W=(1-\tau_W)^{-1}$, the joint survival function is
\begin{align*}
&\mathbb P\{T_{H,ij}^{a}>h_j,T_{D,ij}^{a}>d_j, \ j=1,\ldots,N_i\mid\bm L_i\}\\
&\quad=\mathcal G_{\theta_B}\Bigl[\mathcal G_{\theta_W}\{S_{H,i1}^{a}(h_1),S_{D,i1}^{a}(d_1)\},\ldots,\mathcal G_{\theta_W}\{S_{H,iN_i}^{a}(h_{N_i}),S_{D,iN_i}^{a}(d_{N_i})\}\Bigr].
\end{align*}
We consider conditional Kendall coefficients $\tau_B\in\{0.100,0.300\}$ for event times from different participants in the same cluster and $\tau_W\in\{0.450,0.700\}$ for hospitalization and death in the same participant. These values satisfy $\tau_W\geq\tau_B$. As in the IRT study, $D_{ij}(a)=T_{D,ij}^{a}$, and hospitalization is assigned time $T_{H,ij}^{a}$ if it precedes death and $\infty$ otherwise. Figure \ref{fig:sim-dgp} illustrates the generating structure. Censoring times are generated independently across participants conditional on treatment and the full baseline cluster record and are independent of the event times given this information. Their conditional survival function is $K_{ij}^{a}(t)=\exp\{-\kappa_Ct^{0.650}\exp(\ell_{C,ij}^{a})\}$, where $\ell_{C,ij}^{a}=0.05V_i+0.05X_{2ij}+0.10I_N\ell_i+0.70(2a-1)X_{1ij}$. Administrative follow-up ends at time $2$. The scale is calibrated so that the marginal probability of censoring before death or time $2$ is $c\in\{0.300,0.600\}$. The calibrated values of $\kappa_C$ are $0.223$, $0.684$, $0.228$, and $0.705$ for $(I_N,c)=(0,0.300)$, $(0,0.600)$, $(1,0.300)$, and $(1,0.600)$, respectively.

The censoring and outcome models are fitted separately in each treatment arm. Correctly specified models include $V_i$, $X_{1ij}$, $X_{2ij}$, $X_{2ij}^2-1$, and $\ell_i$ in every configuration. The censoring model uses Cox regression with an estimated baseline cumulative hazard, and the outcome model uses proportional hazards margins with a Gumbel association parameter estimated within each arm. The reduced models retain $V_i$ alone, and the reduced outcome model imposes independence between hospitalization and death. The fitted models describe the event times of an individual participant conditional on baseline covariates. Dependence between participants is accounted for in the cluster-level variance calculation. Pairwise contributions are averaged or summed within cluster pairs according to \eqref{eq:theta-estimators}, and the individual-average estimator is normalized by the squared mean cluster size as in \eqref{eq:lambda-I-estimator}. Exact equal allocation uses the pair assignment probability $m_1m_0/\{m(m-1)\}$, with $m_1=m_0=m/2$. Analytic sandwich standard errors include nuisance model estimation and estimation of mean cluster size, as described in Section \ref{sec:theory}.

We generate $1{,}000$ replications per configuration, analyze each dataset under all four model specifications, and evaluate all three restriction times. For each configuration, true values are approximated using four independent uncensored populations of $100{,}000$ clusters, with $50{,}000$ clusters under each treatment condition. The reference calculations are described in Appendix \ref{subsec:app-crt-metrics}.

\begin{table}[htbp!]
\centering
\caption{Performance of finite-sample estimators for cluster-average and individual-average win ratio estimands ($\mathrm{WR}_C(\tau)$ and $\mathrm{WR}_I(\tau)$) in cluster-randomized trials with $m=50$ clusters, $(\tau_B,\tau_W)=(0.100,0.450)$, and $c=0.300$. The conditional Kendall coefficients $\tau_B$ and $\tau_W$ describe dependence between and within participants, respectively, and $c$ is the probability of censoring before death or time $2$. The indicator $I_N$ distinguishes noninformative ($0$) and informative ($1$) cluster size. RB is signed relative bias. The first model letter refers to censoring and the second to the outcome model; C denotes the aligned specification and M the reduced specification. The reduced models omit prognostic covariates, and the reduced outcome model also imposes independence between the event times. MCSD is the Monte Carlo standard deviation, ASE is the arithmetic average analytic standard error, and COV is the empirical coverage proportion of nominal 95\% normal Wald intervals.}
\label{tab:main-wr-reference}
\resizebox*{!}{0.7\textheight}{%
\begin{tabular}{@{}lrrrrrrrr@{}}
\toprule
& \multicolumn{4}{c}{$\mathrm{WR}_C(\tau)$} & \multicolumn{4}{c}{$\mathrm{WR}_I(\tau)$} \\
\cmidrule(lr){2-5}\cmidrule(l){6-9}
Model & RB (\%) & MCSD & ASE & COV & RB (\%) & MCSD & ASE & COV \\
\midrule
\multicolumn{9}{@{}l}{$I_N=0$} \\
\addlinespace[2pt]
$\tau=1.000$ & \multicolumn{4}{c}{$\mathrm{WR}_C=1.269$} & \multicolumn{4}{c}{$\mathrm{WR}_I=1.268$} \\
$\mathrm{CC}$ & 0.867 & 0.172 & 0.169 & 0.941 & 0.934 & 0.172 & 0.165 & 0.945 \\
$\mathrm{CM}$ & 1.453 & 0.189 & 0.188 & 0.950 & 1.263 & 0.187 & 0.179 & 0.944 \\
$\mathrm{MC}$ & 0.834 & 0.171 & 0.169 & 0.942 & 0.820 & 0.172 & 0.165 & 0.946 \\
$\mathrm{MM}$ & 23.378 & 0.192 & 0.191 & 0.811 & 23.184 & 0.190 & 0.182 & 0.782 \\
\addlinespace[2pt]
$\tau=1.500$ & \multicolumn{4}{c}{$\mathrm{WR}_C=1.243$} & \multicolumn{4}{c}{$\mathrm{WR}_I=1.242$} \\
$\mathrm{CC}$ & 0.494 & 0.147 & 0.143 & 0.948 & 0.623 & 0.145 & 0.140 & 0.956 \\
$\mathrm{CM}$ & 1.099 & 0.167 & 0.165 & 0.949 & 0.969 & 0.163 & 0.157 & 0.942 \\
$\mathrm{MC}$ & 0.434 & 0.147 & 0.143 & 0.949 & 0.463 & 0.145 & 0.140 & 0.952 \\
$\mathrm{MM}$ & 29.727 & 0.172 & 0.168 & 0.660 & 29.589 & 0.168 & 0.160 & 0.638 \\
\addlinespace[2pt]
$\tau=2.000$ & \multicolumn{4}{c}{$\mathrm{WR}_C=1.222$} & \multicolumn{4}{c}{$\mathrm{WR}_I=1.220$} \\
$\mathrm{CC}$ & 0.349 & 0.136 & 0.131 & 0.944 & 0.494 & 0.135 & 0.128 & 0.941 \\
$\mathrm{CM}$ & 0.939 & 0.155 & 0.154 & 0.945 & 0.823 & 0.152 & 0.146 & 0.945 \\
$\mathrm{MC}$ & 0.256 & 0.136 & 0.130 & 0.945 & 0.272 & 0.134 & 0.128 & 0.945 \\
$\mathrm{MM}$ & 34.819 & 0.161 & 0.158 & 0.534 & 34.661 & 0.158 & 0.150 & 0.514 \\
\addlinespace[5pt]
\multicolumn{9}{@{}l}{$I_N=1$} \\
\addlinespace[2pt]
$\tau=1.000$ & \multicolumn{4}{c}{$\mathrm{WR}_C=1.133$} & \multicolumn{4}{c}{$\mathrm{WR}_I=0.979$} \\
$\mathrm{CC}$ & 1.270 & 0.163 & 0.160 & 0.949 & 1.374 & 0.159 & 0.148 & 0.925 \\
$\mathrm{CM}$ & 1.163 & 0.192 & 0.190 & 0.945 & 1.414 & 0.184 & 0.176 & 0.930 \\
$\mathrm{MC}$ & 1.345 & 0.163 & 0.160 & 0.947 & 1.339 & 0.159 & 0.148 & 0.924 \\
$\mathrm{MM}$ & 22.958 & 0.193 & 0.191 & 0.823 & 23.756 & 0.186 & 0.177 & 0.774 \\
\addlinespace[2pt]
$\tau=1.500$ & \multicolumn{4}{c}{$\mathrm{WR}_C=1.152$} & \multicolumn{4}{c}{$\mathrm{WR}_I=0.995$} \\
$\mathrm{CC}$ & 1.001 & 0.145 & 0.143 & 0.945 & 1.114 & 0.140 & 0.131 & 0.931 \\
$\mathrm{CM}$ & 0.936 & 0.173 & 0.173 & 0.944 & 1.178 & 0.166 & 0.161 & 0.933 \\
$\mathrm{MC}$ & 1.058 & 0.145 & 0.142 & 0.942 & 1.030 & 0.140 & 0.131 & 0.927 \\
$\mathrm{MM}$ & 28.241 & 0.175 & 0.176 & 0.732 & 29.027 & 0.168 & 0.163 & 0.657 \\
\addlinespace[2pt]
$\tau=2.000$ & \multicolumn{4}{c}{$\mathrm{WR}_C=1.156$} & \multicolumn{4}{c}{$\mathrm{WR}_I=0.998$} \\
$\mathrm{CC}$ & 0.841 & 0.136 & 0.134 & 0.933 & 0.953 & 0.132 & 0.124 & 0.936 \\
$\mathrm{CM}$ & 0.840 & 0.165 & 0.165 & 0.942 & 1.030 & 0.159 & 0.154 & 0.931 \\
$\mathrm{MC}$ & 0.842 & 0.136 & 0.134 & 0.938 & 0.785 & 0.132 & 0.124 & 0.937 \\
$\mathrm{MM}$ & 32.145 & 0.168 & 0.168 & 0.639 & 32.718 & 0.162 & 0.157 & 0.562 \\
\bottomrule
\end{tabular}%
}
\end{table}

Table \ref{tab:main-wr-reference} reports win ratio results for $m=50$, $(\tau_B,\tau_W)=(0.100,0.450)$, and $30\%$ censoring. Under noninformative cluster size, the cluster-average and individual-average estimands coincide and the two estimators perform similarly. Under $\mathrm{CC}$, relative bias is below $1\%$ at every restriction time and coverage ranges from $0.941$ to $0.956$. Under $\mathrm{CM}$ and $\mathrm{MC}$, relative bias is at most $1.5\%$ and coverage ranges from $0.942$ to $0.952$. At $\tau=2$, the MCSD for the cluster-average win ratio is $0.155$ under $\mathrm{CM}$ and $0.136$ under $\mathrm{MC}$. With informative cluster size, relative bias under $\mathrm{CC}$, $\mathrm{CM}$, and $\mathrm{MC}$ ranges from $0.785\%$ to $1.414\%$ for the two targets, with coverage between $0.924$ and $0.949$.  In the informative cluster-size setting, the treatment hazard ratios for hospitalization and death increase with cluster size, giving less favorable treatment effects on these event hazards in larger clusters. The individual-average estimand gives these larger clusters greater weight than the cluster-average estimand. At $\tau=2$ in the reference configuration, the cluster-average and individual-average win ratios are $1.156$ and $0.998$, respectively. Thus, the comparison favors treatment when clusters receive equal weight and is approximately null when participants receive equal weight. At $60\%$ censoring in the corresponding configuration, relative bias under $\mathrm{CC}$, $\mathrm{CM}$, and $\mathrm{MC}$ ranges from $0.576\%$ to $3.711\%$ and coverage from $0.923$ to $0.953$ across both cluster-size settings; the complete results are in Table \ref{tab:main-wr-c060}. Sampling variability increases with censoring. Under $\mathrm{MM}$ with noninformative cluster size, relative bias for the cluster-average win ratio ranges from about $23\%$ to $35\%$ across restriction times at $30\%$ censoring and from $85.039\%$ to $136.033\%$ at $60\%$ censoring. With informative cluster size and $\tau=2$, relative bias at $30\%$ censoring is $32.145\%$ for the cluster-average win ratio and $32.718\%$ for the individual-average win ratio, with coverage of $0.639$ and $0.562$, respectively. 

Across all configurations, increasing the number of clusters from $50$ to $100$ reduces MCSD and ASE by approximately $30\%$. Greater dependence between participants increases sampling variability, while the effect of changing dependence between the two event times of the same participant is smaller. Under $\mathrm{CC}$, $\mathrm{CM}$, and $\mathrm{MC}$, relative bias for the win ratio ranges from $-0.245\%$ to $6.079\%$, with median coverage of $0.941$, $0.945$, and $0.941$, respectively. Relative bias for the win odds ranges from $-0.291\%$ to $2.764\%$, and absolute bias for the net benefit is at most $0.0080$. Appendix \ref{subsec:app-crt-results} and Figures \ref{fig:fig_appendix_01_wr_m50_loopstyle}--\ref{fig:fig_appendix_06_nb_m100_loopstyle} report all configurations and the additional win summaries.

\section{Illustrative data applications}
\label{sec:application}

We illustrate the proposed methods with two completed trials. The HF-ACTION trial is an IRT with a prioritized composite of death and heart failure hospitalization, and the STRIDE trial is a pragmatic CRT with a prioritized composite of death and fall injury in which the practice sizes vary widely. The two applications illustrate the individual-level and cluster-level methods in turn, and in each case we report the win ratio together with the estimated win, loss, and tie probabilities so that the results can be interpreted on the scale of the underlying pairwise comparisons.

\subsection{Heart Failure: A Controlled Trial Investigating Outcomes of Exercise Training}
\label{subsec:hf-action}

The HF-ACTION trial (Heart Failure: A Controlled Trial Investigating Outcomes of Exercise Training) randomized 2,331 patients with heart failure and left ventricular systolic dysfunction to exercise training plus usual care or to usual care alone, where the exercise program consisted of 36 supervised sessions followed by home based training.\cite{oconnor2009hfaction} Our analysis uses the 2,130 participants, comprising 1,060 assigned to exercise training and 1,070 to usual care. The mean age was 58.6 years, and 28.1\% of participants had recorded female sex.

We consider a prioritized composite endpoint in which death from any cause is compared first and first heart failure (HF) hospitalization second. This composite differs from the primary endpoint of the original trial, which was the time to the first of death from any cause and hospitalization from any cause. The HF hospitalization component was reconstructed from the clinical events committee records together with the HF composite information in the release when its nonfatal component could be identified, so that it is an HF specific endpoint and is neither an all cause hospitalization endpoint nor a committee adjudicated endpoint alone. Later death is preferred, later first HF hospitalization is preferred only when the restricted death times are equal, and both indifference thresholds are zero. A first HF hospitalization does not terminate the follow-up for death. Two years is the principal restriction time, and the win statistics are also evaluated at every distinct observed event time through four years. Before two years, 103 deaths and 204 first HF hospitalizations were observed in the exercise group, compared with 127 deaths and 221 first HF hospitalizations in the usual care group, and observation ended before resolution for 268 exercise participants (25.3\%) and 247 usual care participants (23.1\%), with a median observation time of 2.54 years. The reconstruction of the event histories and additional descriptive summaries are given in \ref{sec:app-hf-action} and Table \ref{tab:hf-descriptive}.

We apply the proposed doubly robust estimator to the win ratio $\mathrm{WR}(\tau)$, with values above one favoring exercise training, and compare it with the IPCW and outcome regression estimators of Section \ref{subsec:irt-estimation} and with an unadjusted estimator that uses only the pairs in which both members are resolved. The unadjusted estimator differs from the conventional win ratio, which compares each pair over its common observed follow-up. The outcome and censoring model specifications, covariates, and treatment of missing baseline covariates are given in \ref{subsec:hf-models}. Pointwise 95\% confidence intervals are constructed on the logarithmic scale from the analytic standard errors of Theorem \ref{thm:irt-inference} and exponentiated.

\begin{figure}[!htbp]
\centering
\includegraphics[width=0.55\textwidth]{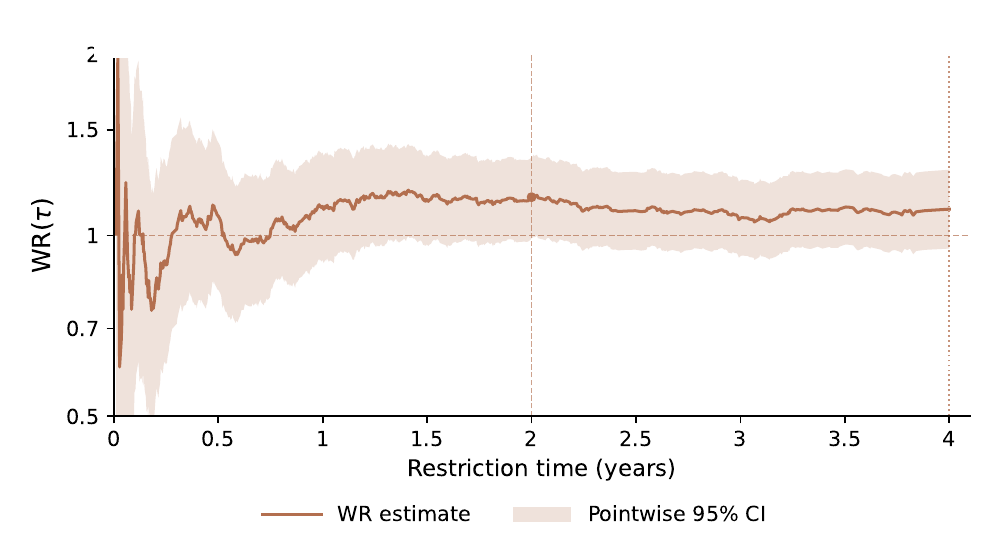}
\caption{Doubly robust win ratio trajectory for HF-ACTION with death from any cause prioritized over first heart failure (HF) hospitalization. The comparison is exercise training versus usual care among 2,130 released participants, with both indifference thresholds equal to zero. The solid curve is the point estimate under the Gumbel outcome model at each of the 626 distinct observed event times and at exactly two and four years, and the shaded region shows pointwise 95\% confidence intervals constructed on the logarithmic scale and exponentiated. The horizontal dashed line marks the null value of one, the dashed vertical line marks the two year restriction time, and the dotted vertical line marks four years.}
\label{fig:hf-wr}
\end{figure}

Figure \ref{fig:hf-wr} shows the doubly robust win ratio trajectory over follow-up. At two years, the doubly robust win ratio is 1.158 (95\% CI 0.991 to 1.354, $p=0.065$), and the IPCW, outcome regression, and unadjusted estimates are 1.171 (95\% CI 0.978 to 1.402), 1.141 (95\% CI 0.974 to 1.337), and 1.160 (95\% CI 0.967 to 1.392), respectively. All point estimates favor exercise training and all confidence intervals include one, so that the results are compatible with a modest benefit of exercise training under the specified hierarchy without establishing it. The agreement among the point estimates does not by itself establish the correct specification of either nuisance model. The doubly robust estimator yields a standard error of 0.080 for the log win ratio, compared with 0.092 for IPCW, a reduction of 13.3\% that is consistent with the efficiency gain from the outcome model observed in the simulation studies. The estimated win and loss probabilities are 0.254 and 0.219, with a tie probability of 0.527 (Table \ref{tab:hf-probabilities}). A win ratio of 1.158 therefore indicates that the win probability exceeds the loss probability by 15.8\% in relative terms and by 3.47 percentage points (95\% CI $-0.21$ to 7.15) on the net benefit scale, and it should not be interpreted as a 15.8\% reduction in the risk of death or hospitalization. At two years, the win odds are 1.072 (95\% CI 0.996 to 1.154), closer to one than the win ratio because half of the tie probability is allocated to each arm, and the intervals for the win odds and the net benefit also include their null values (Table \ref{tab:hf-probabilities}). The three summaries are complementary descriptions of the same hierarchical comparison and not independent evidence of benefit. The corresponding win odds and net benefit trajectories are given in \ref{subsec:hf-results-details} and Figures \ref{fig:hf-wo} and \ref{fig:hf-nb}.

\subsection{Strategies to Reduce Injuries and Develop Confidence in Elders: A cluster-randomized trial}
\label{subsec:stride}

The Strategies to Reduce Injuries and Develop Confidence in Elders (STRIDE) study was a pragmatic CRT of a multifactorial fall injury prevention strategy among community dwelling adults aged 70 years or older at increased risk of fall injury.\cite{bhasin2018stridedesign,bhasin2020stride} Practices were assigned using covariate-constrained randomization stratified by health care system.\cite{bhasin2018stridedesign} The analysis dataset contains 5,451 participants from 86 primary care practices in 10 health care systems, with 2,802 participants in the 43 intervention practices and 2,649 participants in the 43 control practices. The mean age was 79.7 years, and 62.0\% of participants had recorded female sex. The enrolled practice sizes ranged from 10 to 199, with a mean of 63.4 and a coefficient of variation of 0.517. This variation makes the distinction between the cluster-average and individual-average estimands relevant, although unequal practice sizes alone do not establish informative cluster size.

\begin{figure}[!htbp]
\centering
\includegraphics[width=0.8\textwidth]{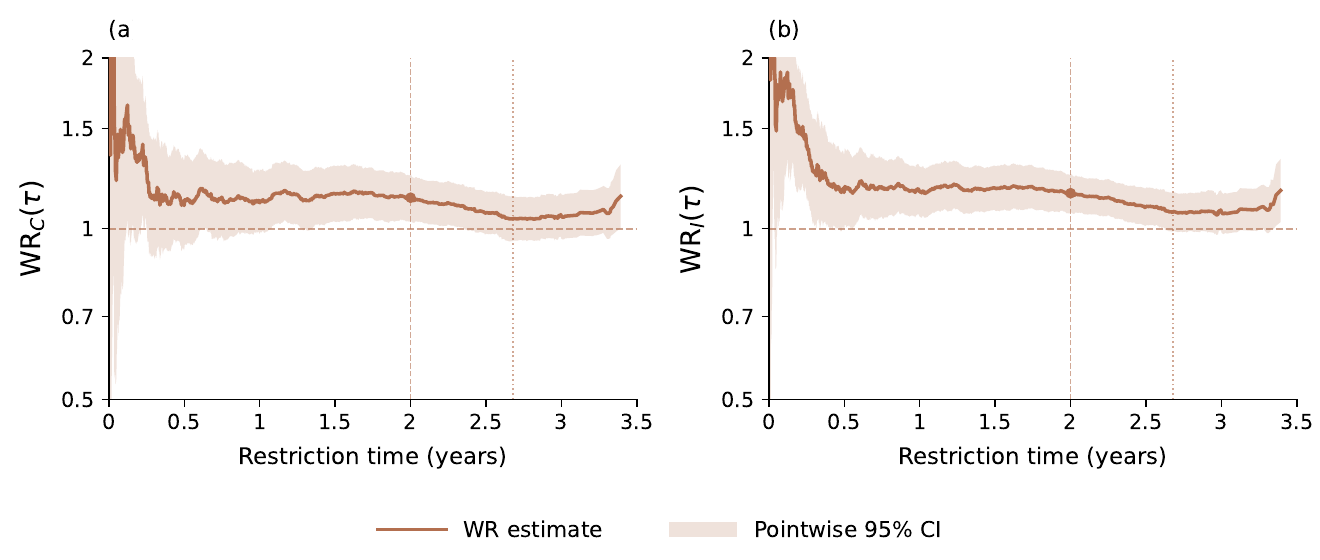}
\caption{Win ratio trajectories for STRIDE with death prioritized over first self-reported fall injury. The comparison is intervention versus enhanced usual care among 5,451 participants from 86 practices, with both indifference thresholds equal to zero. Panel (a) shows the cluster-average win ratio $\mathrm{WR}_C(\tau)$ and panel (b) shows the individual-average win ratio $\mathrm{WR}_I(\tau)$. Solid curves are the doubly robust estimates under the Gumbel outcome model, and shaded regions show pointwise 95\% confidence intervals.}
\label{fig:stride-wr-self-reported}
\end{figure}

We consider a prioritized composite endpoint in which death is compared first and the time to first self-reported fall injury second. Later death is preferred, later first injury is preferred only when the restricted death times are equal, and both indifference thresholds are zero. A death occurring after an injury therefore remains the outcome of highest priority. Injuries recorded at the time of death are not counted as preceding nonfatal events. Time is measured in years from enrollment, two years is the prespecified principal restriction time, and the win statistics are also evaluated at every distinct observed death or first injury time. The median observation time was 2.35 years. Before two years, 75 deaths and 1,004 first self-reported injuries occurred in the intervention group, compared with 72 deaths and 1,090 first self-reported injuries in the control group. Observation ended before resolution for 600 intervention participants (21.4\%) and 572 control participants (21.6\%). Additional participant characteristics and event rates are given in Table \ref{tab:stride-descriptive}.

We apply the proposed doubly robust estimators to the cluster-average and individual-average win ratios, $\mathrm{WR}_C(\tau)$ and $\mathrm{WR}_I(\tau)$, with values above one favoring the intervention. The nuisance models are fitted separately by arm. The outcome model joins proportional hazards margins for self-reported injury and death through a Gumbel survival copula, and the censoring model is a Cox model, both adjusting for health care system, predominantly White practice population, prerandomization practice size tertile, age, recorded sex, number of chronic conditions, recorded race, and surrogate consent. Pointwise 95\% confidence intervals are constructed on the logarithmic scale from the cluster-level analytic covariance, centered within health care system and treatment arm, and exponentiated. Sensitivity analyses replace the Gumbel copula by an independence copula, use a $t$ critical value, and center the covariance within arms. Further details of the fitted models are given in \ref{sec:app-stride}.

Figure \ref{fig:stride-wr-self-reported} shows the doubly robust win ratio trajectories for death prioritized over first self-reported fall injury. At two years, the cluster-average win ratio is 1.134 (95\% CI 1.033 to 1.244) and the individual-average win ratio is 1.155 (95\% CI 1.069 to 1.247). Under both target populations, the estimated probability of an intervention win exceeds the estimated probability of an intervention loss, with pointwise intervals excluding the null value of one. Although the two targets yield similar conclusions at two years, their estimated trajectories differ in magnitude, with the individual-average win ratio generally exceeding the cluster-average win ratio. This pattern is consistent with the STRIDE analysis of Fang et al.\cite{fang2026doubly}, who reported larger individual-average than cluster-average effects for restricted mean time in favor of treatment. These findings illustrate the relevance of distinguishing the two target populations when practice size may be associated with prognosis or treatment response. The estimated tie probability is approximately 32\% under both target populations. The corresponding net benefits are 4.26 (95\% CI 1.15 to 7.37) and 4.90 (95\% CI 2.31 to 7.48) percentage points. The estimated win, loss, and tie probabilities and the additional win summaries are reported in Tables \ref{tab:stride-probabilities} and \ref{tab:stride-wo-nb}. Figures \ref{fig:stride-wo-self-reported} and \ref{fig:stride-nb-self-reported} show the win odds and net benefit trajectories. 

\section{Discussion}
\label{sec:discussion}

We have developed doubly robust estimators of win estimands for prioritized time-to-event outcomes in IRTs and CRTs. The framework defines treatment comparisons through a prespecified clinical hierarchy and restriction time, and uses the available event histories to estimate the corresponding win, loss, and tie probabilities under covariate dependent censoring. Correcting for censoring of both participants in each pair is central to this construction: the joint outcome models predict how partially observed comparisons would be resolved, and the censoring models account for incomplete follow-up. Under the stated assumptions, consistency holds when either set of models is correctly specified for both treatment arms. The analytic variance estimators retain the contribution of fitting these models and account for dependence among comparisons that share a participant or cluster. This provides a computationally convenient alternative to the jackknife calculations used in our previous work on survival and RMT-IF estimands.\cite{fang2026estimands,fang2026doubly} To support practical application, the proposed methods are implemented in the \texttt{DRwinStats} R package, available at \url{https://github.com/fancy575/DRwinStats}. To accelerate likelihood evaluations and pairwise calculations, the package uses compiled C++ routines accessed through R's C interface. A common formula interface accommodates IRTs as a special case of CRTs (where the cluster size is treated as one), with Cox and Kaplan-Meier options and several survival copulas. The \texttt{summary} and \texttt{plot} methods provide the estimated probabilities and win statistics with confidence intervals, and the fitted object also reports censoring rates at the selected restriction times. The example datasets and tutorial in Appendix \ref{sec:app-package} demonstrate the complete analysis for both designs.

In CRTs, the choice between cluster-pair and individual-pair win estimands should be guided by the scientific question and target population. By definition, the cluster-pair estimand represents comparisons between participants drawn from equally weighted practices or communities, whereas the individual-pair estimand represents comparisons between participants drawn from the population pooled across clusters, giving larger clusters greater weight. The former is relevant when the objective is to assess an intervention across settings regarded as equally important; the latter is relevant when interest centers on its consequences for the population of enrolled individuals, including when cluster randomization is primarily a practical design choice. This distinction concerns population weighting, separately from the calculation of uncertainty due to correlated observations. Lee et al.\ \cite{lee2026winning} examined these targets for win comparisons without right censoring and showed that informative cluster size can alter both the magnitude and direction of the treatment comparison. The two targets coincide when cluster size is non-informative for the potential outcome distributions. Unlike average treatment effects on a difference scale, however, their divergence does not require treatment effect heterogeneity by cluster size: changing the population weights changes the outcome distributions being compared, so differences in prognosis associated with cluster size can also matter.\cite{lee2026winning} Our work extends these considerations to censored hierarchical survival outcomes while preserving the chosen population weights. As in our previous survival estimators, cluster size may enter a working model to explain prognosis or follow-up, while the weights used in the population averages determine the target estimand.\cite{fang2026estimands} These considerations are particularly relevant to pragmatic interventions delivered across practices with different enrollment sizes and patient populations. Our informative cluster size simulations illustrate how assigning greater weight to larger practices with different treatment responses can change the treatment comparison. Investigators should therefore prespecify the primary target and consider reporting both estimands when the perspectives of equally weighted practices and individual participants address complementary clinical or policy questions.

The relationship with several previous doubly robust methods concerns both the clinical meaning of the estimand and the information needed to estimate it. For a single survival outcome, earlier work targets contrasts of marginal survival probabilities and restricted mean survival times, describing the probability of remaining event free and the expected event free time through a specified horizon.\cite{fang2026estimands} RMT-IF extends the latter interpretation to progressive multistate outcomes by averaging the time during which a treated participant occupies a more favorable health state than an independently sampled control participant, minus the corresponding time in favor of control. With only alive and dead states, it reduces to the restricted mean survival time difference.\cite{fang2026doubly} With death as the only endpoint, a win probability describes the chance that a treated participant has a longer restricted survival time than an independently sampled control participant. RMT-IF expresses the comparison as a difference in mean restricted survival times. The hierarchical win estimand assigns a win, loss, or tie from the restricted event times according to the prespecified priority order and clinical thresholds. RMT-IF therefore reflects the duration of differences in health states throughout follow-up, including periods before a later death determines the hierarchical comparison. For recurrent events terminated by death, while-alive estimand is the expected clinically weighted number of recurrent events divided by restricted mean survival time.\cite{fang2026whilealive} It describes event burden per unit of population time alive, and reporting its two components clarifies how changes in recurrent morbidity and survival contribute to the treatment contrast. Numerical weights combine event types in this while-alive summary, whereas the clinical priority order determines which outcome resolves a hierarchical win comparison. These differences make the estimands complementary descriptions of treatment benefit. Across these methods, double robustness follows from combining censoring adjustment with predictions of the outcome quantities required by the estimand. The RMT-IF representation uses marginal survival curves for reaching the ordered severity levels, while the while-alive construction combines terminal survival with mean recurrent event increments.\cite{fang2026doubly,fang2026whilealive} The present outcome regression uses joint models for the component event times to calculate the conditional probability of a win, loss, or tie given the observed history. Familiar regression methods remain useful in each setting but are considered as working models that assist the inference about the nonparametric estimands. The survival and RMT-IF estimators use treatment-specific survival predictions from Cox or frailty models, and the while-alive estimator additionally uses proportional rates models for recurrent events.\cite{fang2026estimands,fang2026doubly,fang2026whilealive} The fitted predictions enter censoring-adjusted estimates of the marginal outcome distributions or means, from which the prespecified clinical summaries are calculated. Regression coefficients and hazard ratios describe the working models, however, survival contrasts, time in favor, while-alive rates, and win statistics describe the selected clinical comparisons and provide causal interpretation. This distinction echoes the ICH E9(R1) Estimands Framework, in which the clinical objective determines the treatment conditions, population, outcome, handling of intercurrent events, and summary measure before the statistical model or estimator is chosen.\cite{ich2019e9r1} In particular, incorporating death into a prioritized composite and assessing recurrent event burden during survival represent different choices about the clinical role of death. The proposed methods provide a way to estimate the former comparison while retaining familiar survival models for covariate adjustment and censoring adjustment.

\section*{Acknowledgement}
Research in this article was supported by the National Heart, Lung, and Blood Institute under Award Number 1R01HL178513. The statements presented in this article are solely the responsibility of the authors and do not necessarily represent the official views of the National Institutes of Health.  The illustrative analysis of Heart Failure: A Controlled Trial Investigating Outcomes of Exercise Training (HF-ACTION) in this manuscript was prepared using HF-ACTION Research Materials obtained from the NHLBI Biologic Specimen and Data Repository Information Coordinating Center and does not necessarily reflect the opinions or views of HF-ACTION or the NHLBI. The statements presented in this article are solely the responsibility of the authors and do not necessarily represent the official views of the National Institutes of Health.

\section*{Supplementary Material}
The Appendix provides the supporting methodological, theoretical, and numerical details for the main paper. It contains detailed developments for individual-randomized and cluster-randomized trials, including proofs of identification and double robustness, derivations of the influence functions, regularity conditions, proofs of asymptotic normality and variance consistency, and results for fixed treatment allocation. It also includes the working censoring and joint outcome models, and the observed-data likelihood contributions. Additional sections present the complete simulation designs and supplementary results, further details and sensitivity analyses for the HF-ACTION and STRIDE applications, and a step-by-step tutorial for the \texttt{DRwinStats} R package. 

\section*{Data Availability Statement}

The HF-ACTION study was prepared using the HF-ACTION Research Materials obtained from the NHLBI Biologic Specimen and Data Repository Information Coordinating Center (BioLINCC) and does not necessarily reflect the opinions or views of HF-ACTION or NHLBI. The STRIDE study was funded primarily by the Patient Centered Outcomes Research Institute (PCORI\textsuperscript{\textregistered}), with additional support from the National Institute on Aging (NIA) at NIH. The STRIDE data can be obtained via \url{https://agingresearchbiobank.nia.nih.gov/studies/}.


\clearpage
\appendix
\setcounter{equation}{0}
\setcounter{theorem}{0}
\setcounter{lemma}{0}
\setcounter{proposition}{0}
\setcounter{corollary}{0}
\setcounter{assumption}{0}
\setcounter{condition}{0}
\setcounter{table}{0}
\setcounter{figure}{0}
\renewcommand{\theequation}{S\arabic{equation}}
\renewcommand{\thetheorem}{S\arabic{theorem}}
\renewcommand{\thelemma}{S\arabic{lemma}}
\renewcommand{\theproposition}{S\arabic{proposition}}
\renewcommand{\thecorollary}{S\arabic{corollary}}
\renewcommand{\theassumption}{S\arabic{assumption}}
\renewcommand{\thecondition}{S\arabic{condition}}
\renewcommand{\thetable}{S\arabic{table}}
\renewcommand{\thefigure}{S\arabic{figure}}
\renewcommand{\theHequation}{appendix.\arabic{equation}}
\renewcommand{\theHtheorem}{appendix.\arabic{theorem}}
\renewcommand{\theHlemma}{appendix.\arabic{lemma}}
\renewcommand{\theHproposition}{appendix.\arabic{proposition}}
\renewcommand{\theHcorollary}{appendix.\arabic{corollary}}
\renewcommand{\theHassumption}{appendix.\arabic{assumption}}
\renewcommand{\theHcondition}{appendix.\arabic{condition}}
\renewcommand{\theHtable}{appendix.\arabic{table}}
\renewcommand{\theHfigure}{appendix.\arabic{figure}}
\makeatletter
\def\section@cntformat{Appendix \thesection\quad}
\def\@seccntformat#1{\ifcsname #1@cntformat\endcsname
  \csname #1@cntformat\endcsname
  \else\csname the#1\endcsname\quad\fi}
\makeatother
\etocdepthtag.toc{appendix}
\pdfbookmark[0]{Appendix}{appendix-start}
\begin{center}
{\Large\bfseries Appendix\par}
\medskip
{\bfseries Supplementary material for\par}
\begin{minipage}{0.92\textwidth}
\centering\bfseries Doubly robust win statistics for hierarchical survival endpoints in individual- and cluster-randomized trials\par
\end{minipage}
\end{center}
Appendices A and B provide identification and double robustness proofs for individual- and cluster-randomized trials. Appendix C derives the influence functions, efficiency bounds, asymptotic distributions, and variance estimators. Appendix D collects the working models, Appendices E and F give supplementary simulation and application results, and Appendix G illustrates the DRwinStats package.

\etocsettagdepth{main}{none}
\etocsettagdepth{appendix}{subsection}
\etocsettocstyle{\section*{Appendix contents}}{}
\tableofcontents
\clearpage

\section{Individual-randomized trials}
\label{sec:app-irt-proof}

The notation and assumptions are those of Section \ref{sec:irt}. For the comparison indicators in Section \ref{subsec:irt-estimands}, $w_q(\bm y,\bm y')+w_q(\bm y',\bm y)=1-e_q(\bm y,\bm y')$. With $J_{Q+1}=h_{=}$ and $J_{q+1}=J_qe_q$,
\begin{align}
h_++h_-
&=\sum_{q=0}^QJ_q(1-e_q)
=\sum_{q=0}^Q(J_q-J_{q+1})
=1-h_{=}.
\label{eq:app-partition}
\end{align}
Only the first tier exceeding its indifference threshold contributes a win or loss. Thus $h_+$, $h_-$, and $h_{=}$ partition the possible comparisons. Taking expectations gives $\lambda^++\lambda^-+\lambda^{=}=1$ and $\lambda^+=\sum_{q=0}^Q\lambda^{+,q}$. Interchanging the participants gives $\lambda_{v,a}^-=\lambda_{v,1-a}^+$ for either CRT target. Substitution of the partition identity into the win odds gives $\mathrm{WO}=(1+\mathrm{NB})/(1-\mathrm{NB})$.

\subsection{Identification}
\label{subsec:app-irt-estimands}
\label{sec:hierarchy-identities}
\label{subsec:app-irt-identification}

\begin{proof}[Proof of Theorem \ref{thm:irt-identification}]
Assumption \ref{ass:irt-car} and the definition of $R_i$ imply
\begin{align}
\E\left\{\frac{\Delta_i^R}{K_{0,i}(R_i)}
\,\middle|\,\bm L_i,A_i,\bm Y_i\right\}
&=\frac{\Prb(C_i\geq R_i\mid\bm L_i,A_i,\bm Y_i)}{K_{0,i}(R_i)}\nonumber\\
&=\frac{\left.\Prb(C_i\geq t\mid\bm L_i,A_i)\right|_{t=R_i}}{K_{0,i}(R_i)}
=1.
\label{eq:app-irt-resolution-mean}
\end{align}
Independence of participant records permits separate integration over their censoring times. Consistency and randomization then give
\begin{align}
&\E\left[\frac{\ind(A_i=1,A_k=0)}{\pi_1(\bm L_i)\pi_0(\bm L_k)}
\frac{\Delta_i^R\Delta_k^R h_r(\bm Y_i,\bm Y_k)}
{K_{0,i}(R_i)K_{0,k}(R_k)}
\,\middle|\,\bm L_i,\bm L_k\right]\nonumber\\
&\quad=\frac{\Prb(A_i=1,A_k=0\mid\bm L_i,\bm L_k)}{\pi_1(\bm L_i)\pi_0(\bm L_k)}
\E\left[h_r(\bm Y_i,\bm Y_k)
\prod_{g\in\{i,k\}}\E\left\{\frac{\Delta_g^R}{K_{0,g}(R_g)}\,\middle|\,\bm L_g,A_g,\bm Y_g\right\}
\,\middle|\,\bm L_i,\bm L_k,A_i=1,A_k=0\right]\nonumber\\
&\quad=\E\{h_r(\bm Y_i,\bm Y_k)\mid\bm L_i,\bm L_k,A_i=1,A_k=0\}\nonumber\\
&\quad=\E[h_r\{\bm Y_i(1),\bm Y_k(0)\}\mid\bm L_i,\bm L_k]
=\mu_0^r(\bm L_i,\bm L_k).
\label{eq:app-irt-observed-pair-mean}
\end{align}
Averaging over $(\bm L_i,\bm L_k)$ gives $\lambda^r$. The same calculation with $h_{+,q}$ proves the tier-specific representation. No independence among the event components of one participant is required. Under fixed treatment totals, the joint assignment probability in Appendix \ref{subsec:app-irt-fixed} replaces the product of assignment probabilities.
\end{proof}

\subsection{Censoring adjustment}
\label{subsec:app-irt-estimation}
\label{subsec:app-irt-one-member}

Omit the participant index and let $B$ be an integrable function of the restricted outcome and baseline covariates, determined at $R$. Let $\Lambda^C$ be a continuous predictable working cumulative censoring hazard, and assume that $K=\exp(-\Lambda^C)$ is bounded away from zero. Since $d(K^{-1})=K^{-1}d\Lambda^C$,
\begin{align}
\int_0^U\frac{dM^C(t)}{K(t)}
&=\frac{1-\Delta^R}{K(U)}-\int_0^U\frac{d\Lambda^C(t)}{K(t)}\nonumber\\
&=\frac{1-\Delta^R}{K(U)}-\{K(U)^{-1}-1\}
=1-\frac{\Delta^R}{K(R)}.
\label{eq:app-irt-constant}
\end{align}
Here $U=R$ when $\Delta^R=1$; products with $\Delta^R=0$ follow the convention in the main paper. Hence
\begin{align}
\mathcal A(B;K,q)-B
&=\left\{\frac{\Delta^R}{K(R)}-1\right\}B
+\int_0^U\frac{q(t)}{K(t)}\,dM^C(t)\nonumber\\
&=\int_0^U\frac{q(t)-B}{K(t)}\,dM^C(t).
\label{eq:app-irt-path}
\end{align}
Under Assumption \ref{ass:irt-car}, $M_0^C$ remains a martingale when the complete event trajectory is included at time zero. In that filtration, $B$ is measurable at time zero and the integrand in \eqref{eq:app-irt-path} is predictable. Assume finite absolute expectation of the integrals below. This holds, for example, under bounded inverse weights and cumulative hazards and square-integrable outcome contributions. Substituting $dM^C=dM_0^C+Y^C(d\Lambda_0^C-d\Lambda^C)$ and conditioning on the observed history gives
\begin{align}
\E\{\mathcal A(B;K,q)\}
&=\E(B)+\E\int_0^U\frac{q(t)-B}{K(t)}\,dM^C(t)\nonumber\\
&=\E(B)+\E\int_0^\tau\frac{Y^C(t)\{q(t)-B\}}{K(t)}
\{d\Lambda_0^C(t)-d\Lambda^C(t)\}\nonumber\\
&=\E(B)+\E\int_0^\tau\frac{Y^C(t)\{q(t)-q_0(t)\}}{K(t)}
\{d\Lambda_0^C(t)-d\Lambda^C(t)\},
\label{eq:app-irt-drift}
\end{align}
where $q_0(t)=\E(B\mid\mathcal F_{t^-}^O,R\geq t)$. Thus $\E\{\mathcal A(B;K,q)\}=\E(B)$ if either $\Lambda^C=\Lambda_0^C$ or $q=q_0$ on the observation risk set. Equations \eqref{eq:app-irt-path} and \eqref{eq:app-irt-drift} prove \eqref{eq:true-lift-representation} and \eqref{eq:one-member-bias}, respectively. Each identity also holds conditionally on baseline covariates and treatment.

\subsection{Double robustness}
\label{subsec:app-irt-composition}

Assume that the working censoring survival functions are bounded away from zero and that the integrals below have finite absolute expectations. For consistency, assume $\widehat\eta\to\eta^*$ in probability, continuity of the population kernel mean at $\eta^*$, and a uniform law of large numbers for the fitted kernels in a neighborhood of $\eta^*$. Sufficient conditions are given in Appendix \ref{subsec:app-irt-asymptotics}.

\begin{proof}[Proof of Theorem \ref{thm:irt-dr}]
All working functions in this proof are evaluated at the possibly misspecified limit $\eta^*$. Initially, expectations are conditional on $\bm L_1$, $\bm L_2$, $A_1=1$, and $A_2=0$. Retain $H^r$ and $W_j$ from \eqref{eq:irt-gamma}. Write $V_1^r=\mathcal A_1(H^r;K_1,Q_1^r)$ for the first correction. With $O_1$ fixed, its conditional expectation under the working outcome law of participant 2 is
\begin{align}
q_2^{V_1^r}(s)
&=W_1Q_2^r(\bm Y_1;s)
+\int_0^{U_1}\frac{Q_{12}^r(t,s)}{K_1(t)}\,dM_1^C(t),
\label{eq:app-irt-induced}
\end{align}
since the three regressions in \eqref{eq:irt-gamma} are obtained from the same two working outcome distributions. Applying the second correction therefore gives
\begin{align*}
\mathcal A_2(V_1^r;K_2,q_2^{V_1^r})
&=W_2V_1^r+\int_0^{U_2}\frac{q_2^{V_1^r}(s)}{K_2(s)}\,dM_2^C(s)\\
&=W_1W_2H^r
+W_2\int_0^{U_1}\frac{Q_1^r(t;\bm Y_2)}{K_1(t)}\,dM_1^C(t)
+W_1\int_0^{U_2}\frac{Q_2^r(\bm Y_1;s)}{K_2(s)}\,dM_2^C(s)\\
&\hspace{1em}+\int_0^{U_1}\int_0^{U_2}\frac{Q_{12}^r(t,s)}{K_1(t)K_2(s)}\,dM_2^C(s)\,dM_1^C(t)
=\Gamma_{12}^r.
\end{align*}
Let $Q_{1,0}^r$ be the true conditional expectation of $H^r$ given participant 1's history with $\bm Y_2$ fixed, and let $q_{2,0}^{V_1^r}$ be the true conditional expectation of the same, possibly misspecified, $V_1^r$ given participant 2's history with $O_1$ fixed. Conditional application of \eqref{eq:app-irt-drift} to each participant gives
\begin{align}
\E(\Gamma_{12}^r)
&=\E(H^r)+\E(V_1^r-H^r)
+\E\{\mathcal A_2(V_1^r;K_2,q_2^{V_1^r})-V_1^r\}\nonumber\\
&=\mu_0^r(\bm L_1,\bm L_2)
+\E\int_0^\tau\frac{Y_1^C(t)}{K_1(t)}
\{Q_1^r(t;\bm Y_2)-Q_{1,0}^r(t;\bm Y_2)\}
\{d\Lambda_{0,1}^C(t)-d\Lambda_1^C(t)\}\nonumber\\
&\hspace{1em}+\E\int_0^\tau\frac{Y_2^C(s)}{K_2(s)}
\{q_2^{V_1^r}(s)-q_{2,0}^{V_1^r}(s)\}
\{d\Lambda_{0,2}^C(s)-d\Lambda_2^C(s)\}.
\label{eq:app-irt-pair-drift}
\end{align}
Correct censoring models make both hazard differences zero. Under correct outcome models, $Q_1^r=Q_{1,0}^r$ and $Q_2^r=Q_{2,0}^r$, and iterated conditional expectation gives
\begin{align*}
q_{2,0}^{V_1^r}(s)
&=W_1Q_{2,0}^r(\bm Y_1;s)
+\int_0^{U_1}\frac{
\E\{Q_1^r(t;\bm Y_2)\mid\mathcal F_{2,s^-}^O,R_2\geq s,O_1\}}
{K_1(t)}\,dM_1^C(t)\\
&=W_1Q_2^r(\bm Y_1;s)
+\int_0^{U_1}\frac{Q_{12}^r(t,s)}{K_1(t)}\,dM_1^C(t)
=q_2^{V_1^r}(s).
\end{align*}
The coefficients from participant 1 are fixed in this conditional integration, even if its censoring model is misspecified. Both regression differences in \eqref{eq:app-irt-pair-drift} therefore vanish. Under either model condition, the expectation of the estimating kernel is
\begin{align*}
\E\{g^r(O_i,O_k;\eta^*)\}
&=\E\{G_{ik}^r(\eta^*)\}\\
&=\E\Bigl[\mu^r(\bm L_i,\bm L_k)
+\frac{\Prb(A_i=1,A_k=0\mid\bm L_i,\bm L_k)}{\pi_1(\bm L_i)\pi_0(\bm L_k)}\\
&\qquad\qquad \{\E(\Gamma_{ik}^r\mid\bm L_i,\bm L_k,A_i=1,A_k=0)-\mu^r(\bm L_i,\bm L_k)\}\Bigr]\\
&=\E\{\mu^r(\bm L_i,\bm L_k)+\mu_0^r(\bm L_i,\bm L_k)-\mu^r(\bm L_i,\bm L_k)\}\\
&=\E\{\mu_0^r(\bm L_i,\bm L_k)\}
=\lambda^r.
\end{align*}
The first equality follows by exchangeability of the two participant records. The baseline pair regression need not be correct. If $Pg_\eta^r$ denotes the population kernel mean, the assumed uniform law and continuity give
\begin{align*}
\widehat\lambda_{\mathrm{DR}}^r-\lambda^r
&=(U_n-P)g_{\widehat\eta}^r+Pg_{\widehat\eta}^r-Pg_{\eta^*}^r
=\op(1),
\end{align*}
which proves consistency.
\end{proof}

\section{Cluster-randomized trials}
\label{sec:app-crt}

Theorem \ref{thm:pair-dr} applies to the participant-specific histories defined in Section \ref{subsec:crt-estimation}. The history $\mathcal F_{ij,t}^O$ contains the complete baseline record $\bm L_i$, treatment $A_i$, and participant $j$'s observed event and censoring history. The full observed cluster record is used for inference and the efficiency calculations in Appendix \ref{sec:app-theory}.

\subsection{Cluster sampling and target populations}
\label{subsec:app-crt-estimands}
\label{subsec:app-crt-observation}
\label{subsec:app-weighting}

Let $\mathcal Y$ collect the restricted outcomes in a cluster. Write $P_s^F(d\bm l,d\mathscr y)=P_L(d\bm l)P_s(d\mathscr y\mid\bm l)$ for the distribution of a population cluster under treatment $s$, and let $\E_s$ denote expectation under this distribution. Consistency and randomization imply
\begin{align}
\E\left\{\frac{\ind(A=s)}{\pi_s(\bm L)}f(\bm L,\mathcal Y)\right\}
&=\E\left[\frac{\E\{\ind(A=s)\mid\bm L,\mathcal Y(s)\}}{\pi_s(\bm L)}f\{\bm L,\mathcal Y(s)\}\right]\nonumber\\
&=\E[f\{\bm L,\mathcal Y(s)\}]=\E_s(f)
\label{eq:app-randomization}
\end{align}
for every integrable $f$. The identity also holds conditional on the baseline record of an independent comparison cluster.

Under Assumption \ref{ass:car}, the marginal censoring residual for participant $j$ remains a martingale after the complete event trajectories are revealed. Its enlarged filtration is
\begin{equation}
\mathcal G_{ij,t}=\mathcal F_{ij,t}^O\vee\sigma(\mathcal Y_i).
\label{eq:app-enlarged-filtration}
\end{equation}
The censoring vector is conditionally independent of these trajectories given $\bm L_i,A_i$, so revealing them does not change $K_{0,ij}$. Dependence among censoring times within a cluster is permitted. The marginal residual need not be a martingale after other members' censoring histories are also revealed; the participant corrections therefore use $\mathcal F_{ij,t}^O$, as specified in the main paper. The continuous censoring hazards in Section \ref{subsec:irt-estimands} are used throughout.

\begin{figure}[!htbp]
\centering
\includegraphics[width=0.8\textwidth]{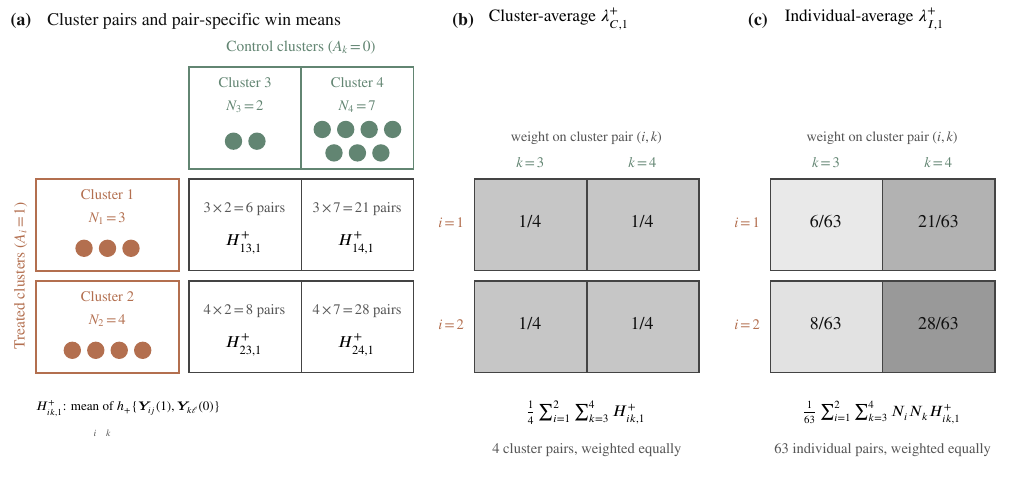}
\caption{Cluster-average and individual-average target populations. Panel (a) shows participant comparisons within cluster pairs; panels (b) and (c) show equal cluster and equal participant weighting.}
\label{fig:target-populations}
\end{figure}

Independence between clusters gives $\E(N_iN_k)=\nu^2$ and
\begin{align}
\lambda_{I,a}^r-\lambda_{C,a}^r
&=\frac{\E(N_iN_kH_{ik,a}^r)-\E(N_iN_k)\E(H_{ik,a}^r)}{\nu^2}\nonumber\\
&=\frac{\Cov\{N_iN_k,H_{ik,a}^r\}}{\nu^2}.
\label{eq:app-target-difference}
\end{align}
The targets coincide when the conditional mean comparison is constant across cluster sizes. Sampling clusters in proportion to size and then sampling participants uniformly within those clusters gives the individual-average target. Figure \ref{fig:target-populations} illustrates the distinction. The treated cluster sizes are 3 and 4, and the control cluster sizes are 2 and 7. Equal cluster weighting assigns weight $1/4$ to each cluster pair; equal participant weighting assigns weights $6/63$, $21/63$, $8/63$, and $28/63$.

\subsection{Proof of Theorem \ref{thm:identification}}
\label{subsec:app-identification}

Assumption \ref{ass:car} and positivity give
\begin{align}
\E\left\{\frac{\Delta_{ij}^R}{K_{0,ij}(R_{ij})}\,\middle|\,\bm L_i,\mathcal Y_i,A_i\right\}
&=\frac{\Prb(C_{ij}\geq R_{ij}\mid\bm L_i,\mathcal Y_i,A_i)}{K_{0,ij}(R_{ij})}=1.
\label{eq:app-ipw-conditional}
\end{align}
For participants from independent clusters assigned to $a$ and $b=1-a$, integration over the two censoring variables therefore yields
\begin{align*}
&\E\left\{\frac{\Delta_{ij}^R\Delta_{k\ell}^R h_r(\bm Y_{ij},\bm Y_{k\ell})}{K_{0,ij}(R_{ij})K_{0,k\ell}(R_{k\ell})}\,\middle|\,\bm L_i,\mathcal Y_i,\bm L_k,\mathcal Y_k,A_i=a,A_k=b\right\}\\
&\quad=h_r(\bm Y_{ij},\bm Y_{k\ell})
\E\left\{\frac{\Delta_{ij}^R}{K_{0,ij}(R_{ij})}\,\middle|\,\bm L_i,\mathcal Y_i,A_i=a\right\}
\E\left\{\frac{\Delta_{k\ell}^R}{K_{0,k\ell}(R_{k\ell})}\,\middle|\,\bm L_k,\mathcal Y_k,A_k=b\right\}\\
&\quad=h_r(\bm Y_{ij},\bm Y_{k\ell}).
\end{align*}
The factorization uses independence between clusters only. Averaging over participant pairs and applying \eqref{eq:app-randomization} gives
\begin{align}
&\E\Bigg[\frac{\ind(A_i=a,A_k=b)}{\pi_a(\bm L_i)\pi_b(\bm L_k)}
\frac1{N_iN_k}\sum_{j=1}^{N_i}\sum_{\ell=1}^{N_k}
\frac{\Delta_{ij}^R\Delta_{k\ell}^R h_r(\bm Y_{ij},\bm Y_{k\ell})}{K_{0,ij}(R_{ij})K_{0,k\ell}(R_{k\ell})}\,\Bigg|\,\bm L_i,\bm L_k\Bigg]\nonumber\\
&\quad=\E\left[\frac1{N_iN_k}\sum_{j=1}^{N_i}\sum_{\ell=1}^{N_k}
h_r\{\bm Y_{ij}(a),\bm Y_{k\ell}(b)\}\,\middle|\,\bm L_i,\bm L_k\right]\nonumber\\
&\quad=\E(H_{ik,a}^r\mid\bm L_i,\bm L_k)
=\mu_{C,a,0}^r(\bm L_i,\bm L_k).
\label{eq:app-crt-observed-pair-mean}
\end{align}
Taking expectations gives $\lambda_{C,a}^r$. Because the cluster sizes are baseline measurable,
\begin{align*}
\E\{N_iN_k\mu_{C,a,0}^r(\bm L_i,\bm L_k)\}
&=\E\big[\E\{N_iN_kH_{ik,a}^r\mid\bm L_i,\bm L_k\}\big]\\
&=\E\{\mu_{I,a,0}^r(\bm L_i,\bm L_k)\}
=\theta_{I,a}^r.
\end{align*}
The observed baseline distribution identifies $\nu>0$, so division by $\nu^2$ identifies $\lambda_{I,a}^r$. Bounded comparison indicators and inverse censoring weights, together with $\E(N_iN_k)=\nu^2<\infty$, justify these expectations. The argument applies to each tier indicator and reduces to the IRT proof when $N_i\equiv1$ and $a=1$.

\subsection{Double robustness}
\label{subsec:app-crt-estimation}
\label{subsec:app-general-drift}
\label{sec:pair-score-proof}
\label{subsec:app-crt-augmentation}

The conditional form of \eqref{eq:app-irt-drift} applies to an outcome contribution $B_{ij}$ determined at $R_{ij}$. With $q_{ij,0}(t)=\E(B_{ij}\mid\mathcal F_{ij,t^-}^O,R_{ij}\geq t)$ and generic nuisance limits $K_{ij}^*,q_{ij}^*$,
\begin{align}
&\E\{\mathcal A_{ij}(B_{ij};K_{ij}^*,q_{ij}^*)-B_{ij}\mid\bm L_i,A_i\}\nonumber\\
&\quad=\E\left[\int_0^{U_{ij}}\frac{q_{ij}^*(t)-B_{ij}}{K_{ij}^*(t)}\,dM_{ij}^C(t;\Lambda_{ij}^{C,*})\,\middle|\,\bm L_i,A_i\right]\nonumber\\
&\quad=\E\left[\int_0^\tau\frac{Y_{ij}^C(t)}{K_{ij}^*(t)}
\{q_{ij}^*(t)-q_{ij,0}(t)\}
\{d\Lambda_{0,ij}^C(t)-d\Lambda_{ij}^{C,*}(t)\}\,\middle|\,\bm L_i,A_i\right].
\label{eq:app-general-drift}
\end{align}
The martingale term has conditional mean zero in \eqref{eq:app-enlarged-filtration}; conditioning the remaining integral on $\mathcal F_{ij,t^-}^O$ replaces $B_{ij}$ by $q_{ij,0}(t)$. Either correct nuisance function makes the final integrand zero.

\begin{proof}[Proof of Theorem \ref{thm:pair-dr}]
Fix $\bm L_i,\bm L_k,A_i=a,A_k=b$. Identify participants 1 and 2 in Appendix \ref{subsec:app-irt-composition} with $(i,j)$ and $(k,\ell)$. Write $\mathcal A_{ij}$ and $\mathcal A_{k\ell}$ for their corrections at $\eta^*$, with the second correction using the induced regression in \eqref{eq:app-irt-induced}. Independence between clusters permits conditional integration over either participant while the other cluster record is held fixed. Under either model condition of Theorem \ref{thm:pair-dr},
\begin{align*}
&\E\{\Gamma_{ij,k\ell,a}^r(\eta^*)\mid\bm L_i,\bm L_k,A_i=a,A_k=b\}\\
&\quad=\E\big[\mathcal A_{k\ell}\{\mathcal A_{ij}h_r(\bm Y_{ij},\bm Y_{k\ell})\}\mid\bm L_i,\bm L_k,A_i=a,A_k=b\big]\\
&\quad=\E\{\mathcal A_{ij}h_r(\bm Y_{ij},\bm Y_{k\ell})\mid\bm L_i,\bm L_k,A_i=a,A_k=b\}\\
&\quad=\E\{h_r(\bm Y_{ij},\bm Y_{k\ell})\mid\bm L_i,\bm L_k,A_i=a,A_k=b\}\\
&\quad=\E[h_r\{\bm Y_{ij}(a),\bm Y_{k\ell}(b)\}\mid\bm L_i,\bm L_k].
\end{align*}
When both censoring models are correct, each correction preserves its conditional mean by \eqref{eq:app-general-drift}. When both outcome models are correct, the conditional regressions for both corrections are correct by \eqref{eq:app-irt-induced}, including the conditional expectation of the first augmentation term. The same mean identities then hold for arbitrary working censoring limits. No independence between participants in the same cluster is required.
\end{proof}

For any working baseline regression, averaging the pair identity for $v=C$ or summing it for $v=I$ gives
\begin{align*}
&\E\left[\mu_{v,a}^r(\bm L_i,\bm L_k;\eta^*)+
\frac{\ind(A_i=a,A_k=b)}{\pi_a(\bm L_i)\pi_b(\bm L_k)}
\{\overline\Gamma_{v,ik,a}^r(\eta^*)-\mu_{v,a}^r(\bm L_i,\bm L_k;\eta^*)\}\,\middle|\,\bm L_i,\bm L_k\right]\\
&\quad=\mu_{v,a}^r(\bm L_i,\bm L_k;\eta^*)+
\frac{\Prb(A_i=a,A_k=b\mid\bm L_i,\bm L_k)}{\pi_a(\bm L_i)\pi_b(\bm L_k)}
\{\mu_{v,a,0}^r(\bm L_i,\bm L_k)-\mu_{v,a}^r(\bm L_i,\bm L_k;\eta^*)\}\\
&\quad=\mu_{v,a,0}^r(\bm L_i,\bm L_k).
\end{align*}
Thus the baseline regression cancels in expectation. Averaging over baseline records and symmetrizing gives $\E\{g_{C,a}^r(\eta^*)\}=\lambda_{C,a}^r$ and $\E\{g_{I,a}^r(\eta^*)\}=\theta_{I,a}^r$. The uniform law in Appendix \ref{sec:parametric-asymptotic-proof} establishes consistency of the fitted kernel averages. Since $\widehat\nu\to\nu>0$, Slutsky's theorem gives $U_mg_{I,a}^r(\widehat\eta)/\widehat\nu^2\to\theta_{I,a}^r/\nu^2=\lambda_{I,a}^r$ in probability, proving the consistency assertion of Theorem \ref{thm:dr-efficiency}.

\section{Asymptotic properties}
\label{sec:app-theory}

This section establishes the results in Section \ref{sec:theory}. Unless the true nuisance functions are specified, expectations and derivatives are evaluated at the probability limits of the fitted working models.

\subsection{Efficient influence function in individual-randomized trials}
\label{subsec:app-irt-theory}
\label{subsec:app-irt-efficiency}

Let $\E_s$ denote expectation under the arm $s$ outcome and observation laws, with the population baseline distribution. For an independent participant $k$, define $\kappa_1^r(\bm y,\bm l)=\E_0\{h_r(\bm y,\bm Y_k)\}$ and $\kappa_0^r(\bm y,\bm l)=\E_1\{h_r(\bm Y_k,\bm y)\}$. Their conditional means are $\bar\mu_s^r(\bm L_i)=\E_s\{\kappa_s^r(\bm Y_i,\bm L_i)\mid\bm L_i\}$. Set $\Xi_{s,i}^r=\mathcal A_i(\kappa_s^r;K_{0,i},q_{s,0,i}^r)$, where $q_{s,0,i}^r(t)$ is the conditional expectation of $\kappa_s^r$ given $\mathcal F_{i,t^-}^O$ and $R_i\geq t$.

\begin{proposition}[Efficient influence function in IRTs]
\label{prop:irt-efficiency}
Under the sampling, randomization, and censoring assumptions of Section \ref{sec:irt}, the efficient influence function for $\lambda^r$ in the nonparametric outcome model is
\begin{equation}
\phi^r(O_i)=\sum_{s=0}^1\frac{\ind(A_i=s)}{\pi_s(\bm L_i)}
\{\Xi_{s,i}^r-\bar\mu_s^r(\bm L_i)\}
+\bar\mu_1^r(\bm L_i)+\bar\mu_0^r(\bm L_i)-2\lambda^r.
\label{eq:irt-eif}
\end{equation}
\end{proposition}

\begin{proof}
Fix $r$ and omit its superscript. Consider a regular parametric submodel $P_\epsilon$ through the true distribution, with the randomization probabilities fixed. Write $P_{s,\epsilon}(d\bm l,d\bm y)=P_{L,\epsilon}(d\bm l)P_{s,\epsilon}(d\bm y\mid\bm l)$ and let $S_L$ and $S_{Y,s}$ denote its baseline and conditional outcome scores, respectively. Then $\E(S_L)=0$ and $\E_s(S_{Y,s}\mid\bm L)=0$. Differentiating the target functional gives
\begin{align*}
\dot\lambda
&=\left.\frac{d}{d\epsilon}\int h_r(\bm y_1,\bm y_0)
\,dP_{1,\epsilon}(\bm l_1,\bm y_1)\,dP_{0,\epsilon}(\bm l_0,\bm y_0)\right|_{\epsilon=0}\\
&=\sum_{s=0}^1\E_s\{\kappa_s(S_L+S_{Y,s})\}\\
&=\E\{(\bar\mu_1+\bar\mu_0-2\lambda)S_L\}
+\sum_{s=0}^1\E_s\{(\kappa_s-\bar\mu_s)S_{Y,s}\}\\
&=\E\left[\left\{\bar\mu_1+\bar\mu_0-2\lambda
+\sum_{s=0}^1\frac{\ind(A=s)}{\pi_s(\bm L)}(\kappa_s-\bar\mu_s)\right\}
\left\{S_L+\sum_{s=0}^1\ind(A=s)S_{Y,s}\right\}\right].
\end{align*}
The second factor is the complete-data score. The first is centered and belongs to the complete-data tangent space, since its baseline component is centered and each outcome component has conditional mean zero given $\bm L$.

Under Assumption \ref{ass:irt-car}, the observed-data score decomposes as
$S^O=S_L+\sum_{s=0}^1\ind(A=s)\{\E_s(S_{Y,s}\mid O)+S_{C,s}^O\}$,
where $S_{C,s}^O$ is the observed censoring score. To identify the observed-data gradient, take $B=\kappa_s$ and let $\mathscr H_t$ denote the uncensored event history. The stopped martingale $\mathcal M_B(t)=\E_s(B\mid\mathscr H_t,\bm L)$ satisfies $\mathcal M_B(0)=\bar\mu_s$ and $\mathcal M_B(R)=B$. Conditional independent censoring gives $q_{s,0}(t)=\mathcal M_B(t^-)$ on the observation risk set. With $J(t)=\ind(C>t)$ and $K_0(t)=K_{0,s}(t\mid\bm L)$, integration by parts gives
\begin{align*}
d\{J\mathcal M_B/K_0\}(t)
&=\frac{J(t^-)}{K_0(t)}\,d\mathcal M_B(t)
+\frac{\mathcal M_B(t^-)J(t^-)}{K_0(t)}\,d\Lambda_0^C(t)
-\frac{\mathcal M_B(t^-)}{K_0(t)}\,dN^C(t)\\
&=\frac{J(t^-)}{K_0(t)}\,d\mathcal M_B(t)
-\frac{q_{s,0}(t)}{K_0(t)}\,dM_0^C(t).
\end{align*}
The censoring law is continuous, so censoring and the event martingale have no common jumps and $J(R)=\Delta^R$ almost surely. Integration through $R$ therefore yields
\begin{align}
\Xi_s-\bar\mu_s
&=\frac{\Delta^R B}{K_0(R)}
+\int_0^U\frac{q_{s,0}(t)}{K_0(t)}\,dM_0^C(t)-\bar\mu_s\nonumber\\
&=\int_0^U\frac{d\mathcal M_B(t)}{K_0(t)}.
\label{eq:app-irt-event-lift}
\end{align}
By the conditional version of \eqref{eq:app-irt-path}, $\E_s(\Xi_s\mid\bm L,\bm Y)=B$. Thus
\begin{align*}
\E_s\{(\Xi_s-\bar\mu_s)\E_s(S_{Y,s}\mid O)\}
&=\E_s\{(\Xi_s-\bar\mu_s)S_{Y,s}\}\\
&=\E_s[\{\E_s(\Xi_s\mid\bm L,\bm Y)-\bar\mu_s\}S_{Y,s}]\\
&=\E_s\{(\kappa_s-\bar\mu_s)S_{Y,s}\}.
\end{align*}
For a submodel varying only the censoring distribution, hold the functions $K_0$ and $q_{s,0}$ in $\Xi_s$ fixed. The regression $q_{s,0}$ remains correct, and \eqref{eq:app-irt-drift} gives $\E_{s,\epsilon}(\Xi_s\mid\bm L)=\bar\mu_s$. Consequently,
\begin{align*}
\E_s\{(\Xi_s-\bar\mu_s)S_{C,s}^O\mid\bm L\}
&=\left.\frac{d}{d\epsilon}\E_{s,\epsilon}(\Xi_s-\bar\mu_s\mid\bm L)\right|_{\epsilon=0}
=0,\\
\E(\phi S^O)
&=\E\{(\bar\mu_1+\bar\mu_0-2\lambda)S_L\}
+\sum_{s=0}^1\E_s\{(\Xi_s-\bar\mu_s)\E_s(S_{Y,s}\mid O)\}\\
&=\E\{(\bar\mu_1+\bar\mu_0-2\lambda)S_L\}
+\sum_{s=0}^1\E_s\{(\kappa_s-\bar\mu_s)S_{Y,s}\}
=\dot\lambda.
\end{align*}

For tangent-space membership, define $u_B=\int_0^R K_0(t)^{-1}\,d\mathcal M_B(t)$. Positivity and the martingale isometry give $\E_s(u_B^2)<\infty$ and $\E_s(u_B\mid\bm L)=0$. Conditional independence and the martingale property imply
\begin{align*}
\E_s(u_B\mid O)
&=\int_0^U K_0(t)^{-1}\,d\mathcal M_B(t)
+\E_s\left\{\int_U^R K_0(t)^{-1}\,d\mathcal M_B(t)\,\middle|\,O\right\}\\
&=\int_0^U K_0(t)^{-1}\,d\mathcal M_B(t)
=\Xi_s-\bar\mu_s.
\end{align*}
Bounded, conditionally centered truncations approximate $u_B$ in $L^2$. Their conditional expectations are observed outcome scores and converge to $\Xi_s-\bar\mu_s$, since conditional expectation is an $L^2$ contraction. Multiplication by the bounded baseline function $1/\pi_s(\bm L)$ preserves this tangent-space membership. The baseline component also belongs to the observed tangent space. Hence $\phi$ represents every pathwise derivative and belongs to its tangent closure, proving that it is the efficient influence function.
\end{proof}

Let $\mathcal T^O$ denote the observed-data tangent closure, let $\bm\phi=(\phi^+,\phi^-)^{\mathsf T}$, and let $\widetilde{\bm\phi}$ be any regular influence vector for $\bm\lambda$. The pathwise derivative identities imply
\begin{align*}
\E\{(\widetilde{\bm\phi}-\bm\phi)S^O\}
&=\dot{\bm\lambda}-\dot{\bm\lambda}=\bm0,\\
\Pi_{\mathcal T^O}(\widetilde{\bm\phi})
&=\Pi_{\mathcal T^O}(\bm\phi)=\bm\phi,\\
\Var(\widetilde{\bm\phi})
&=\Var(\bm\phi)+\E\{(\widetilde{\bm\phi}-\bm\phi)(\widetilde{\bm\phi}-\bm\phi)^{\mathsf T}\}
\succeq\Var(\bm\phi).
\end{align*}
The projection acts coordinatewise. Write $\bm\Xi_s=(\Xi_s^+,\Xi_s^-)^{\mathsf T}$ and $\overline{\bm\mu}_s=(\bar\mu_s^+,\bar\mu_s^-)^{\mathsf T}$. The conditional outcome residuals are centered given $\bm L$, and the cross-arm product is zero. The semiparametric efficiency bound for the $\sqrt n$-scaled estimator is therefore
\begin{align}
\bm\Sigma_{\mathrm{IRT}}^{\mathrm{eff}}
&=\Var(\bm\phi)
=\Var\{\E(\bm\phi\mid\bm L)\}+\E\{\Var(\bm\phi\mid\bm L)\}\nonumber\\
&=\Var\{\overline{\bm\mu}_1(\bm L)+\overline{\bm\mu}_0(\bm L)\}
+\E\left\{\sum_{s=0}^1\frac{\Var_s(\bm\Xi_s\mid\bm L)}{\pi_s(\bm L)}\right\}.
\label{eq:app-irt-efficiency-bound}
\end{align}
The first-order covariance bound for $\widehat{\bm\lambda}$ is $\bm\Sigma_{\mathrm{IRT}}^{\mathrm{eff}}/n$, with equality precisely when $\widetilde{\bm\phi}=\bm\phi$ almost surely. For a differentiable summary $f(\bm\lambda)$, the efficient influence function is $\nabla f(\bm\lambda)^{\mathsf T}\bm\phi$ and its variance bound is $\nabla f(\bm\lambda)^{\mathsf T}\bm\Sigma_{\mathrm{IRT}}^{\mathrm{eff}}\nabla f(\bm\lambda)/n$. The gradients and interior conditions are those in Section \ref{subsec:irt-theory}.

At the true nuisance functions, integration of \eqref{eq:irt-gamma} over the independent opposing participant eliminates the terms containing that participant's censoring martingale. The remaining terms give $\Xi_s$, and integration of the correct baseline pair regression gives $\bar\mu_s$. The projection calculation in Appendix \ref{sec:kernel-projection-proof}, with $N_i=1$, thus gives
\begin{align*}
2[\E\{g^r(O_i,O_k;\eta_0)\mid O_i\}-\lambda^r]
&=\phi^r(O_i).
\end{align*}
The kernel mean is constant in each nuisance block when the other block is correct, and its derivative with respect to the baseline pair regression is zero under known randomization. Thus $\dot{\bm\Psi}_{\eta_0}=0$. Under the regularity conditions below, the practical estimator attains the bound when both nuisance blocks and the baseline pair regression are correct.

\subsection{Asymptotic inference in individual-randomized trials}
\label{subsec:app-irt-asymptotics}
\label{subsec:app-irt-variance}

Use $\bm g$, $\bm\Psi$, and $\bm\lambda$ from Section \ref{subsec:irt-theory}, and write $\overline{\bm g}(o;\eta)=\E\{\bm g(o,O_k;\eta)\}$. Sufficient conditions for Theorem \ref{thm:irt-inference} are as follows.
\begin{enumerate}[label=(\roman*)]
\item The nuisance estimator converges to an interior limit $\eta^*$ in a normed space and satisfies $\widehat\eta-\eta^*=n^{-1}\sum_i\bm\psi_i+\op(n^{-1/2})$, where $\E(\bm\psi_i)=\bm0$, $\E\|\bm\psi_i\|^2<\infty$, and $\|\widehat\eta-\eta^*\|=\Op(n^{-1/2})$.
\item The kernel mean is Fr\'echet differentiable at $\eta^*$, with bounded linear derivative $\dot{\bm\Psi}_{\eta^*}$ and a locally quadratic remainder.
\item The kernels have a square-integrable envelope near $\eta^*$, and $\|\bm g_\eta-\bm g_{\eta^*}\|_{L^2(P^2)}\to0$ as $\eta\to\eta^*$.
\item For some $c_n\to\infty$ with $c_n/\sqrt n\to0$ and $\Prb\{\sqrt n\|\widehat\eta-\eta^*\|\leq c_n\}\to1$,
\begin{equation}
\sup_{\|\eta-\eta^*\|\leq c_n/\sqrt n}
\sqrt n\|(U_n-P)(\bm g_\eta-\bm g_{\eta^*})\|\xrightarrow{p}0,
\label{eq:app-irt-uprocess}
\end{equation}
where $P$ denotes expectation over an independent pair.
\end{enumerate}
The true and working censoring survival functions satisfy the positivity bounds in Appendix \ref{subsec:app-irt-composition}. Every reported standardized contrast has positive limiting variance. These conditions allow regular parametric and semiparametric nuisance estimators; their interpretation for estimated cumulative hazards is given in Appendix \ref{subsec:app-functional-nuisance}.

Under either model condition in Theorem \ref{thm:irt-dr}, let $\bm\chi_i=2\{\overline{\bm g}(O_i;\eta^*)-\bm\Psi(\eta^*)\}+\dot{\bm\Psi}_{\eta^*}[\bm\psi_i]$. Then
\begin{equation}
\sqrt n(\widehat{\bm\lambda}-\bm\lambda)
=\frac1{\sqrt n}\sum_{i=1}^n\bm\chi_i+\op(1),
\label{eq:irt-al}
\end{equation}
with covariance $\bm\Sigma_{\mathrm{IRT}}=\E(\bm\chi_i\bm\chi_i^{\mathsf T})$.

\begin{proof}[Proof of Theorem \ref{thm:irt-inference}]
For a scalar coordinate, write $g_1(o)=\E\{g_{\eta^*}(o,O_k)\}-\Psi(\eta^*)$ and $g_2(o,o')=g_{\eta^*}(o,o')-\Psi(\eta^*)-g_1(o)-g_1(o')$. The Hoeffding decomposition gives
\begin{align*}
U_ng_{\eta^*}-\Psi(\eta^*)
&=\binom n2^{-1}\sum_{i<k}\{g_1(O_i)+g_1(O_k)+g_2(O_i,O_k)\}\\
&=\frac2n\sum_{i=1}^n g_1(O_i)+U_ng_2.
\end{align*}
The second kernel has conditional mean zero given either argument. Disjoint pairs are independent, and, for distinct $i,k,\ell$,
\begin{align*}
\E\{g_2(O_i,O_k)g_2(O_i,O_\ell)\}
&=\E[\E\{g_2(O_i,O_k)\mid O_i\}\E\{g_2(O_i,O_\ell)\mid O_i\}]
=0,\\
\E\{(\sqrt nU_ng_2)^2\}
&=n\binom n2^{-1}\E(g_2^2)
=\frac{2\E(g_2^2)}{n-1}\longrightarrow0.
\end{align*}
The fitted-kernel expansion is
\begin{align*}
U_ng_{\widehat\eta}-\Psi(\eta^*)
&=U_ng_{\eta^*}-\Psi(\eta^*)
+(U_n-P)(g_{\widehat\eta}-g_{\eta^*})
+\Psi(\widehat\eta)-\Psi(\eta^*)\\
&=\frac2n\sum_i g_1(O_i)
+\dot\Psi_{\eta^*}[\widehat\eta-\eta^*]
+\Op(\|\widehat\eta-\eta^*\|^2)+\op(n^{-1/2})\\
&=\frac1n\sum_i\{2g_1(O_i)+\dot\Psi_{\eta^*}[\psi_i]\}
+\op(n^{-1/2}).
\end{align*}
Here \eqref{eq:app-irt-uprocess} controls the empirical-process term, and $\|\widehat\eta-\eta^*\|^2=\Op(n^{-1})$. Theorem \ref{thm:irt-dr} gives $\bm\Psi(\eta^*)=\bm\lambda$, proving \eqref{eq:irt-al}. Jensen's inequality and boundedness of the derivative give $\E\|\bm\chi_i\|^2<\infty$, while $\E(\bm\chi_i)=\bm0$. The multivariate central limit theorem yields the stated normal limit. Equation \eqref{eq:irt-al} also gives $\widehat{\bm\lambda}-\bm\lambda=\Op(n^{-1/2})$. A vector that includes the tie probability may have singular covariance; positive variance is needed only for each reported standardized contrast.
\end{proof}

For variance consistency, assume in addition that $U_n\|\bm g_{\widehat\eta}-\bm g_{\eta^*}\|^2=\op(1)$ and $n^{-1}\sum_i\|\widehat{\dot{\bm\Psi}}[\widehat{\bm\psi}_i]-\dot{\bm\Psi}_{\eta^*}[\bm\psi_i]\|^2=\op(1)$. For $\widetilde{\bm g}_i=(n-1)^{-1}\sum_{k\ne i}\bm g(O_i,O_k;\eta^*)$, conditional independence of its partners implies
\begin{align*}
\E\|\widetilde{\bm g}_i-\overline{\bm g}(O_i;\eta^*)\|^2
&=\frac1{n-1}\E[\operatorname{tr}\{\Var(\bm g(O_i,O_k;\eta^*)\mid O_i)\}]\\
&\leq\frac{\E\|\bm g_{\eta^*}\|^2}{n-1}.
\end{align*}
If $\widehat{\bm g}_i$ denotes the fitted row average, Jensen's inequality gives
\begin{align*}
\frac1n\sum_i\|\widehat{\bm g}_i-\widetilde{\bm g}_i\|^2
&\leq\frac1{n(n-1)}\sum_{i\ne k}\|\bm g(O_i,O_k;\widehat\eta)-\bm g(O_i,O_k;\eta^*)\|^2\\
&=U_n\|\bm g_{\widehat\eta}-\bm g_{\eta^*}\|^2
=\op(1).
\end{align*}
The preceding expectation bound, Markov's inequality, and the assumed nuisance consistency yield $n^{-1}\sum_i\|\widehat{\bm\chi}_i-\bm\chi_i\|^2=\op(1)$. Expanding the outer products and applying Cauchy--Schwarz gives
\begin{align*}
&\left\|\frac1n\sum_i(\widehat{\bm\chi}_i\widehat{\bm\chi}_i^{\mathsf T}-\bm\chi_i\bm\chi_i^{\mathsf T})\right\|_{\mathrm F}\\
&\quad\leq2\left\{\frac1n\sum_i\|\widehat{\bm\chi}_i-\bm\chi_i\|^2\right\}^{1/2}
\left\{\frac1n\sum_i\|\bm\chi_i\|^2\right\}^{1/2}
+\frac1n\sum_i\|\widehat{\bm\chi}_i-\bm\chi_i\|^2
=\op(1).
\end{align*}
The law of large numbers and the same empirical mean-square bound imply $n^{-1}\sum_i\widehat{\bm\chi}_i\widehat{\bm\chi}_i^{\mathsf T}\xrightarrow{p}\bm\Sigma_{\mathrm{IRT}}$ and $\overline{\bm\chi}\xrightarrow{p}\bm0$. Hence
\begin{align*}
n\widehat{\bm V}
&=\frac{n}{n-1}\left\{\frac1n\sum_i\widehat{\bm\chi}_i\widehat{\bm\chi}_i^{\mathsf T}
-\overline{\bm\chi}\,\overline{\bm\chi}^{\mathsf T}\right\}
\xrightarrow{p}\bm\Sigma_{\mathrm{IRT}}.
\end{align*}
The delta method yields consistent variance estimates for the win summaries, and Slutsky's theorem gives their asymptotic Wald coverage under the stated interior and positive-variance conditions.

\subsection{Fixed treatment totals in individual-randomized trials}
\label{subsec:app-irt-fixed}

Under simple complete randomization, the joint assignment probability for distinct participants is $\Prb(A_i=1,A_k=0)=\{n_1/n\}\{n_0/(n-1)\}=n_1n_0/\{n(n-1)\}$. The same identity holds conditional on their baseline covariates. Substitution of this probability for $\pi_1\pi_0$ preserves the randomization augmentation. Its reciprocal is $(n-1)/n$ times the reciprocal Bernoulli probability based on the observed treatment fractions. Under the kernel moment conditions, the resulting difference in the estimator is $\Op(n^{-1})$.

Conditional on the assignment vector, the participant records form two independent samples from the arm-specific distributions. If $n_s/n\to p_s\in(0,1)$, let $\bm\chi_s$ be the limiting influence expression in arm $s$, including nuisance estimation under this design. Expansion in the two empirical arm distributions gives
\begin{equation*}
\sqrt n(\widehat{\bm\lambda}-\bm\lambda)
=\frac1{\sqrt n}\sum_{s=0}^1\sum_{i:A_i=s}
\{\bm\chi_s(O_i)-\E_s\bm\chi_s\}+\op(1).
\end{equation*}
Finite-sample treatment fractions in the influence expression are replaced by their $L^2$ limits. The degenerate remainder is negligible by the shared-index calculation above. Appendix \ref{sec:fixed-allocation-proof} gives the corresponding calculation for the full augmented kernel; setting $N_i=1$ yields this expansion. The two arm sums are independent, so
\begin{align*}
\bm\Sigma_{\mathrm{IRT,fix}}
&=\lim_{n\to\infty}\Var\left[\frac1{\sqrt n}\sum_{s=0}^1\sum_{i:A_i=s}
\{\bm\chi_s(O_i)-\E_s\bm\chi_s\}\right]\\
&=\lim_{n\to\infty}\sum_{s=0}^1\frac{n_s}{n}\Var_s(\bm\chi_s)
=\sum_{s=0}^1p_s\Var_s(\bm\chi_s).
\end{align*}
The empirical mean-square argument applies within each arm, since both arm sample sizes diverge. Multiplying the consistent arm covariance estimates by $n_s/n^2$ and adding them gives \eqref{eq:app-irt-fixed-cov}, including the factors $n_s/(n_s-1)$. The finite-sample projection adjustment in the main paper tends to one and preserves this consistency. At the true nuisance functions and baseline pair regression, $\E(\bm\phi\mid A=s)=\bm0$ for constant allocation probabilities, so arm centering leaves \eqref{eq:app-irt-efficiency-bound} unchanged. These conclusions concern simple complete randomization; stratified or covariate-adaptive assignment requires a design-specific analysis.

\subsection{Complete-data influence functions in CRTs}
\label{subsec:app-crt-theory}
\label{sec:notation}
\label{sec:complete-C-proof}

\begin{assumption}[Moment conditions]
\label{ass:moments}
For some $\delta>0$, $\E(N_i^{4+\delta})<\infty$. The first projections of the complete-data kernels are square integrable. For each reported win ratio and win odds, $\min(\lambda_{v,1}^+,\lambda_{v,1}^-)\geq\epsilon_\lambda$ and $|\mathrm{NB}_v|\leq1-\epsilon_\lambda$ for some $\epsilon_\lambda>0$.
\end{assumption}
Condition \ref{cond:nondegeneracy} gives the additional variance requirement for standardized inference. The cluster-size moment condition controls the unnormalized individual-average kernels.

Let $F=(\bm L,\mathcal Y)$ and $O^F=(\bm L,A,\mathcal Y)$ denote the complete event records, and use $P_s^F$ and $\E_s$ as defined in Appendix \ref{subsec:app-crt-observation}. Consider a dominated parametric submodel $\{P_\epsilon:|\epsilon|<c\}$ through the true law, differentiable in quadratic mean, with the randomization probabilities fixed. Its complete-data score is
\begin{equation*}
S^F=S_L(\bm L)+\sum_{s=0}^1\ind(A=s)S_{Y,s}(F),
\end{equation*}
where $\E(S_L)=0$ and $\E_s(S_{Y,s}\mid\bm L)=0$. All derivatives are evaluated at $\epsilon=0$. Assume square integrability and domination sufficient to interchange differentiation and integration. The conditional outcome tangent space in arm $s$ is $\mathcal S_{Y,s}=\{u\in L^2(P_s^F):\E_s(u\mid\bm L)=0\}$.

Fix the treatment comparison $a$ versus $b=1-a$ and the category $r$. The first projections are
\begin{align*}
\kappa_{C,a,a}^r(F_i)&=\E_b(H_{ik,a}^r\mid F_i),\\
\kappa_{C,a,b}^r(F_i)&=\E_a(H_{ki,a}^r\mid F_i),\\
\kappa_{I,a,a}^r(F_i)&=\E_b(N_iN_kH_{ik,a}^r\mid F_i),\\
\kappa_{I,a,b}^r(F_i)&=\E_a(N_iN_kH_{ki,a}^r\mid F_i).
\end{align*}
Write $\bar\mu_{v,a,s}^r(\bm L)=\E_s(\kappa_{v,a,s}^r\mid\bm L)$; the last subscript identifies the arm of the retained cluster. Each cluster-average projection has expectation $\lambda_{C,a}^r$, and each individual-average numerator projection has expectation $\theta_{I,a}^r$.

\begin{theorem}[Complete-data efficient influence functions]
\label{thm:complete-eif}
In the nonparametric complete-data model with known randomization probabilities, the efficient influence functions for the cluster-average probability and the individual-average numerator are
\begin{align}
\varphi_{C,a}^r
&=\bar\mu_{C,a,a}^r+\bar\mu_{C,a,b}^r-2\lambda_{C,a}^r
+\sum_{s\in\{a,b\}}\frac{\ind(A=s)}{\pi_s(\bm L)}
(\kappa_{C,a,s}^r-\bar\mu_{C,a,s}^r),
\label{eq:complete-eif-C}\\
\varphi_{\theta I,a}^r
&=\bar\mu_{I,a,a}^r+\bar\mu_{I,a,b}^r-2\theta_{I,a}^r
+\sum_{s\in\{a,b\}}\frac{\ind(A=s)}{\pi_s(\bm L)}
(\kappa_{I,a,s}^r-\bar\mu_{I,a,s}^r).
\label{eq:complete-eif-I-numerator}
\end{align}
The efficient influence function for $\lambda_{I,a}^r=\theta_{I,a}^r/\nu^2$ is
\begin{equation}
\varphi_{I,a}^r=\nu^{-2}\varphi_{\theta I,a}^r
-2\lambda_{I,a}^r\nu^{-1}(N-\nu).
\label{eq:complete-eif-I}
\end{equation}
\end{theorem}

\begin{proof}
Suppress $a$ and $r$ in the target and projection subscripts. Differentiating the two independent cluster distributions gives
\begin{align*}
\dot\lambda_C
&=\left.\frac{\partial}{\partial\epsilon}
\iint H_a(f_1,f_2)P_{a,\epsilon}^F(df_1)P_{b,\epsilon}^F(df_2)\right|_{\epsilon=0}\\
&=\iint H_a(f_1,f_2)
\{S_L(\bm l_1)+S_{Y,a}(f_1)+S_L(\bm l_2)+S_{Y,b}(f_2)\}
P_a^F(df_1)P_b^F(df_2)\\
&=\E\{(\bar\mu_{C,a}+\bar\mu_{C,b}-2\lambda_C)S_L\}
+\sum_{s\in\{a,b\}}\E_s\{(\kappa_{C,s}-\bar\mu_{C,s})S_{Y,s}\}\\
&=\E\{(\bar\mu_{C,a}+\bar\mu_{C,b}-2\lambda_C)S_L\}
+\sum_{s\in\{a,b\}}\E\left\{\frac{\ind(A=s)}{\pi_s(\bm L)}
(\kappa_{C,s}-\bar\mu_{C,s})S_{Y,s}\right\}\\
&=\E(\varphi_CS^F).
\end{align*}
The third equality uses conditional centering of the outcome scores; the fourth uses \eqref{eq:app-randomization}. The baseline term belongs to the baseline tangent space, and each inverse probability weighted residual belongs to its conditional outcome tangent space. Their sum is centered and belongs to the complete-data tangent space, identifying \eqref{eq:complete-eif-C} as the canonical gradient.

Retaining the cluster-size factors inside the same calculation yields
\begin{align*}
\dot\theta_I
&=\left.\frac{\partial}{\partial\epsilon}
\iint N_1N_2H_a(f_1,f_2)P_{a,\epsilon}^F(df_1)P_{b,\epsilon}^F(df_2)\right|_{\epsilon=0}\\
&=\E\{(\bar\mu_{I,a}+\bar\mu_{I,b}-2\theta_I)S_L\}
+\sum_{s\in\{a,b\}}\E_s\{(\kappa_{I,s}-\bar\mu_{I,s})S_{Y,s}\}\\
&=\E(\varphi_{\theta I}S^F).
\end{align*}
This centered gradient belongs to the same tangent space. Since cluster size is observed at baseline,
\begin{align*}
\dot\nu&=\left.\frac{\partial}{\partial\epsilon}\E_\epsilon(N)\right|_{\epsilon=0}
=\E(NS_L)=\E\{(N-\nu)S^F\},\\
\dot\lambda_I
&=\nu^{-2}\dot\theta_I-2\theta_I\nu^{-3}\dot\nu
=\E\left[\left\{\nu^{-2}\varphi_{\theta I}
-2\lambda_I\nu^{-1}(N-\nu)\right\}S^F\right]
=\E(\varphi_IS^F).
\end{align*}
The gradient $N-\nu$ and the stated linear combination belong to the tangent space, proving \eqref{eq:complete-eif-I}.
\end{proof}

The normalization contributes both variance and covariance terms:
\begin{align*}
\Var(\varphi_I)
&=\Var\{\nu^{-2}\varphi_{\theta I}-2\lambda_I\nu^{-1}(N-\nu)\}\\
&=\nu^{-4}\Var(\varphi_{\theta I})
+4\lambda_I^2\nu^{-2}\Var(N)
-4\lambda_I\nu^{-3}\Cov(\varphi_{\theta I},N).
\end{align*}
Thus the numerator and estimated mean cluster size cannot be treated as independent.

\subsection{Observed-data influence functions in CRTs}
\label{sec:observed-eif-proof}

For $u\in\mathcal S_{Y,s}$, define the coarsening score operator by $\mathcal K_su=\E_s(u\mid O)$. Conditional Jensen's inequality gives $\|\mathcal K_su\|_2\leq\|u\|_2$. To obtain the observed score, let $g_{s,\epsilon}(o\mid f,\bm l)$ denote the observation density and $p_{s,\epsilon}(f\mid\bm l)$ the conditional event density. Holding the censoring law fixed gives
\begin{align*}
\left.\partial_\epsilon\log p_{s,\epsilon}^O(o\mid\bm l)\right|_0
&=\frac{\int g_{s,0}(o\mid f,\bm l)p_{s,0}(f\mid\bm l)S_{Y,s}(f)\,df}
{\int g_{s,0}(o\mid f,\bm l)p_{s,0}(f\mid\bm l)\,df}\\
&=\E_s\{S_{Y,s}(F)\mid O=o,\bm L=\bm l\}
=\mathcal K_sS_{Y,s}(o).
\end{align*}
Under Assumption \ref{ass:car}, the observation density is constant over event records compatible with the observed record. The likelihood factors into event and censoring contributions. If $S_{C,s}$ is the observed score obtained by varying only the joint conditional censoring distribution, normalization gives
\begin{align*}
\E_s(S_{C,s}\mid F=f)
&=\int S_{C,s}(o)g_{s,0}(o\mid f,\bm L)\,do\\
&=\left.\partial_\epsilon\int g_{s,\epsilon}(o\mid f,\bm L)\,do\right|_0
=0.
\end{align*}
Hence the observed score decomposes as
\begin{equation*}
S^O=S_L+\sum_{s=0}^1\ind(A=s)\{\mathcal K_sS_{Y,s}+S_{C,s}\}.
\end{equation*}
Let $\mathcal T_{Y,s}^O=\overline{\operatorname{ran}(\mathcal K_s)}$ and let $\mathcal T_{C,s}$ be the closed linear span of the censoring scores. These spaces are orthogonal, since
\begin{align*}
\E_s\{(\mathcal K_su)S_{C,s}\}
&=\E_s\{\E_s(u\mid O)S_{C,s}\}
=\E_s(uS_{C,s})\\
&=\E_s\{u\E_s(S_{C,s}\mid F)\}
=0.
\end{align*}
Both spaces are conditionally centered given $\bm L$ and are orthogonal to baseline functions. The known randomization probabilities contribute no score. These statements allow dependence among censoring times within a cluster; they do not require a marginal censoring residual to remain a martingale after other participants' censoring histories are added.

For a retained cluster in arm $a$, define
\begin{align*}
B_{C,a,a,ij}^r
&=\E_b\left\{N_k^{-1}\sum_{\ell=1}^{N_k}
h_r(\bm Y_{ij},\bm Y_{k\ell})\mid\bm Y_{ij},\bm L_i\right\},\\
B_{I,a,a,ij}^r
&=\E_b\left\{\sum_{\ell=1}^{N_k}
h_r(\bm Y_{ij},\bm Y_{k\ell})\mid\bm Y_{ij},\bm L_i\right\}.
\end{align*}
Reverse the arguments of $h_r$ for a retained cluster in arm $b$. Then $\kappa_{C,a,s}^r=N_i^{-1}\sum_jB_{C,a,s,ij}^r$ and $\kappa_{I,a,s}^r=\sum_jB_{I,a,s,ij}^r$. Each contribution depends on the retained participant's restricted outcome and the complete baseline record. Its true observed history regression is
\begin{equation}
q_{v,a,s,ij,0}^r(t)
=\E_s\{B_{v,a,s,ij}^r\mid\mathcal F_{ij,t^-}^O,R_{ij}\geq t\}.
\label{eq:marginal-q}
\end{equation}
The filtration is the participant history specified in Appendix \ref{subsec:app-crt-observation}. Put $\xi_{v,a,s,ij}^r=\mathcal A_{ij}(B_{v,a,s,ij}^r;K_{0,ij},q_{v,a,s,ij,0}^r)$ and define
\begin{align}
\Xi_{C,a,s}^r&=N_i^{-1}\sum_j\xi_{C,a,s,ij}^r,\nonumber\\
\Xi_{I,a,s}^r&=\sum_j\xi_{I,a,s,ij}^r.
\label{eq:xi-participant}
\end{align}
The corresponding observed-data influence functions are
\begin{align}
\phi_{C,a}^r
&=\bar\mu_{C,a,a}^r+\bar\mu_{C,a,b}^r-2\lambda_{C,a}^r
+\sum_{s\in\{a,b\}}\frac{\ind(A=s)}{\pi_s(\bm L)}
(\Xi_{C,a,s}^r-\bar\mu_{C,a,s}^r),
\label{eq:observed-eif-C}\\
\phi_{\theta I,a}^r
&=\bar\mu_{I,a,a}^r+\bar\mu_{I,a,b}^r-2\theta_{I,a}^r
+\sum_{s\in\{a,b\}}\frac{\ind(A=s)}{\pi_s(\bm L)}
(\Xi_{I,a,s}^r-\bar\mu_{I,a,s}^r),
\label{eq:observed-eif-I-numerator}\\
\phi_{I,a}^r
&=\nu^{-2}\phi_{\theta I,a}^r-2\lambda_{I,a}^r\nu^{-1}(N-\nu).
\label{eq:observed-eif-I}
\end{align}
These are gradients in the full observed cluster model. Their projections onto its tangent space give the efficient influence functions.

\begin{proof}
For one participant, suppress indices and write $\xi_B=\mathcal A(B;K_0,q_0)$. The identity in Appendix \ref{subsec:app-general-drift}, applied conditionally on the complete cluster event record, gives
\begin{align}
\E_s(\xi_B\mid F)
&=B+\E_s\left\{\int_0^U\frac{q_0(t)-B}{K_0(t)}\,dM_0^C(t)\,\middle|\,F\right\}
=B,\nonumber\\
\E_s(\xi_B\mid\bm L)
&=\E_s\{\E_s(\xi_B\mid F)\mid\bm L\}
=\E_s(B\mid\bm L).
\label{eq:app-lift-conditional}
\end{align}
The censoring integral has conditional mean zero in the participant filtration enlarged by $F$. For every $u\in\mathcal S_{Y,s}$,
\begin{align*}
\E_s\{\xi_B\mathcal K_su\}
&=\E_s\{\xi_B\E_s(u\mid O)\}
=\E_s(\xi_Bu)\\
&=\E_s\{u\E_s(\xi_B\mid F)\}
=\E_s(Bu).
\end{align*}
Thus the observed contribution preserves every outcome derivative, including those that change dependence among participants.

For censoring orthogonality, vary the joint conditional censoring distribution while holding the event law and the functions $K_0,q_0$ in $\xi_B$ fixed. The regression $q_0$ remains correct because its event history and baseline conditioning are unchanged. The outcome-correct part of \eqref{eq:app-general-drift} therefore yields
\begin{align*}
\E_{s,\epsilon}(\xi_B\mid\bm L)&=\E_s(B\mid\bm L),\\
\E_s(\xi_BS_{C,s}\mid\bm L)
&=\left.\partial_\epsilon\E_{s,\epsilon}(\xi_B\mid\bm L)\right|_0
=0.
\end{align*}
The identity extends to $\mathcal T_{C,s}$ by continuity of the $L^2$ inner product. Summing the participant contributions with the baseline weights $N^{-1}$ or one gives
\begin{align*}
\E_s(\Xi_{v,a,s}^r\mid F)&=\kappa_{v,a,s}^r,\\
\E_s(\Xi_{v,a,s}^r\mid\bm L)&=\bar\mu_{v,a,s}^r,\\
\E_s\{(\Xi_{v,a,s}^r-\bar\mu_{v,a,s}^r)\mathcal K_su\}
&=\E_s\{(\kappa_{v,a,s}^r-\bar\mu_{v,a,s}^r)u\},\\
\E_s\{(\Xi_{v,a,s}^r-\bar\mu_{v,a,s}^r)S_{C,s}\}&=0.
\end{align*}
Substitution into the complete-data derivative calculation gives $\E(\phi_{C,a}^rS^O)=\dot\lambda_{C,a}^r$ and $\E(\phi_{\theta I,a}^rS^O)=\dot\theta_{I,a}^r$. Conditional centering gives zero means. Since $N$ is fully observed, the quotient rule gives $\E(\phi_{I,a}^rS^O)=\dot\lambda_{I,a}^r$.
\end{proof}

\subsection{Semiparametric efficiency bounds in CRTs}
\label{subsec:app-canonical}

On the conditional mean-zero spaces, the adjoint of $\mathcal K_s$ is $\mathcal K_s^*v=\E_s(v\mid F)$, since
\begin{align*}
\langle v,\mathcal K_su\rangle
&=\E_s\{v\E_s(u\mid O)\}
=\E_s(vu)
=\E_s\{u\E_s(v\mid F)\}
=\langle\mathcal K_s^*v,u\rangle.
\end{align*}
The range--null-space identity is therefore
\begin{equation}
(\mathcal T_{Y,s}^O)^\perp=\ker(\mathcal K_s^*).
\label{eq:app-range-null}
\end{equation}
For either numerator target, write $v_{v,a,s}^r=\Xi_{v,a,s}^r-\bar\mu_{v,a,s}^r$ and $b_{v,a,s}^r=\kappa_{v,a,s}^r-\bar\mu_{v,a,s}^r$. The preceding proof gives $\mathcal K_s^*v_{v,a,s}^r=b_{v,a,s}^r$ and $v_{v,a,s}^r\perp\mathcal T_{C,s}$. Define
\begin{align}
w_{v,a,s}^r&=\Pi_{\mathcal T_{Y,s}^O}v_{v,a,s}^r,\nonumber\\
d_{v,a,s}^r&=v_{v,a,s}^r-w_{v,a,s}^r.
\label{eq:app-canonical-projection}
\end{align}
Projection onto a closed subspace is unique. By \eqref{eq:app-range-null},
\begin{align*}
\mathcal K_s^*d_{v,a,s}^r&=0,\\
\mathcal K_s^*w_{v,a,s}^r
&=\mathcal K_s^*v_{v,a,s}^r-\mathcal K_s^*d_{v,a,s}^r
=b_{v,a,s}^r.
\end{align*}
Both $w_{v,a,s}^r$ and $d_{v,a,s}^r$ remain orthogonal to $\mathcal T_{C,s}$. For any bounded baseline function $c$, $c(\bm L)\mathcal K_su=\mathcal K_s\{c(\bm L)u\}$. Thus $\mathcal T_{Y,s}^O$ and its orthogonal complement are stable under baseline multiplication, and the projection commutes with multiplication by $1/\pi_s(\bm L)$.

Fix $a$ and stack the win and loss coordinates in bold vectors. Define the centered baseline components
\begin{align*}
\bm h_C(\bm L)&=\bar{\bm\mu}_{C,a,a}(\bm L)+\bar{\bm\mu}_{C,a,b}(\bm L)-2\bm\lambda_{C,a},\\
\bm h_I(\bm L)&=\nu^{-2}\{\bar{\bm\mu}_{I,a,a}(\bm L)+\bar{\bm\mu}_{I,a,b}(\bm L)-2\bm\theta_{I,a}\}
-2\nu^{-1}\bm\lambda_{I,a}(N-\nu).
\end{align*}

\begin{theorem}[Efficiency bounds in CRTs]
\label{thm:observed-eif}
Under the assumptions of Section \ref{sec:crt}, Assumption \ref{ass:moments}, and the square integrability and differentiability conditions above, the efficient influence functions for the cluster-average and individual-average probability vectors are
\begin{align}
\bm\phi_C^{\mathrm{eff}}
&=\bm h_C(\bm L)+\sum_{s\in\{a,b\}}\frac{\ind(A=s)}{\pi_s(\bm L)}\bm w_{C,a,s},
\label{eq:app-canonical-C}\\
\bm\phi_I^{\mathrm{eff}}
&=\bm h_I(\bm L)+\nu^{-2}\sum_{s\in\{a,b\}}\frac{\ind(A=s)}{\pi_s(\bm L)}\bm w_{I,a,s}.
\label{eq:app-canonical-I}
\end{align}
The semiparametric efficiency bounds at the $\sqrt m$ scale are
\begin{align}
\bm{\mathcal B}_C
&=\E(\bm h_C\bm h_C^{\mathsf T})
+\sum_{s\in\{a,b\}}\E_s\left\{\frac{\bm w_{C,a,s}\bm w_{C,a,s}^{\mathsf T}}{\pi_s(\bm L)}\right\},
\label{eq:app-bound-C}\\
\bm{\mathcal B}_I
&=\E(\bm h_I\bm h_I^{\mathsf T})
+\nu^{-4}\sum_{s\in\{a,b\}}\E_s\left\{\frac{\bm w_{I,a,s}\bm w_{I,a,s}^{\mathsf T}}{\pi_s(\bm L)}\right\}.
\label{eq:app-bound-I}
\end{align}
For joint estimation, the cross-covariance block is
\begin{equation}
\bm{\mathcal B}_{CI}
=\E(\bm h_C\bm h_I^{\mathsf T})
+\nu^{-2}\sum_{s\in\{a,b\}}\E_s\left\{\frac{\bm w_{C,a,s}\bm w_{I,a,s}^{\mathsf T}}{\pi_s(\bm L)}\right\}.
\label{eq:app-bound-CI}
\end{equation}
The first-order covariance lower bound is the corresponding block matrix divided by $m$. For a differentiable scalar summary $f$, it is $\nabla f^{\mathsf T}\bm{\mathcal B}_v\nabla f/m$.
\end{theorem}

\begin{proof}
Let $\mathcal T^O$ be the closed observed-data tangent space generated by $S^O$. The baseline component, the arm-specific observed outcome spaces, and the censoring spaces are mutually orthogonal. By \eqref{eq:app-canonical-projection} and baseline multiplication,
\begin{align*}
\Pi_{\mathcal T^O}\bm\phi_C
&=\bm h_C+\sum_{s\in\{a,b\}}\frac{\ind(A=s)}{\pi_s(\bm L)}
\Pi_{\mathcal T_{Y,s}^O}\bm v_{C,a,s}
=\bm\phi_C^{\mathrm{eff}},\\
\Pi_{\mathcal T^O}\bm\phi_I
&=\bm h_I+\nu^{-2}\sum_{s\in\{a,b\}}\frac{\ind(A=s)}{\pi_s(\bm L)}
\Pi_{\mathcal T_{Y,s}^O}\bm v_{I,a,s}
=\bm\phi_I^{\mathrm{eff}},\\
\E_s\{w_{v,a,s}^r\mathcal K_sS_{Y,s}\}
&=\E_s\{(\mathcal K_s^*w_{v,a,s}^r)S_{Y,s}\}
=\E_s(b_{v,a,s}^rS_{Y,s})
=\E_s\{v_{v,a,s}^r\mathcal K_sS_{Y,s}\}.
\end{align*}
The projected gradients preserve every outcome derivative and remain orthogonal to censoring scores. Thus $\E(\bm\phi_v^{\mathrm{eff}}S^O)=\dot{\bm\lambda}_{v,a}$ for every submodel, and $\bm\phi_v^{\mathrm{eff}}\in\mathcal T^O$ coordinatewise. This identifies the efficient influence functions.

Let $\widetilde{\bm\phi}_v$ be the influence function of any regular asymptotically linear estimator of the same probability vector. Both influence functions represent the same pathwise derivative, so
\begin{align*}
\E\{(\widetilde{\bm\phi}_v-\bm\phi_v^{\mathrm{eff}})S^O\}
&=\dot{\bm\lambda}_{v,a}-\dot{\bm\lambda}_{v,a}=\bm0,\\
\Pi_{\mathcal T^O}\widetilde{\bm\phi}_v
&=\bm\phi_v^{\mathrm{eff}},\\
\Var(\widetilde{\bm\phi}_v)
&=\Var\{\bm\phi_v^{\mathrm{eff}}+(\widetilde{\bm\phi}_v-\bm\phi_v^{\mathrm{eff}})\}\\
&=\Var(\bm\phi_v^{\mathrm{eff}})
+\E\{(\widetilde{\bm\phi}_v-\bm\phi_v^{\mathrm{eff}})
(\widetilde{\bm\phi}_v-\bm\phi_v^{\mathrm{eff}})^{\mathsf T}\}\\
&\succeq\Var(\bm\phi_v^{\mathrm{eff}}).
\end{align*}
Conditional centering removes baseline--outcome cross terms, and different treatment indicators have zero product. Consequently,
\begin{align*}
\Var(\bm\phi_C^{\mathrm{eff}})
&=\E(\bm h_C\bm h_C^{\mathsf T})
+\sum_{s\in\{a,b\}}\E\left\{
\frac{\ind(A=s)}{\pi_s(\bm L)^2}\bm w_{C,a,s}\bm w_{C,a,s}^{\mathsf T}\right\}\\
&=\E(\bm h_C\bm h_C^{\mathsf T})
+\sum_{s\in\{a,b\}}\E_s\left\{
\frac{\bm w_{C,a,s}\bm w_{C,a,s}^{\mathsf T}}{\pi_s(\bm L)}\right\}
=\bm{\mathcal B}_C,\\
\Var(\bm\phi_I^{\mathrm{eff}})
&=\E(\bm h_I\bm h_I^{\mathsf T})
+\nu^{-4}\sum_{s\in\{a,b\}}\E_s\left\{
\frac{\bm w_{I,a,s}\bm w_{I,a,s}^{\mathsf T}}{\pi_s(\bm L)}\right\}
=\bm{\mathcal B}_I,\\
\Cov(\bm\phi_C^{\mathrm{eff}},\bm\phi_I^{\mathrm{eff}})
&=\E(\bm h_C\bm h_I^{\mathsf T})
+\nu^{-2}\sum_{s\in\{a,b\}}\E_s\left\{
\frac{\bm w_{C,a,s}\bm w_{I,a,s}^{\mathsf T}}{\pi_s(\bm L)}\right\}
=\bm{\mathcal B}_{CI}.
\end{align*}
The same projection argument applies to the stacked vector. The chain rule gives the scalar-summary bound.
\end{proof}

For the individual-average target, the equivalent variance decomposition is
\begin{align*}
\bm{\mathcal B}_I
&=\Var\{\nu^{-2}\bm\phi_{\theta I}^{\mathrm{eff}}-2\nu^{-1}\bm\lambda_{I,a}(N-\nu)\}\\
&=\nu^{-4}\Var(\bm\phi_{\theta I}^{\mathrm{eff}})
+4\nu^{-2}\bm\lambda_{I,a}\bm\lambda_{I,a}^{\mathsf T}\Var(N)\\
&\quad-2\nu^{-3}\left[\E\{\bm\phi_{\theta I}^{\mathrm{eff}}(N-\nu)\}\bm\lambda_{I,a}^{\mathsf T}
+\bm\lambda_{I,a}\E\{(N-\nu)(\bm\phi_{\theta I}^{\mathrm{eff}})^{\mathsf T}\}\right].
\end{align*}
Noninformative cluster size does not remove the normalization term. With $N=\nu=1$, the targets coincide, and Appendix \ref{subsec:app-irt-efficiency} gives $\bm w_s=\bm v_s$, reducing the bounds to the IRT bound.

When the censoring, outcome, and baseline pair models are correct, Appendix \ref{sec:kernel-projection-proof} identifies the practical estimator's first projection as \eqref{eq:observed-eif-C} or \eqref{eq:observed-eif-I}. Let $\bm\Sigma_v^{\mathrm{pr}}$ be its covariance. The projection residual is orthogonal to every coordinate of the efficient influence function, giving
\begin{align}
\bm\Sigma_C^{\mathrm{pr}}-\bm{\mathcal B}_C
&=\E\{(\bm\phi_C-\bm\phi_C^{\mathrm{eff}})(\bm\phi_C-\bm\phi_C^{\mathrm{eff}})^{\mathsf T}\}\nonumber\\
&=\sum_{s\in\{a,b\}}\E_s\left\{\frac{\bm d_{C,a,s}\bm d_{C,a,s}^{\mathsf T}}{\pi_s(\bm L)}\right\}\succeq\bm0,
\label{eq:app-gap-C}\\
\bm\Sigma_I^{\mathrm{pr}}-\bm{\mathcal B}_I
&=\E\{(\bm\phi_I-\bm\phi_I^{\mathrm{eff}})(\bm\phi_I-\bm\phi_I^{\mathrm{eff}})^{\mathsf T}\}\nonumber\\
&=\nu^{-4}\sum_{s\in\{a,b\}}\E_s\left\{\frac{\bm d_{I,a,s}\bm d_{I,a,s}^{\mathsf T}}{\pi_s(\bm L)}\right\}\succeq\bm0.
\label{eq:app-gap-I}
\end{align}
The same identities hold blockwise for joint estimation of both targets. Thus the practical probability vector is efficient if and only if, for both arms and both categories,
\begin{equation}
v_{v,a,s}^r=\Xi_{v,a,s}^r-\bar\mu_{v,a,s}^r
\in\overline{\operatorname{ran}(\mathcal K_s)}.
\label{eq:app-canonical-range}
\end{equation}
A scalar contrast requires only that its corresponding linear combination of projection residuals vanish. A related conditional information condition is
\begin{align}
&\E_s\{B_{v,a,s,ij}^r\mid\sigma(O_i),\Delta_{ij}^R=0\}\nonumber\\
&\qquad=\E_s\{B_{v,a,s,ij}^r\mid\mathcal F_{ij,U_{ij}}^O,
R_{ij}>U_{ij},\Delta_{ij}^R=0\}.
\label{eq:participantwise-information}
\end{align}
Under this condition, other cluster members' records provide no additional conditional mean information about the censored participant's contribution. The exact efficiency condition is \eqref{eq:app-canonical-range}; the conditional information identity alone does not establish tangent-space membership. Appendix \ref{subsec:app-product-efficiency} gives sufficient conditions for both.

If $\sum_s\E_s\|\bm d_{v,a,s}\|^2\leq\varepsilon^2$, positivity gives
\begin{align*}
\operatorname{tr}(\bm\Sigma_C^{\mathrm{pr}}-\bm{\mathcal B}_C)
&=\sum_s\E_s\{\|\bm d_{C,a,s}\|^2/\pi_s(\bm L)\}
\leq\epsilon_\pi^{-1}\varepsilon^2,\\
\operatorname{tr}(\bm\Sigma_I^{\mathrm{pr}}-\bm{\mathcal B}_I)
&=\nu^{-4}\sum_s\E_s\{\|\bm d_{I,a,s}\|^2/\pi_s(\bm L)\}
\leq\nu^{-4}\epsilon_\pi^{-1}\varepsilon^2,
\end{align*}
using the corresponding residual condition for each target. A small projection residual therefore implies a small absolute variance reduction. Correct participant models and unrestricted within-cluster dependence do not give a general bound on the relative reduction. The simulation comparisons do not estimate this joint-cluster efficiency bound.

For a scalar coordinate, suppress the target and arm subscripts and write $v$, $b=\mathcal K^*v$, and $w=\Pi_{\overline{\operatorname{ran}(\mathcal K)}}v$. Minimizing $\|v-\mathcal Ku\|_2^2+\epsilon\|u\|_2^2$ over the complete-data conditional mean-zero space gives
\begin{align*}
0&=\left.\partial_t\{\|v-\mathcal K(u+th)\|_2^2+\epsilon\|u+th\|_2^2\}\right|_{t=0}\\
&=2\langle\mathcal K^*(\mathcal Ku-v)+\epsilon u,h\rangle
\end{align*}
for every direction $h$. Since $\mathcal K^*\mathcal K+\epsilon I$ is bounded below by $\epsilon I$, the unique solution and its observed projection are
\begin{align}
u_\epsilon&=(\mathcal K^*\mathcal K+\epsilon I)^{-1}b,\nonumber\\
w_\epsilon&=\mathcal Ku_\epsilon.
\label{eq:app-regularized-projection}
\end{align}
Choose $u_\zeta$ with $\|\mathcal Ku_\zeta-w\|_2\leq\zeta$. Optimality and orthogonality give
\begin{align*}
\|w_\epsilon-w\|_2^2+\epsilon\|u_\epsilon\|_2^2
&=\|v-w_\epsilon\|_2^2-\|v-w\|_2^2+\epsilon\|u_\epsilon\|_2^2\\
&\leq\|v-\mathcal Ku_\zeta\|_2^2-\|v-w\|_2^2+\epsilon\|u_\zeta\|_2^2\\
&=\|w-\mathcal Ku_\zeta\|_2^2+\epsilon\|u_\zeta\|_2^2
\leq\zeta^2+\epsilon\|u_\zeta\|_2^2.
\end{align*}
Letting $\epsilon\downarrow0$ and then $\zeta\downarrow0$ proves $w_\epsilon\to w$ in $L^2$. No inverse of $\mathcal K^*\mathcal K$ is assumed. An exact equation $\mathcal K^*\mathcal Ku=b$ requires additional range conditions. Marginal censoring positivity need not bound the probability of observing an entire cluster away from zero when cluster size varies. These representations follow from the projection characterization of semiparametric efficiency.\cite{bickel1993efficient,tsiatis2006semiparametric}

The following result gives sufficient conditions for attaining the bound by estimating the projection residual. Let $P$ denote the observed cluster law, with $Pf=\int f(o)P(do)$, and let $\bm\vartheta$ be either probability vector or their stacked vector. Write $\bm\delta=\bm\phi^{\mathrm{pr}}-\bm\phi^{\mathrm{eff}}$. Split the $m$ independent clusters into a fixed number $K$ of folds $\mathcal I_k$, with $m_k/m\to\rho_k\in(0,1)$. All participants in a cluster remain in the same fold. Estimate $\bm\delta$ outside fold $k$ and denote the fitted function by $\widehat{\bm\delta}_{-k}$.

\begin{proposition}[Attainment of the efficiency bound]
\label{prop:app-joint-attainment}
Suppose the practical estimator is regular and asymptotically linear in the stated observed-data model at a law where the censoring, outcome, and baseline pair models are correct:
\begin{equation*}
\widehat{\bm\vartheta}_{\mathrm{pr}}-\bm\vartheta
=\frac1m\sum_{i=1}^m\bm\phi^{\mathrm{pr}}(O_i)+o_P(m^{-1/2}).
\end{equation*}
Assume $\|\widehat{\bm\delta}_{-k}-\bm\delta\|_{L^2(P)}=o_P(1)$ for every fold and
\begin{equation}
\sqrt m\left\|\sum_{k=1}^K\frac{m_k}{m}P\widehat{\bm\delta}_{-k}\right\|=o_P(1).
\label{eq:app-projection-drift}
\end{equation}
Then the corrected estimator
\begin{equation}
\widehat{\bm\vartheta}_{\mathrm{eff}}
=\widehat{\bm\vartheta}_{\mathrm{pr}}
-\frac1m\sum_{k=1}^K\sum_{i\in\mathcal I_k}\widehat{\bm\delta}_{-k}(O_i)
\label{eq:app-efficient-correction}
\end{equation}
satisfies
\begin{equation*}
\sqrt m(\widehat{\bm\vartheta}_{\mathrm{eff}}-\bm\vartheta)
=\frac1{\sqrt m}\sum_{i=1}^m\bm\phi^{\mathrm{eff}}(O_i)+o_P(1)
\end{equation*}
and attains the bound in Theorem \ref{thm:observed-eif}.
\end{proposition}

\begin{proof}
Let $P_{m,k}$ be the empirical law in fold $k$. Conditional on its training clusters, the evaluation clusters are independent of $\widehat{\bm\delta}_{-k}$, and
\begin{align*}
&\E\left[\left\|\sqrt m\,\frac{m_k}{m}(P_{m,k}-P)
(\widehat{\bm\delta}_{-k}-\bm\delta)\right\|^2
\,\middle|\,\{O_i:i\notin\mathcal I_k\}\right]\\
&\qquad\leq\frac{m_k}{m}\|\widehat{\bm\delta}_{-k}-\bm\delta\|_{L^2(P)}^2
=o_P(1).
\end{align*}
Conditional Chebyshev's inequality and fixed $K$ give a total empirical remainder of $o_P(m^{-1/2})$; independence between fitted functions from different folds is unnecessary. Since $P\bm\delta=\bm0$,
\begin{align*}
\frac1m\sum_{k=1}^K\sum_{i\in\mathcal I_k}\widehat{\bm\delta}_{-k}(O_i)
&=\frac1m\sum_{i=1}^m\bm\delta(O_i)
+\sum_{k=1}^K\frac{m_k}{m}P\widehat{\bm\delta}_{-k}
+o_P(m^{-1/2})\\
&=\frac1m\sum_{i=1}^m\bm\delta(O_i)+o_P(m^{-1/2}),\\
\sqrt m(\widehat{\bm\vartheta}_{\mathrm{eff}}-\bm\vartheta)
&=\frac1{\sqrt m}\sum_{i=1}^m
\{\bm\phi^{\mathrm{pr}}(O_i)-\bm\delta(O_i)\}+o_P(1)\\
&=\frac1{\sqrt m}\sum_{i=1}^m\bm\phi^{\mathrm{eff}}(O_i)+o_P(1).
\end{align*}
The central limit theorem gives the stated covariance bound. Orthogonality of $\bm\delta$ to every observed score gives zero derivative for the correction. The displayed expansion and contiguity under local submodels therefore preserve regularity.
\end{proof}

Condition \eqref{eq:app-projection-drift} includes error from estimating the joint event and censoring distributions, the operator and its projections, and any regularization or numerical approximation. Consistency alone does not imply this condition. The $L^2$ condition also requires the relevant residual to be identifiable and consistently estimable on the analysis horizon. Marginal censoring positivity does not ensure these properties for unrestricted joint cluster distributions.

For a correctly specified finite-dimensional joint model, suppose $\widehat\eta-\eta_0=O_P(m^{-1/2})$ and $\eta\mapsto\bm\delta_\eta$ has an integrable differentiable expansion with a locally quadratic remainder. Centering and score orthogonality give
\begin{align*}
P\dot{\bm\delta}_{\eta_0}
&=\left.\partial_\eta(P_\eta\bm\delta_\eta)\right|_{\eta_0}
-P(\bm\delta_{\eta_0}\bm S_{\eta_0}^{\mathsf T})
=\bm0,\\
P\bm\delta_{\widehat\eta}
&=P\bm\delta_{\eta_0}
+P\dot{\bm\delta}_{\eta_0}(\widehat\eta-\eta_0)
+O_P(\|\widehat\eta-\eta_0\|^2)
=O_P(m^{-1}).
\end{align*}
A projection calculation of sufficient accuracy then establishes the drift condition. Smoothness of the likelihood alone does not imply smoothness of the projection. Flexible joint models require separate remainder control. This additional estimation is not part of the practical estimator used in the main paper.

\subsection{Information from other cluster members}
\label{subsec:app-eif-computation}
\label{subsec:app-product-efficiency}

The operator $\mathcal K_s$ conditions on the full observed cluster record. The efficient influence function may therefore use other members' observed outcomes to predict a censored participant's contribution. Products of different members' observation indicators can arise within one cluster's influence function, as for marginal survival estimands in CRTs.\cite{fang2026estimands}

Consider two outcomes $Z_1,Z_2$, conditional on baseline and treatment. Let $J_1,J_2$ be independent observation indicators, independent of the outcomes, with probabilities $p_1,p_2>0$. For a centered complete-data score $u(Z_1,Z_2)$, write $u_j=\E(u\mid Z_j)$. The four observation patterns give
\begin{align*}
\mathcal Ku
&=J_1(1-J_2)u_1+(1-J_1)J_2u_2+J_1J_2u\\
&=J_1u_1+J_2u_2+J_1J_2(u-u_1-u_2),\\
\mathcal K^*\mathcal Ku
&=\E(\mathcal Ku\mid Z_1,Z_2)\\
&=p_1p_2u+p_1(1-p_2)u_1+(1-p_1)p_2u_2,\\
\langle u,\mathcal K^*\mathcal Ku\rangle
&=p_1p_2\E(u^2)+p_1(1-p_2)\E(u_1^2)
+(1-p_1)p_2\E(u_2^2)\\
&\geq p_1p_2\|u\|_2^2.
\end{align*}
Thus $\mathcal K^*\mathcal K$ has a bounded inverse, and the efficient contribution for a complete-data gradient $b$ is $\mathcal Ku$, with $u=(\mathcal K^*\mathcal K)^{-1}b$. Even when $b$ is additive, the product term in $\mathcal Ku$ can be nonzero.

For an explicit example, let $Z_1,Z_2\in\{-1,1\}$ with $\Prb(Z_1=z_1,Z_2=z_2)=(1+\rho z_1z_2)/4$, $|\rho|<1$, and let $p_1=p_2=p\in(0,1)$. For the mean of $(Z_1+Z_2)/2$, the complete-data gradient is $b=(Z_1+Z_2)/2$. Put $c=1+(1-p)\rho>0$. Since $\E(Z_2\mid Z_1)=\rho Z_1$ and conversely,
\begin{align*}
\mathcal K^*\mathcal Kb
&=p^2b+p(1-p)\{\E(b\mid Z_1)+\E(b\mid Z_2)\}\\
&=p^2b+p(1-p)(1+\rho)b=pc\,b.
\end{align*}
Hence $u=b/(pc)$, and the efficient and participant contributions are
\begin{align*}
w=\mathcal Ku
&=\frac{(1+\rho)(J_1Z_1+J_2Z_2)-\rho J_1J_2(Z_1+Z_2)}{2pc},\\
v&=\frac{J_1Z_1+J_2Z_2}{2p}.
\end{align*}
Although $\mathcal K^*v=b$, the $J_1J_2$ term is generally required for efficiency. The variances and their difference are
\begin{align*}
\E(v^2)
&=\frac{\E\{(J_1Z_1+J_2Z_2)^2\}}{4p^2}
=\frac{2p+2p^2\rho}{4p^2}
=\frac{1+p\rho}{2p},\\
\E(w^2)
&=\E\{u\mathcal K^*\mathcal Ku\}
=\E(ub)
=\frac{\E(b^2)}{pc}
=\frac{1+\rho}{2pc},\\
\E\{(v-w)^2\}
&=\E(v^2)-\E(w^2)
=\frac{1+p\rho}{2p}-\frac{1+\rho}{2pc}
=\frac{(1-p)\rho^2}{2c},\\
\frac{\E\{(v-w)^2\}}{\E(v^2)}
&=\frac{p(1-p)\rho^2}{\{1+(1-p)\rho\}(1+p\rho)}.
\end{align*}
For fixed $p$, both gains are quadratic near $\rho=0$, but neither is uniformly small over the permitted dependence model. This calculation illustrates the projection; it does not bound the gains for the simulated win estimators.

With larger clusters, analogous terms can involve three or more members, together with their partially observed event histories. They remain functions of a single cluster and can contribute to first-order variance at rate $m^{-1/2}$. In contrast, the double censoring integral in \eqref{eq:crt-gamma} combines participants from independent clusters. At the true censoring models it has zero first projection and a U-statistic average of order $O_P(m^{-1})$, as shown in Appendix \ref{sec:kernel-projection-proof}. Its degeneracy does not remove the within-cluster contributions above. The practical estimator retains that double integral but uses each participant's own history for prediction. General attainment of the efficiency bound requires the additional joint conditional predictions and projection correction described above.

A sufficient condition for efficiency is conditional independence of participant event histories and conditional independence of participant censoring times, given $\bm L$ and treatment, in addition to Assumption \ref{ass:car}. Dependence among the component event times of one participant remains unrestricted. These conditions concern efficiency at the true law and are not imposed on consistency or variance estimation in the main paper.

For a contribution $B_j$, let $\mathscr H_{j,t}$ be the uncensored event filtration and let $\mathcal M_{B_j}(t)=\E_s(B_j\mid\mathscr H_{j,t},\bm L)$, stopped at $R_j$. Then $\mathcal M_{B_j}(R_j)=B_j$, $\mathcal M_{B_j}(0)=\E_s(B_j\mid\bm L)$, and $q_{j,0}(t)=\mathcal M_{B_j}(t^-)$ on the observation risk set. The integration-by-parts identity from Appendix \ref{subsec:app-irt-efficiency} gives
\begin{align*}
\xi_{B_j}-\E_s(B_j\mid\bm L)
&=\frac{\Delta_j^RB_j}{K_{0,j}(R_j)}
+\int_0^{U_j}\frac{\mathcal M_{B_j}(t^-)}{K_{0,j}(t)}\,dM_{0,j}^C(t)
-\mathcal M_{B_j}(0)\\
&=\int_0^{U_j}\frac{d\mathcal M_{B_j}(t)}{K_{0,j}(t)}.
\end{align*}
Define $u_j=\int_0^{R_j}K_{0,j}(t)^{-1}\,d\mathcal M_{B_j}(t)$. It has conditional mean zero and finite second moment by positivity and the martingale isometry. The remaining event-martingale increments after censoring have conditional mean zero, and conditional independence of participants gives
\begin{align*}
\mathcal K_su_j
&=\E_s(u_j\mid O_i)
=\E_s(u_j\mid\mathcal F_{ij,U_{ij}}^O)\\
&=\int_0^{U_j}\frac{d\mathcal M_{B_j}(t)}{K_{0,j}(t)}
=\xi_{B_j}-\E_s(B_j\mid\bm L).
\end{align*}
With the contributions appropriate to each target,
\begin{align*}
\mathcal K_s\left(N^{-1}\sum_ju_j\right)
&=N^{-1}\sum_j\{\xi_{B_j}-\E_s(B_j\mid\bm L)\}
=\Xi_{C,a,s}-\bar\mu_{C,a,s},\\
\mathcal K_s\left(\sum_ju_j\right)
&=\sum_j\{\xi_{B_j}-\E_s(B_j\mid\bm L)\}
=\Xi_{I,a,s}-\bar\mu_{I,a,s}.
\end{align*}
Assumption \ref{ass:moments} ensures square integrability of these sums. This proves \eqref{eq:app-canonical-range}; the same conditional independence proves \eqref{eq:participantwise-information}. All residuals in \eqref{eq:app-gap-C} and \eqref{eq:app-gap-I} therefore vanish. Under correct working models and the stated regularity conditions, the practical estimator attains the efficiency bounds. Correct participant models or a cluster-level variance calculation alone do not imply this sufficient independence condition.

\subsection{Projection of the pairwise estimator}
\label{sec:kernel-projection-proof}
\label{subsec:app-baseline-optimality}

Let $\mathcal T_0$, $\mathcal T_1$, $\mathcal T_2$, and $\mathcal T_{12}$ be the four terms in \eqref{eq:crt-gamma}, with the comparison indices suppressed. At the true nuisance functions, condition on the observed cluster record $O_1$ containing the first participant. Write $B_1=\E_2(H\mid\bm Y_1,\bm L_1)$, where $\E_2$ integrates an independent comparison cluster under its assigned treatment, and let $q_{1,0}^{B_1}(t)$ be its observed history regression. The censoring identity and iterated expectation give
\begin{align*}
\E_2(\mathcal T_0\mid O_1)
&=\frac{\Delta_1^R}{K_{0,1}(R_1)}
\E_2\left\{\frac{\Delta_2^RH}{K_{0,2}(R_2)}\,\middle|\,O_1\right\}
=\frac{\Delta_1^RB_1}{K_{0,1}(R_1)},\\
\E_2(\mathcal T_1\mid O_1)
&=\int_0^{U_1}
\frac{\E_2\{\Delta_2^RQ_{1,0}(t;\bm Y_2)/K_{0,2}(R_2)\mid O_1\}}{K_{0,1}(t)}
\,\mathrm dM_{0,1}^C(t)\\
&=\int_0^{U_1}
\frac{\E_2\{Q_{1,0}(t;\bm Y_2)\mid O_1\}}{K_{0,1}(t)}
\,\mathrm dM_{0,1}^C(t)
=\int_0^{U_1}\frac{q_{1,0}^{B_1}(t)}{K_{0,1}(t)}
\,\mathrm dM_{0,1}^C(t).
\end{align*}
An unobserved $\bm Y_1$ is not evaluated when $\Delta_1^R=0$. Conditional on $O_1$, the integrands in the two remaining terms are predictable for participant 2. Independence of the clusters and the martingale property therefore imply
\begin{align*}
\E_2(\mathcal T_2\mid O_1)
&=\frac{\Delta_1^R}{K_{0,1}(R_1)}
\E_2\left\{\int_0^{U_2}\frac{Q_{2,0}(\bm Y_1;s)}{K_{0,2}(s)}
\,\mathrm dM_{0,2}^C(s)\,\middle|\,O_1\right\}=0,\\
\E_2(\mathcal T_{12}\mid O_1)
&=\E_2\left[\int_0^{U_2}\frac{1}{K_{0,2}(s)}
\left\{\int_0^{U_1}\frac{Q_{12,0}(t,s)}{K_{0,1}(t)}\,\mathrm dM_{0,1}^C(t)\right\}
\mathrm dM_{0,2}^C(s)\,\middle|\,O_1\right]=0.
\end{align*}
Square integrability justifies the interchanges of conditional expectation and integration. Adding these identities yields
\begin{align}
\E_2(\Gamma_{12}\mid O_1)
&=\frac{\Delta_1^RB_1}{K_{0,1}(R_1)}
+\int_0^{U_1}\frac{q_{1,0}^{B_1}(t)}{K_{0,1}(t)}\,\mathrm dM_{0,1}^C(t)
=\mathcal A_1(B_1;K_{0,1},q_{1,0}^{B_1}).
\label{eq:app-participant-projection}
\end{align}
Interchanging the participants gives the projection conditional on $O_2$.

Aggregation over participants in the retained cluster gives
\begin{align*}
\E_b(\overline\Gamma_{C,ik,a}^r\mid O_i)
&=N_i^{-1}\sum_j\xi_{C,a,a,ij}^r=\Xi_{C,a,a}^r,\\
\E_b(\overline\Gamma_{I,ik,a}^r\mid O_i)
&=\sum_j\xi_{I,a,a,ij}^r=\Xi_{I,a,a}^r.
\end{align*}
The opposite orientation gives $\Xi_{v,a,b}^r$. At the correct baseline regression, its two projections are $\bar\mu_{v,a,a}^r$ and $\bar\mu_{v,a,b}^r$. Applying \eqref{eq:app-randomization} gives
\begin{align*}
2\E\{g_{v,a}^r(O_i,O_k;\eta_0)\mid O_i\}
&=\bar\mu_{v,a,a}^r+\bar\mu_{v,a,b}^r
+\frac{\ind(A_i=a)}{\pi_a(\bm L_i)}(\Xi_{v,a,a}^r-\bar\mu_{v,a,a}^r)\\
&\quad+\frac{\ind(A_i=b)}{\pi_b(\bm L_i)}(\Xi_{v,a,b}^r-\bar\mu_{v,a,b}^r).
\end{align*}
Subtracting twice the kernel mean proves
\begin{align}
2[\E\{g_{C,a}^r(O_i,O_k;\eta_0)\mid O_i\}-\lambda_{C,a}^r]
&=\phi_{C,a}^r(O_i),\label{eq:kernel-projection-C}\\
2[\E\{g_{I,a}^r(O_i,O_k;\eta_0)\mid O_i\}-\theta_{I,a}^r]
&=\phi_{\theta I,a}^r(O_i).
\label{eq:kernel-projection-I}
\end{align}
The individual-average probability also includes the normalization contribution in \eqref{eq:observed-eif-I}. These identities retain the population baseline distribution when assignment probabilities depend on covariates.

For the double correction, correct censoring models alone give conditional mean zero given either cluster. Let $D(O_i,O_k)$ be its symmetric aggregation with inverse randomization weights. For pairs sharing one cluster,
\begin{align*}
\E\{D(O_1,O_2)D(O_1,O_3)\}
&=\E[\E\{D(O_1,O_2)\mid O_1\}\E\{D(O_1,O_3)\mid O_1\}]=0.
\end{align*}
Disjoint pairs are independent, so
\begin{align}
\E\{(U_mD)^2\}
&=\binom m2^{-2}\sum_{i<k}\E\{D(O_i,O_k)^2\}
=\frac{2\E(D^2)}{m(m-1)}.
\label{eq:app-degenerate-variance}
\end{align}
Thus $U_mD=\Op(m^{-1})$ and $\sqrt mU_mD=\op(1)$. Under either model condition, the variance estimator in \eqref{eq:hi-hat} uses the full pairwise kernel.

To compare baseline adjustments, suppose that the censoring and outcome models are correct, and let $d_s$ be the difference between a working baseline projection in arm $s$ and its true value. With $I_a=\ind(A=a)$ and $\pi_b=1-\pi_a$,
\begin{align*}
\phi^\mu-\phi^0
&=\left(1-\frac{I_a}{\pi_a}\right)d_a
+\left(1-\frac{1-I_a}{\pi_b}\right)d_b
=-(I_a-\pi_a)\left(\frac{d_a}{\pi_a}-\frac{d_b}{\pi_b}\right).
\end{align*}
Its conditional mean given $\bm L$ is zero. Its covariance with the outcome residuals is also zero because $\E_s(\Xi_s-\bar\mu_s^0\mid\bm L)=0$. Hence
\begin{align}
\Var(\phi^\mu)-\Var(\phi^0)
&=2\E\{\phi^0(\phi^\mu-\phi^0)\}+\E\{(\phi^\mu-\phi^0)^2\}\nonumber\\
&=\E\{\pi_a\pi_b(d_a/\pi_a-d_b/\pi_b)^2\}\geq0.
\label{eq:app-augmentation-gap}
\end{align}
For the individual-average probability, the influence difference is multiplied by $\nu^{-2}$ and the variance difference by $\nu^{-4}$. The mean-size contribution is baseline measurable and orthogonal to this difference. The true baseline projections therefore minimize variance within this class of adjustments.

\subsection{Influence functions for nuisance estimators}
\label{subsec:app-nuisance-equations}
\label{subsec:app-cox}

Write $P_mf=m^{-1}\sum_i f(O_i)$. Suppose that $P_m\bm S(O;\widehat\eta)=\bm0$, $\widehat\eta\to\eta^*$, $\E\{\bm S(O;\eta^*)\}=\bm0$, and $\bm A=\E\{\partial\bm S(O;\eta^*)/\partial\eta^{\mathsf T}\}$ is nonsingular. Assume a square integrable score and a locally uniform law of large numbers for its derivative. With $\bm A_m=\int_0^1P_m\dot{\bm S}_{\eta^*+t(\widehat\eta-\eta^*)}\,\mathrm dt$, Taylor expansion gives $\bm0=P_m\bm S(\eta^*)+\bm A_m(\widehat\eta-\eta^*)$. The uniform derivative law gives $\bm A_m\to\bm A$. Since $P_m\bm S(\eta^*)=\Op(m^{-1/2})$,
\begin{align*}
\widehat\eta-\eta^*
&=-\bm A_m^{-1}P_m\bm S(\eta^*)
=-\frac1m\sum_i\bm A^{-1}\bm S(O_i;\eta^*)+\op(m^{-1/2}).
\end{align*}
Thus $\bm\psi_i=-\bm A^{-1}\bm S(O_i;\eta^*)$.

For a symmetric pairwise equation $U_m\bm s(\widehat\eta)=\bm0$, let $\bm s_1(O_i)=\E\{\bm s(O_i,O_k;\eta^*)\mid O_i\}$ and let $\bm A$ be the population derivative. The population equation implies $\E\{\bm s_1(O_i)\}=\bm0$. The remainder $\bm s_2(O_i,O_k)=\bm s(O_i,O_k;\eta^*)-\bm s_1(O_i)-\bm s_1(O_k)$ has conditional mean zero given either cluster. Equation \eqref{eq:app-degenerate-variance} gives $U_m\bm s_2=\Op(m^{-1})$, and therefore
\begin{align*}
\bm0
&=U_m\bm s(\eta^*)+\bm A_m(\widehat\eta-\eta^*)\\
&=\frac2m\sum_i\bm s_1(O_i)+\bm A_m(\widehat\eta-\eta^*)+\op(m^{-1/2}),\\
\widehat\eta-\eta^*
&=-\frac2m\sum_i\bm A^{-1}\bm s_1(O_i)+\op(m^{-1/2}).
\end{align*}
Thus $\bm\psi_i=-2\bm A^{-1}\bm s_1(O_i)$. Stacking cluster scores and twice the pairwise score projections, and using their joint derivative matrix, retains the covariance among nuisance estimators. The same argument applies to jointly estimated outcome regression and association parameters. Function-valued components are treated below.

For the censoring model in \eqref{eq:crt-working-censoring}, fix treatment arm $s$ and allow predictable covariates $\bm Z_{ij}(t)$. Define $r_i^{(k)}(\bm\gamma_s,t)=\ind(A_i=s)\sum_jY_{ij}^C(t)e^{\bm\gamma_s^{\mathsf T}\bm Z_{ij}(t)}\bm Z_{ij}(t)^{\otimes k}$ for $k=0,1,2$, where $\bm Z^{\otimes0}=1$, $\bm Z^{\otimes1}=\bm Z$, and $\bm Z^{\otimes2}=\bm Z\bm Z^{\mathsf T}$. Write $s^{(k)}=\E(r_i^{(k)})$, $n_i^C(t)=\ind(A_i=s)\sum_jN_{ij}^C(t)$, $F_C(t)=\E\{n_i^C(t)\}$, $\bm e=s^{(1)}/s^{(0)}$, and $\bm v=s^{(2)}/s^{(0)}-\bm e\bm e^{\mathsf T}$. Let $\bm\gamma_s^*$ solve the population partial-likelihood equation and let $\Lambda_{0s}^{C,*}(t)=\int_0^t\mathrm dF_C(u)/s^{(0)}(\bm\gamma_s^*,u)$ be the limit of the Breslow estimator. With $\bm e_m=(P_mr^{(1)})/(P_mr^{(0)})$, the estimating function is
\begin{align*}
U_m^{\mathrm{Cox}}(\bm\gamma_s)
&=P_m\left\{\ind(A_i=s)\sum_j\int\bm Z_{ij}(t)\,\mathrm dN_{ij}^C(t)\right\}
-\int\bm e_m(\bm\gamma_s,t)\,\mathrm d(P_mn_i^C)(t).
\end{align*}
Assume the regularity conditions for clustered Cox estimation,\cite{lin1994cox} including a consistent interior solution, positive limiting risk denominators, a nonsingular derivative, and empirical-process conditions giving score and baseline-hazard remainders of order $\op(m^{-1/2})$. All quantities below are evaluated at $\bm\gamma_s^*$. The ratio derivative is $\delta\bm e=(\delta s^{(1)}-\bm e\delta s^{(0)})/s^{(0)}$. Since $s^{(1)}-\bm e s^{(0)}=\bm0$, differentiation of the empirical risk and event measures gives the centered cluster score
\begin{align*}
\bm\zeta_i
&=\ind(A_i=s)\sum_j\int\{\bm Z_{ij}(t)-\bm e(t)\}\,\mathrm dN_{ij}^C(t)
-\int\frac{r_i^{(1)}(t)-\bm e(t)r_i^{(0)}(t)}{s^{(0)}(t)}\,\mathrm dF_C(t)\\
&=\ind(A_i=s)\sum_j\int\{\bm Z_{ij}(t)-\bm e(t)\}
\{\mathrm dN_{ij}^C(t)-Y_{ij}^C(t)e^{\bm\gamma_s^{*\mathsf T}\bm Z_{ij}(t)}
\,\mathrm d\Lambda_{0s}^{C,*}(t)\}.
\end{align*}
Its expectation is zero by the population score equation. Since $\partial\bm e/\partial\bm\gamma_s^{\mathsf T}=\bm v$, the negative population derivative is $\bm J=\int\bm v(t)\,\mathrm dF_C(t)$, and
\begin{align*}
\bm0
&=P_m\bm\zeta_i-\bm J(\widehat{\bm\gamma}_s-\bm\gamma_s^*)+\op(m^{-1/2}),\\
\widehat{\bm\gamma}_s-\bm\gamma_s^*
&=\frac1m\sum_i\bm J^{-1}\bm\zeta_i+\op(m^{-1/2}).
\end{align*}
Thus $\IF_{\bm\gamma_s,i}=\bm J^{-1}\bm\zeta_i$, including under misspecification.

At fixed $\bm\gamma_s^*$, differentiation of the numerator and denominator of the Breslow estimator gives
\begin{align*}
&\int_0^t\frac{\mathrm dn_i^C(u)-\mathrm dF_C(u)}{s^{(0)}(u)}
-\int_0^t\frac{r_i^{(0)}(u)-s^{(0)}(u)}{s^{(0)}(u)}\,\mathrm d\Lambda_{0s}^{C,*}(u)\\
&\qquad=\int_0^t\frac{\mathrm dn_i^C(u)-r_i^{(0)}(u)\,\mathrm d\Lambda_{0s}^{C,*}(u)}{s^{(0)}(u)}.
\end{align*}
Here $\mathrm dF_C=s^{(0)}\mathrm d\Lambda_{0s}^{C,*}$. Incorporating estimation of $\bm\gamma_s$ gives
\begin{align*}
\IF_{\Lambda_{0s}^C,i}(t)
&=\int_0^t\frac{\mathrm dn_i^C(u)-r_i^{(0)}(u)\,\mathrm d\Lambda_{0s}^{C,*}(u)}{s^{(0)}(u)}
-\left\{\int_0^t\bm e(u)\,\mathrm d\Lambda_{0s}^{C,*}(u)\right\}^{\mathsf T}\IF_{\bm\gamma_s,i}.
\end{align*}
Both terms have mean zero. For a covariate path $\bm z(t)$, write $K_s^*(t)=\exp\{-\int_0^t e^{\bm\gamma_s^{*\mathsf T}\bm z(u)}\,\mathrm d\Lambda_{0s}^{C,*}(u)\}$. The chain rule gives
\begin{align*}
\IF_{K_s,i}(t)
&=-K_s^*(t)\Bigg[\int_0^t e^{\bm\gamma_s^{*\mathsf T}\bm z(u)}\,\mathrm d\IF_{\Lambda_{0s}^C,i}(u)\\
&\qquad+\left\{\int_0^t e^{\bm\gamma_s^{*\mathsf T}\bm z(u)}\bm z(u)\,\mathrm d\Lambda_{0s}^{C,*}(u)\right\}^{\mathsf T}\IF_{\bm\gamma_s,i}\Bigg].
\end{align*}
These regression and baseline-hazard contributions enter \eqref{eq:hi-hat}. Summing participant scores within clusters before computing their covariance accounts for within-cluster dependence.

\subsection{Consistency and asymptotic normality}
\label{sec:parametric-asymptotic-proof}
\label{subsec:app-functional-nuisance}

Let $\Psi_{v,a}^r(\eta)=\E\{g_{v,a}^r(O_i,O_k;\eta)\}$ and $\overline g_{v,a}^r(o;\eta)=\E\{g_{v,a}^r(o,O_k;\eta)\}$. The following conditions, together with Assumption \ref{ass:moments}, suffice for the results in Section \ref{subsec:crt-theory}.

\begin{condition}[Positivity]
\label{cond:positivity}
The randomization and censoring positivity assumptions hold. For some $\epsilon_K>0$, $K_{ij,s}(\tau;\eta)\geq\epsilon_K$ almost surely for all $\eta$ in a neighborhood of $\eta^*$.
\end{condition}

\begin{condition}[Regular nuisance estimation]
\label{cond:nuisance}
The nuisance estimator is consistent for an interior limit $\eta^*$ and satisfies
\begin{equation}
\sqrt m(\widehat\eta-\eta^*)=\frac1{\sqrt m}\sum_{i=1}^m\bm\psi_i+\op(1),
\label{eq:nuisance-linear}
\end{equation}
where the cluster contributions are independent and identically distributed, mean zero, and square integrable.
\end{condition}

\begin{condition}[Smoothness]
\label{cond:smoothness}
For finite-dimensional nuisance parameters, the kernel and its derivative are continuous in a compact neighborhood of $\eta^*$ and have square integrable envelopes there. The kernel mean has derivative $\bm D_{v,a}^r=\partial\Psi_{v,a}^r(\eta^*)/\partial\eta$. For function-valued components, the additional conditions below apply.
\end{condition}

\begin{condition}[Positive variance]
\label{cond:nondegeneracy}
The joint influence covariance is finite, and every contrast used for a standardized confidence interval has positive limiting variance.
\end{condition}

Under either model condition, the cluster influence contribution is
\begin{equation}
\chi_{\Psi,v,a,i}^r
=2\{\overline g_{v,a}^r(O_i;\eta^*)-\Psi_{v,a}^r(\eta^*)\}
+(\bm D_{v,a}^r)^{\mathsf T}\bm\psi_i.
\label{eq:influence-decomposition}
\end{equation}
Together with $\chi_{\nu,i}=N_i-\nu$, these form the influence vector with covariance $\bm\Sigma_{\mathrm{CRT}}$ in the main paper.

\begin{proof}[Proof of Theorems \ref{thm:dr-efficiency} and \ref{thm:asymptotic-normality}]
Set $L=-\log\epsilon_K$. Positivity implies $\Lambda^C(\tau)\leq L$, and the total variation of $M^C$ on $[0,U]$ is at most $1+L$. Since the working conditional comparison probabilities lie in $[0,1]$, \eqref{eq:crt-gamma} gives $|\Gamma_{12}|\leq\epsilon_K^{-2}(2+L)^2$. Suppose that the baseline regressions are bounded by a constant for the cluster-average target and by a constant times $N_iN_k$ for the individual-average numerator. Then $|g_C|\leq C$ and $|g_I|\leq CN_iN_k$ for a finite $C$. Both kernels are square integrable because $\E(N_i^2N_k^2)=\{\E(N_i^2)\}^2<\infty$. Condition \ref{cond:smoothness} supplies the derivative envelopes.

For a square integrable symmetric kernel $f$, let $f_1(o)=\E\{f(o,O_k)\}-\E(f)$ and $f_2(o,o')=f(o,o')-\E(f)-f_1(o)-f_1(o')$. The Hoeffding decomposition is
\begin{equation}
U_mf-\E(f)=\frac2m\sum_i f_1(O_i)+U_mf_2.
\label{eq:app-hoeffding}
\end{equation}
The calculation in Appendix \ref{subsec:app-irt-asymptotics} gives $\Var(U_mf)=4\E(f_1^2)/m+2\E(f_2^2)/\{m(m-1)\}$. Thus $U_mf\to\E(f)$ and $\sqrt mU_mf_2=\op(1)$.\cite{hoeffding1948class,vandervaart1998asymptotic}

For a kernel or derivative coordinate $f_\eta$, let $\mathcal N$ be the compact parameter neighborhood. Continuity and domination give, for any $\epsilon>0$, a finite cover with centers $\eta_l$ and neighborhoods $B_l$ such that $\omega_l=\sup_{\eta\in B_l}|f_\eta-f_{\eta_l}|$ satisfies $\E(\omega_l)<\epsilon$. Hence
\begin{align*}
\sup_{\eta\in\mathcal N}|U_mf_\eta-\E(f_\eta)|
&\leq\max_l|U_mf_{\eta_l}-\E(f_{\eta_l})|
+\max_lU_m\omega_l+\max_l\E(\omega_l)\\
&\leq2\epsilon+\op(1).
\end{align*}
Letting $\epsilon\downarrow0$ proves locally uniform convergence for the kernels and their derivatives. Under either model condition, Appendix \ref{subsec:app-crt-augmentation} gives $\Psi_{C,a}^r(\eta^*)=\lambda_{C,a}^r$ and $\Psi_{I,a}^r(\eta^*)=\theta_{I,a}^r$. Uniform convergence and $\widehat\eta\to\eta^*$ establish consistency of the two kernel averages. Since $\widehat\nu\to\nu>0$, their normalized estimators are also consistent, proving Theorem \ref{thm:dr-efficiency}.

Suppress target indices. The integrated Taylor expansion gives
\begin{align*}
U_mg_{\widehat\eta}-U_mg_{\eta^*}
&=\left\{\int_0^1U_m\dot g_{\eta^*+t(\widehat\eta-\eta^*)}\,\mathrm dt\right\}^{\mathsf T}(\widehat\eta-\eta^*)\\
&=\{\bm D+\op(1)\}^{\mathsf T}(\widehat\eta-\eta^*)
=\bm D^{\mathsf T}(\widehat\eta-\eta^*)+\op(m^{-1/2}).
\end{align*}
At $\eta^*$, the remainder $g_2(O_i,O_k)=g(O_i,O_k;\eta^*)-\overline g(O_i;\eta^*)-\overline g(O_k;\eta^*)+\Psi(\eta^*)$ has conditional mean zero given either cluster. Its mean square is $\E\{(U_mg_2)^2\}=2\E(g_2^2)/\{m(m-1)\}$. Combining these facts with \eqref{eq:app-hoeffding} and \eqref{eq:nuisance-linear} yields
\begin{align}
\sqrt m\{U_mg_{\widehat\eta}-\Psi(\eta^*)\}
&=\sqrt m\{U_mg_{\eta^*}-\Psi(\eta^*)\}
+\sqrt m\bm D^{\mathsf T}(\widehat\eta-\eta^*)+\op(1)\nonumber\\
&=\frac1{\sqrt m}\sum_i[2\{\overline g(O_i;\eta^*)-\Psi(\eta^*)\}
+\bm D^{\mathsf T}\bm\psi_i]+\op(1).
\label{eq:app-target-AL}
\end{align}
The multivariate central limit theorem applies jointly to the four kernel averages and $\widehat\nu$. For the individual-average probability, the map $(\theta,\nu)\mapsto\theta/\nu^2$ has gradient $(\nu^{-2},-2\theta\nu^{-3})^{\mathsf T}$ and a quadratic remainder of order $\Op(m^{-1})$. Thus
\begin{align}
\sqrt m(\widehat\lambda_{I,a}^r-\lambda_{I,a}^r)
&=\frac{\sqrt m(\widehat\theta_{I,a}^r-\theta_{I,a}^r)}{\nu^2}
-\frac{2\theta_{I,a}^r}{\nu^3}\sqrt m(\widehat\nu-\nu)+\op(1)\nonumber\\
&=\frac1{\sqrt m}\sum_i\left\{\frac{\chi_{\Psi,I,a,i}^r}{\nu^2}
-\frac{2\lambda_{I,a}^r}{\nu}(N_i-\nu)\right\}+\op(1).
\label{eq:app-I-AL}
\end{align}
This proves Theorem \ref{thm:asymptotic-normality}.
\end{proof}

The derivative with respect to the baseline regression is zero because randomization probabilities are known. For the censoring and outcome components, double robustness gives $\Psi(\eta_{C,0},\eta_O)=\vartheta$ and $\Psi(\eta_C,\eta_{O,0})=\vartheta$ throughout the relevant neighborhoods. Differentiation shows that the derivative for the misspecified component vanishes under either model condition. Estimation of the correctly specified component can contribute to \eqref{eq:influence-decomposition}. When both models are correct, both derivatives vanish. If the baseline regression is also correct, \eqref{eq:kernel-projection-C} and \eqref{eq:kernel-projection-I} identify the influence function of the practical estimator. It attains the bounds in Theorem \ref{thm:observed-eif} if and only if the corresponding residuals in \eqref{eq:app-gap-C} and \eqref{eq:app-gap-I} vanish, as specified in \eqref{eq:app-canonical-range}.

For estimated cumulative hazards, interpret \eqref{eq:nuisance-linear} in a normed space containing coefficients and hazard paths, and assume $\|\widehat\eta-\eta^*\|=\Op(m^{-1/2})$. Impose the differentiability, square integrability, and local U-process conditions of Appendix \ref{subsec:app-irt-asymptotics}, with $n$ replaced by $m$. In particular, $\dot\Psi_{\eta^*}$ is a bounded linear map, and locally
\begin{align*}
|\Psi(\eta)-\Psi(\eta^*)-\dot\Psi_{\eta^*}[\eta-\eta^*]|
&\leq C\|\eta-\eta^*\|^2.
\end{align*}
Writing $P^{\otimes2}g=\E\{g(O_1,O_2)\}$, require $\sqrt m(U_m-P^{\otimes2})(g_{\widehat\eta}-g_{\eta^*})=\op(1)$. Then
\begin{align*}
U_mg_{\widehat\eta}-U_mg_{\eta^*}
&=(U_m-P^{\otimes2})(g_{\widehat\eta}-g_{\eta^*})
+\Psi(\widehat\eta)-\Psi(\eta^*)\\
&=\dot\Psi_{\eta^*}[\widehat\eta-\eta^*]+\Op(m^{-1})+\op(m^{-1/2})
=\dot\Psi_{\eta^*}[\widehat\eta-\eta^*]+\op(m^{-1/2}).
\end{align*}
Equation \eqref{eq:app-target-AL} follows with $\bm D^{\mathsf T}\bm\psi_i$ interpreted as $\dot\Psi_{\eta^*}[\bm\psi_i]$, including the baseline-hazard contributions derived above. These regularity conditions supplement correct specification of either working model.

\subsection{Variance estimation and win summaries}
\label{sec:variance-proof}
\label{sec:app-derived-inference}

In addition to the preceding conditions, assume $U_m\|\bm g_{\widehat\eta}-\bm g_{\eta^*}\|^2=\op(1)$ and empirical mean square convergence of the propagated nuisance contributions. For finite-dimensional parameters, sufficient conditions are $\widehat{\bm D}\to\bm D$ and
\begin{equation}
\frac1m\sum_i\|\widehat{\bm\psi}_i-\bm\psi_i\|^2=\op(1).
\label{eq:app-nuisance-IF-consistency}
\end{equation}
For function-valued parameters, require $m^{-1}\sum_i\|\widehat{\dot{\bm\Psi}}[\widehat{\bm\psi}_i]-\dot{\bm\Psi}_{\eta^*}[\bm\psi_i]\|^2=\op(1)$ directly.

Conditional on $O_i$, the terms $g(O_i,O_k;\eta^*)-\overline g(O_i;\eta^*)$, $k\ne i$, are independent and centered. Consequently,
\begin{align}
\E\left[\frac1m\sum_i\left\{\frac1{m-1}\sum_{k\ne i}g(O_i,O_k;\eta^*)
-\overline g(O_i;\eta^*)\right\}^2\right]
&=\frac{\E[\Var\{g(O_i,O_k;\eta^*)\mid O_i\}]}{m-1}\nonumber\\
&\leq\frac{\E(g_{\eta^*}^2)}{m-1}.
\label{eq:app-row-L2}
\end{align}
For $h_{ik}=g(O_i,O_k;\widehat\eta)-g(O_i,O_k;\eta^*)$, Jensen's inequality gives
\begin{align*}
\frac1m\sum_i\left(\frac1{m-1}\sum_{k\ne i}h_{ik}\right)^2
&\leq\frac1{m(m-1)}\sum_i\sum_{k\ne i}h_{ik}^2
=U_mh^2=\op(1).
\end{align*}
This inequality does not require independence between $\widehat\eta$ and the observations. In finite dimensions, the nuisance contribution satisfies
\begin{align*}
P_m|\widehat{\bm D}^{\mathsf T}\widehat{\bm\psi}_i-\bm D^{\mathsf T}\bm\psi_i|^2
&\leq2\|\widehat{\bm D}-\bm D\|^2P_m\|\bm\psi_i\|^2
+2\|\widehat{\bm D}\|^2P_m\|\widehat{\bm\psi}_i-\bm\psi_i\|^2
=\op(1).
\end{align*}
Together with point-estimator consistency, these bounds give $P_m\|\widehat{\bm\chi}_i-\bm\chi_i\|^2=\op(1)$ for the vector in \eqref{eq:hi-hat}. For score estimators, \eqref{eq:app-nuisance-IF-consistency} follows from consistency of the estimated inverse derivative and a square integrable local envelope for score differences. For pairwise nuisance equations, \eqref{eq:app-row-L2} also controls the estimated score projection.

For the individual-average contribution, suppress $a,r$ and put $a_0=\nu^{-2}$ and $b_0=2\lambda_I/\nu$, with fitted values $\widehat a_0,\widehat b_0$. Direct subtraction and the inequality $(x_1+\cdots+x_4)^2\leq4\sum_{k=1}^4x_k^2$ give
\begin{align*}
\widehat\chi_{I,i}-\chi_{I,i}
&=(\widehat a_0-a_0)\chi_{\Psi,I,i}
+\widehat a_0(\widehat\chi_{\Psi,I,i}-\chi_{\Psi,I,i})
-(\widehat b_0-b_0)(N_i-\nu)+\widehat b_0(\widehat\nu-\nu),\\
P_m(\widehat\chi_{I,i}-\chi_{I,i})^2
&\leq4|\widehat a_0-a_0|^2P_m\chi_{\Psi,I,i}^2
+4\widehat a_0^2P_m(\widehat\chi_{\Psi,I,i}-\chi_{\Psi,I,i})^2\\
&\quad+4|\widehat b_0-b_0|^2P_m(N_i-\nu)^2
+4\widehat b_0^2(\widehat\nu-\nu)^2
=\op(1).
\end{align*}
The last equality follows from consistency, the numerator result, and finite second moments.

Set $\bm e_i=\widehat{\bm\chi}_i-\bm\chi_i$. Expansion and the Cauchy--Schwarz inequality give
\begin{align*}
\left\|P_m(\widehat{\bm\chi}_i\widehat{\bm\chi}_i^{\mathsf T}-\bm\chi_i\bm\chi_i^{\mathsf T})\right\|
&\leq2\{P_m\|\bm\chi_i\|^2\}^{1/2}\{P_m\|\bm e_i\|^2\}^{1/2}
+P_m\|\bm e_i\|^2=\op(1).
\end{align*}
The law of large numbers therefore yields
\begin{align}
\widehat{\bm\Sigma}_{\mathrm{CRT}}
&=P_m\widehat{\bm\chi}_i\widehat{\bm\chi}_i^{\mathsf T}
=P_m\bm\chi_i\bm\chi_i^{\mathsf T}+\op(1)
\longrightarrow\E(\bm\chi_i\bm\chi_i^{\mathsf T})
=\bm\Sigma_{\mathrm{CRT}}.
\label{eq:app-sandwich-consistency}
\end{align}
The same argument applies to the normalized probability vector. Moreover, $\|P_m\widehat{\bm\chi}_i\|\leq\{P_m\|\bm e_i\|^2\}^{1/2}+\|P_m\bm\chi_i\|=\op(1)$, so centering removes only an $\op(1)$ outer product. A finite sample multiplier tending to one also preserves consistency.

For either target, write $\bm\lambda=(\lambda^+,\lambda^-)^{\mathsf T}$ and let $(\chi_i^+,\chi_i^-)^{\mathsf T}$ be its influence vector. Differentiating the summaries in the main paper gives
\begin{align}
\chi_i^{=}&=-\chi_i^+-\chi_i^-,\nonumber\\
\chi_{\mathrm{NB},i}&=\chi_i^+-\chi_i^-,\nonumber\\
\chi_{\log\mathrm{WR},i}&=\chi_i^+/\lambda^+-\chi_i^-/\lambda^-,\nonumber\\
\chi_{\log\mathrm{WO},i}&=\frac{2(\chi_i^+-\chi_i^-)}{1-(\lambda^+-\lambda^-)^2}.
\label{eq:app-summary-IF}
\end{align}
For the last identity, $\mathrm{WO}=(1+\mathrm{NB})/(1-\mathrm{NB})$ and
\begin{align*}
\frac{\mathrm d}{\mathrm db}\{\log(1+b)-\log(1-b)\}
&=\frac1{1+b}+\frac1{1-b}=\frac2{1-b^2}.
\end{align*}
The joint covariance gives \eqref{eq:win-summary-variance}. At interior probabilities and positive limiting contrast variance, the delta method and Slutsky's theorem give nominal asymptotic coverage for the logarithmic intervals. Exponentiation preserves the coverage event.

For the individual-average win ratio, $\widehat{\mathrm{WR}}_I=\widehat\theta_I^+/\widehat\theta_I^-$. Substitution of \eqref{eq:app-I-AL} gives
\begin{align*}
\chi_{\log\mathrm{WR},I,i}
&=\frac{\chi_{\Psi,I,i}^+}{\nu^2\lambda_I^+}
-\frac{2(N_i-\nu)}{\nu}
-\frac{\chi_{\Psi,I,i}^-}{\nu^2\lambda_I^-}
+\frac{2(N_i-\nu)}{\nu}
=\frac{\chi_{\Psi,I,i}^+}{\theta_I^+}-\frac{\chi_{\Psi,I,i}^-}{\theta_I^-}.
\end{align*}
Mean-size normalization remains in the influence functions for the individual-average probabilities, net benefit, and win odds.

The tie probability is estimated by the complement. With $s_N^2=(m-1)^{-1}\sum_i(N_i-\widehat\nu)^2$,
\begin{align*}
\frac{U_m(N_iN_k)}{\widehat\nu^2}
&=\frac{(\sum_iN_i)^2-\sum_iN_i^2}{m(m-1)\widehat\nu^2}
=1-\frac{s_N^2}{m\widehat\nu^2}.
\end{align*}
Since $s_N^2=\Op(1)$ and $\widehat\nu\to\nu>0$, $s_N^2/(m\widehat\nu^2)=\Op(m^{-1})$.

\subsection{Fixed treatment totals in cluster-randomized trials}
\label{sec:fixed-allocation-proof}

Suppose that simple complete randomization assigns exactly $m_s$ clusters to arm $s$, independently of their baseline and potential outcome records, with $p_{s,m}=m_s/m\to p_s\in(0,1)$. The ordered cross-arm assignment probability is $m_am_b/\{m(m-1)\}$. Conditional on the assignment vector, the records in each arm are independent samples from their arm-specific distributions.

The augmented kernel includes baseline regression terms for same-arm and cross-arm pairs. Let $g_m$ use the products $p_{s,m}p_{t,m}$ of marginal assignment probabilities. Define $k_{st,m}(x)=\E_t\{g_m(x,O_t)\}$, $G_{st,m}=\E_s\{k_{st,m}(O_s)\}$, and $b_{s,m}(x)=\sum_t p_{t,m}k_{st,m}(x)$. The mixture mean is $\Psi_m=\sum_{s,t}p_{s,m}p_{t,m}G_{st,m}$, which equals the target under either model condition. Counting pairs gives
\begin{align*}
\E(U_mg_m\mid\bm A)
&=\frac{\sum_sm_s(m_s-1)G_{ss,m}+2m_0m_1G_{01,m}}{m(m-1)}\\
&=\Psi_m+\frac{\Psi_m-\sum_sp_{s,m}G_{ss,m}}{m-1}
=\Psi_m+O(m^{-1}),
\end{align*}
where the last equality follows from the bounded second-moment envelopes.

For an observation in arm $s$, its projection over the available partners is $\widetilde b_{s,m}(x)=\{mb_{s,m}(x)-k_{ss,m}(x)\}/(m-1)$. For an arm-$s$, arm-$t$ pair, set $r_{st,m}(x,y)=g_m(x,y)-k_{st,m}(x)\allowbreak-k_{ts,m}(y)+G_{st,m}$. Symmetry gives $G_{st,m}=G_{ts,m}$, and integration over either argument gives zero. Conditional centering makes remainder terms for pairs sharing one cluster uncorrelated; disjoint pairs are independent. Their total variance is $O(m^2)/\binom m2^2=O(m^{-2})$. The linear projections therefore give
\begin{align*}
U_mg_m-\E(U_mg_m\mid\bm A)
&=\frac2m\sum_s\sum_{i:A_i=s}
\{\widetilde b_{s,m}(O_i)-\E_s(\widetilde b_{s,m})\}+\op(m^{-1/2})\\
&=\frac2m\sum_s\sum_{i:A_i=s}
\{b_{s,m}(O_i)-\E_s(b_{s,m})\}+\op(m^{-1/2}),
\end{align*}
since $\widetilde b_{s,m}-b_{s,m}=(b_{s,m}-k_{ss,m})/(m-1)$.

Assume $b_{s,m}\to b_s$ in $L^2(P_s^O)$, uniformly integrable squared envelopes, and within-arm nuisance expansions with centered contributions $\bm\psi_s$. The independent within-arm central limit theorems then give influence contributions $2\{b_s-\E_s(b_s)\}+\dot\Psi[\bm\psi_s]$ and covariance $\sum_sp_s\Var_s(\bm\chi_s)$ for the joint vector. Randomization gives the same baseline distribution in both arms, so the mean-size contribution remains $N_i-\nu$. The individual-average transformation applies to this joint vector.

For the exact ordered-pair probability,
\begin{align*}
q_{ab,m}^{-1}
&=\frac{m(m-1)}{m_am_b}
=(1-m^{-1})(p_{a,m}p_{b,m})^{-1}.
\end{align*}
Thus replacing the product probability changes the inverse weight by $O(m^{-1})$. The square integrable kernel envelope makes the corresponding estimator difference $\Op(m^{-1})$, preserving the limiting distribution while restoring the exact finite sample randomization identity.

The empirical mean square arguments above establish consistency of the estimated within-arm covariances. Hence \eqref{eq:app-irt-fixed-cov}, with $n,n_s$ replaced by $m,m_s$ and full cluster influence vectors, consistently estimates the covariance of $\widehat{\bm\vartheta}$. By the law of total variance, fixing the arm counts removes variation in the arm means and retains $\sum_sp_s\Var_s(\bm\chi_s)$. Other restricted randomization designs require their own joint assignment probabilities and covariance calculation.

\section{Working models}
\label{sec:app-working-models}

This section collects the likelihood contributions, conditional distributions, and application-specific choices for the working models in Sections \ref{subsec:irt-working-models} and \ref{subsec:crt-working-models}. Contributions from nuisance estimation are derived in Appendix \ref{subsec:app-nuisance-equations}.

\subsection{Joint outcome models}
\label{subsec:app-continuation}

Consider death and one nonfatal component $q$, with joint survival function $S_{Dq,a}(d,u\mid\bm L)$ in \eqref{eq:irt-working-copula}. The first argument is the death time and the second is the latent nonfatal event time. Under conditional independent censoring, the likelihood factors into outcome and censoring contributions. Suppressing the participant index, the outcome contribution is
\begin{align*}
L^O(\eta)
&=\begin{cases}
S_{Dq,a}(t,t\mid\bm L), & \text{neither event observed by }t,\\
-\partial_u S_{Dq,a}(t,u\mid\bm L), & \text{nonfatal event at }u<t\text{ and survival through }t,\\
-\partial_d S_{Dq,a}(d,u\mid\bm L)\big|_{u=d}, & \text{death at }d\text{ without a prior nonfatal event},\\
\partial_d\partial_u S_{Dq,a}(d,u\mid\bm L), & \text{nonfatal event at }u<d\text{ followed by death at }d.
\end{cases}
\end{align*}
For death without a prior nonfatal event, differentiation with respect to $d$ precedes evaluation at $u=d$. This integrates over latent nonfatal event times beyond death. The regression coefficients, baseline cumulative hazards, and association parameter in arm $a$ jointly maximize $\ell_a^O(\eta)=\sum_{i:A_i=a}\log L_i^O(\eta)$.

Before either event occurs, for $d,u\geq t$,
\begin{align*}
\Prb_\eta(T_D>d,T_q>u\mid T_D>t,T_q>t,A=a,\bm L)
&=\frac{\Prb_\eta(T_D>d,T_q>u\mid A=a,\bm L)}{\Prb_\eta(T_D>t,T_q>t\mid A=a,\bm L)}\\
&=\frac{S_{Dq,a}(d,u\mid\bm L)}{S_{Dq,a}(t,t\mid\bm L)}.
\end{align*}
After a nonfatal event at $u<t$, the conditional survival probability for $d\geq t$ is $\Prb_\eta(T_D>d\mid T_q=u,T_D>t,A=a,\bm L)=\{-\partial_u S_{Dq,a}(d,u\mid\bm L)\}/\{-\partial_u S_{Dq,a}(t,u\mid\bm L)\}$. The fitted distributions are restricted at $\tau$, with $T_i^q=\infty$ when death precludes the nonfatal event, as in Section \ref{subsec:irt-estimands}. The baseline and observed history regressions follow by integrating the comparison indicator against these distributions, as specified in Section \ref{subsec:irt-working-models}. In particular,
\begin{align*}
Q_{12}^r(t,s)
&=\int\!\int h_r(\bm y_1,\bm y_2)\,
 dF_1(\bm y_1\mid\mathcal F_{1,t^-}^{O},R_1\geq t)\,
 dF_0(\bm y_2\mid\mathcal F_{2,s^-}^{O},R_2\geq s)\\
&=\int Q_1^r(t;\bm y_2)\,
 dF_0(\bm y_2\mid\mathcal F_{2,s^-}^{O},R_2\geq s)\\
&=\int Q_2^r(\bm y_1;s)\,
 dF_1(\bm y_1\mid\mathcal F_{1,t^-}^{O},R_1\geq t).
\end{align*}
Here $F_a$ is the restricted outcome distribution under the working model in arm $a$. The equality shows why all three regressions must use the same two outcome models. For additional nonfatal components, the multivariate model in Section \ref{subsec:irt-working-models} is fitted by integrating over latent event times consistent with each observed history. The same restriction and comparison rules then give the required conditional expectations.

The copula families are listed in Table \ref{tab:main-copula-choices} of the main paper. The reduced outcome models in the simulation studies retain $V_i$ alone; the CRT reduced outcome model also imposes conditional independence between hospitalization and death. The remaining simulation model specifications are given in Section \ref{sec:simulation}.

\subsection{Censoring models}
\label{subsec:app-censoring-models}

The arm-specific Cox models and their partial-likelihood and Breslow estimators are given in Section \ref{subsec:irt-working-models}. Their response is $(U_i,1-\Delta_i^R)$: censoring before resolution is an event, and death or survival through $\tau$ ends the censoring risk period. A nonfatal event does not end observation for death. The fitted survival function and cumulative hazard determine the weights and censoring residuals in \eqref{eq:irt-gamma}. Under censoring independent of outcomes and baseline covariates within arms, the Kaplan--Meier alternative in that section may be used. The influence contributions for the Cox coefficients and baseline cumulative hazards are derived in Appendix \ref{subsec:app-cox}.

\subsection{Models for participants in a cluster}
\label{subsec:app-crt-working-models}

For CRTs, the outcome likelihood contributions above use $w_{ij,a}(d,u\mid\bm L_i)$ from \eqref{eq:crt-working-copula}. The full cluster model in \eqref{eq:crt-cluster-copula} distinguishes within-participant association from between-participant association; the participant margins supply the working likelihood used here. The working log likelihood is $\ell_a^O(\eta)=\sum_{i:A_i=a}\sum_{j=1}^{N_i}\log L_{ij}^O(\eta)$, and maximization fits the participant outcome models under working independence.\cite{lin1994cox,fang2026estimands} Censoring models use the covariates and risk sets in Section \ref{subsec:crt-working-models}. Under Assumption \ref{ass:car}, each participant's censoring time and event outcomes are conditionally independent given treatment and the complete baseline cluster record. Conditional predictions therefore use the participant's observed history $\mathcal F_{ij,t}^{O}$ together with that baseline information. All nuisance influence contributions are aggregated within clusters for variance estimation.

\subsection{HF-ACTION}
\label{subsec:hf-models}

The outcome and censoring models are fitted separately by treatment arm. Both include age, recorded sex, New York Heart Association functional class, HF etiology, left ventricular ejection fraction, systolic blood pressure, estimated glomerular filtration rate, peak oxygen uptake, Kansas City Cardiomyopathy Questionnaire overall score, diabetes, atrial fibrillation or flutter, HF hospitalization in the preceding six months, and enrollment outside the United States. Missing baseline values are replaced by fixed values, with indicators for missing estimated glomerular filtration rate, missing peak oxygen uptake, and any other missing baseline covariate. All released participants are retained.

The outcome model combines proportional hazards margins for hospitalization and death through a Gumbel survival copula. The working Cox censoring model uses the baseline covariates listed above. The fitted Gumbel parameters are 1.262 in usual care and 1.278 in exercise training, corresponding to conditional Kendall coefficients of 0.208 and 0.218. The event histories and supplementary results are given in Appendix \ref{sec:app-hf-action}.

\subsection{STRIDE}
\label{subsec:stride-models}

The models use the baseline covariates listed in Section \ref{subsec:stride}, with indicators for health care system and prerandomization practice size tertile. They also include urban or rural practice location and indicators for missing recorded sex and unknown, refused, or missing recorded race. The outcome model combines proportional hazards margins for first self-reported fall injury and death through a Gumbel survival copula. Models are fitted under working independence across participants, and conditional expectations use each participant's observed history and the baseline information specified above.

The censoring model is a Cox regression fitted separately by arm. Death ends the censoring risk period, whereas observation for mortality continues after a first injury. The fitted Gumbel parameters for enhanced usual care and intervention are 1.0093 and 1.0014. The independence sensitivity model fixes this parameter at one and retains the same margins and covariates. Inference and sensitivity results are given in Appendix \ref{sec:app-stride}.

\section{Simulation studies}
\label{sec:app-simulation}

The designs and working models are specified in Section \ref{sec:simulation}. This section reports calibrated censoring parameters, reference values, and supplementary results. Summaries use replications with available estimates and confidence intervals. The tables use RB for signed relative bias in percent, AB for absolute bias, MCSD for Monte Carlo standard deviation, ASE for average analytic standard error, and COV for coverage of nominal $95\%$ normal Wald intervals. Standard deviations and standard errors refer to the logarithms of the win ratio and win odds, and to the difference in win and loss probabilities for net benefit. Model abbreviations follow Section \ref{sec:simulation}. WR, WO, and NB denote the win ratio, win odds, and net benefit.

In the summary figures, rows A to E give the true value, bias, MCSD, ASE/MCSD, and COV. Bias is relative for WR and WO and absolute for NB. The solid, dashed, and dotted lines in row A identify $\tau=1$, $1.5$, and $2$, respectively. Reference lines mark zero bias, ASE/MCSD equal to one, and coverage of $0.95$.

\subsection{Individual-randomized trials}
\label{sec:app-irt-simulation}
\label{subsec:app-irt-generation}
\label{subsec:app-irt-reference}
\label{subsec:app-irt-supp-results}

The Gumbel dependence in \eqref{eq:sim-irt-copula} is generated using a positive stable variable shared by the two event times, followed by the inverse Weibull distributions in \eqref{eq:sim-irt-margins}. Censoring times are generated independently of the event times conditional on treatment and baseline covariates. The Weibull censoring scale $\kappa_C$ is calibrated by Monte Carlo to obtain $\Prb(C_i<D_i\wedge2)=c$ and is held fixed across replications. Table \ref{tab:irt-calibration} reports the calibrated values for each conditional Kendall coefficient $\tau_W$ and censoring probability $c$.

\begin{table}[!htbp]
\centering
\caption{Calibrated censoring parameters for the IRT simulations.}
\label{tab:irt-calibration}
\begin{tabular}{@{}rrr@{}}
\toprule
$\tau_W$ & Target $c$ & $\kappa_C$ \\
\midrule
0.450 & 0.300 & 0.21170352 \\
0.450 & 0.600 & 0.65263586 \\
0.700 & 0.300 & 0.21161268 \\
0.700 & 0.600 & 0.65217070 \\
\bottomrule
\end{tabular}
\end{table}

The win and loss probabilities are averaged across the four uncensored reference populations in Section \ref{subsec:sim-irt} before calculating the win statistics. Table \ref{tab:irt-reference} reports the values at each restriction time, with $p_W$, $p_L$, and $p_T$ denoting the win, loss, and tie probabilities. The reference populations do not depend on the censoring setting.

\begin{table}[!htbp]
\centering
\caption{Reference probabilities and win estimands for the IRT simulations.}
\label{tab:irt-reference}
\begingroup
\begin{tabular}{@{}rrrrrrrr@{}}
\toprule
$\tau_W$ & $\tau$ & $p_W$ & $p_L$ & $p_T$ & WR & WO & NB \\
\midrule
0.450 & 1.0 & 0.252896 & 0.197900 & 0.549204 & 1.277901 & 1.116395 & 0.054997 \\
0.450 & 1.5 & 0.325336 & 0.257518 & 0.417147 & 1.263352 & 1.145503 & 0.067818 \\
0.450 & 2.0 & 0.372457 & 0.299106 & 0.328437 & 1.245232 & 1.158313 & 0.073351 \\
\addlinespace[4pt]
0.700 & 1.0 & 0.243492 & 0.188748 & 0.567760 & 1.290040 & 1.115830 & 0.054744 \\
0.700 & 1.5 & 0.315315 & 0.247187 & 0.437497 & 1.275614 & 1.146218 & 0.068128 \\
0.700 & 2.0 & 0.363170 & 0.288492 & 0.348338 & 1.258857 & 1.161410 & 0.074678 \\
\bottomrule
\end{tabular}
\endgroup
\end{table}

\begin{figure}[!htbp]
\centering
\includegraphics[width=0.97\textwidth,height=0.67\textheight,keepaspectratio]{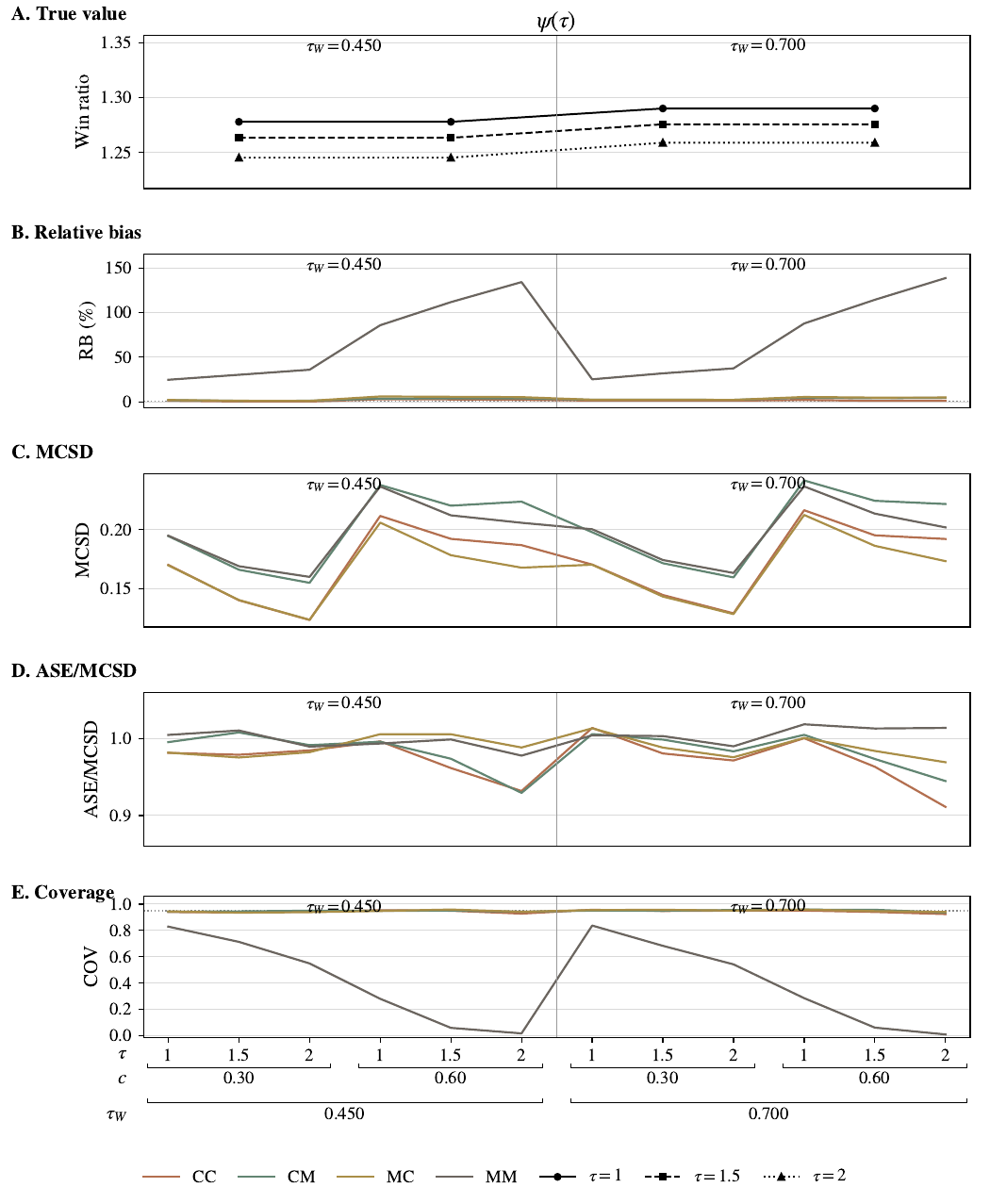}
\caption{Win ratio results for IRTs with $n=500$, equal treatment allocation, and death prioritized over first hospitalization. Both indifference thresholds are zero.}
\label{fig:irt-wr-all}
\end{figure}

Table \ref{tab:irt-app-wr-high} reports the win ratio results for $\tau_W=0.700$, with the two censoring settings presented side by side. Figure \ref{fig:irt-wr-all} summarizes the win ratio results across all IRT configurations. The horizontal axes of the IRT figures group $\tau\in\{1,1.5,2\}$ within $\tau_W\in\{0.450,0.700\}$ and $c\in\{0.300,0.600\}$.

\begin{table}[!htbp]
\centering
\caption{Doubly robust win ratio results for IRTs with $n=500$ and $\tau_W=0.700$.}
\label{tab:irt-app-wr-high}
\begingroup
\begin{tabular*}{\textwidth}{@{\extracolsep{\fill}}lrrrr@{\hspace{14pt}}rrrr@{}}
\toprule
& \multicolumn{4}{c}{$c=0.300$} & \multicolumn{4}{c}{$c=0.600$} \\
\cmidrule(lr){2-5}\cmidrule(l){6-9}
Model & RB (\%) & MCSD & ASE & COV & RB (\%) & MCSD & ASE & COV \\
\midrule
\multicolumn{9}{@{}l}{$\tau=1.0$, $\mathrm{WR}_0=1.290$} \\
$\mathrm{CC}$ & 1.418 & 0.170 & 0.173 & 0.957 & 2.073 & 0.217 & 0.217 & 0.951 \\
$\mathrm{CM}$ & 2.254 & 0.198 & 0.199 & 0.951 & 4.560 & 0.242 & 0.243 & 0.958 \\
$\mathrm{MC}$ & 2.339 & 0.170 & 0.173 & 0.955 & 5.421 & 0.213 & 0.213 & 0.959 \\
$\mathrm{MM}$ & 25.220 & 0.200 & 0.201 & 0.837 & 87.776 & 0.237 & 0.241 & 0.285 \\
\addlinespace[5pt]
\multicolumn{9}{@{}l}{$\tau=1.5$, $\mathrm{WR}_0=1.276$} \\
$\mathrm{CC}$ & 1.448 & 0.144 & 0.142 & 0.948 & 1.096 & 0.195 & 0.188 & 0.940 \\
$\mathrm{CM}$ & 2.252 & 0.172 & 0.171 & 0.950 & 4.044 & 0.225 & 0.219 & 0.957 \\
$\mathrm{MC}$ & 2.289 & 0.143 & 0.142 & 0.957 & 4.731 & 0.186 & 0.183 & 0.947 \\
$\mathrm{MM}$ & 31.894 & 0.174 & 0.175 & 0.683 & 114.350 & 0.214 & 0.217 & 0.061 \\
\addlinespace[5pt]
\multicolumn{9}{@{}l}{$\tau=2.0$, $\mathrm{WR}_0=1.259$} \\
$\mathrm{CC}$ & 1.336 & 0.129 & 0.125 & 0.951 & 1.267 & 0.192 & 0.175 & 0.925 \\
$\mathrm{CM}$ & 2.153 & 0.159 & 0.157 & 0.955 & 4.527 & 0.222 & 0.210 & 0.936 \\
$\mathrm{MC}$ & 2.153 & 0.128 & 0.125 & 0.951 & 4.847 & 0.173 & 0.168 & 0.940 \\
$\mathrm{MM}$ & 37.465 & 0.163 & 0.162 & 0.542 & 138.643 & 0.202 & 0.205 & 0.009 \\
\bottomrule
\end{tabular*}
\endgroup
\end{table}

\FloatBarrier

\begin{figure}[!htbp]
\centering
\includegraphics[width=0.97\textwidth,height=0.67\textheight,keepaspectratio]{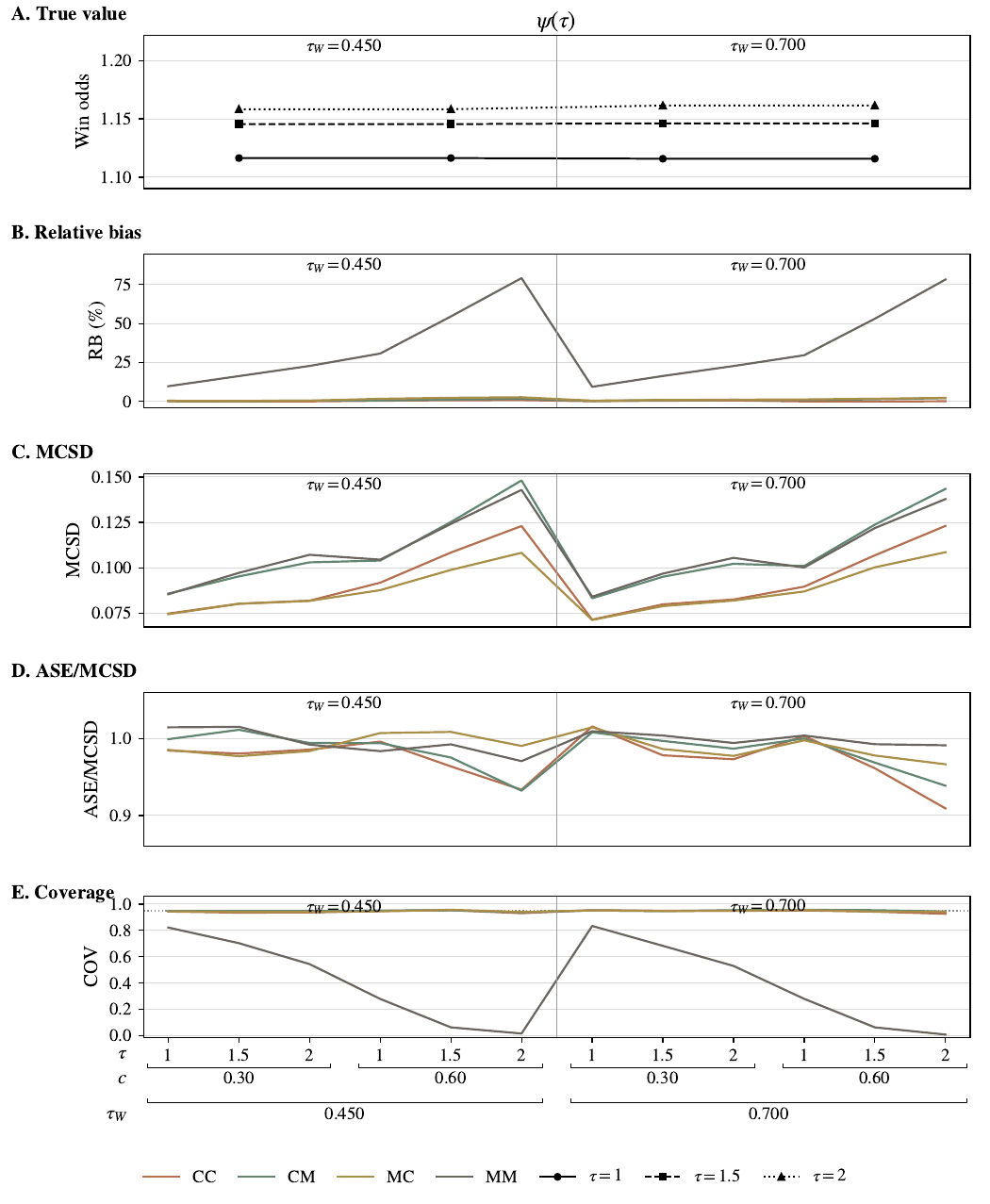}
\caption{Win odds results for IRTs with $n=500$ and equal treatment allocation.}
\label{fig:irt-wo-all}
\end{figure}

Tables \ref{tab:irt-app-wo-low} and \ref{tab:irt-app-wo-high} report the win odds results. Relative bias under $\mathrm{CC}$, $\mathrm{CM}$, and $\mathrm{MC}$ ranges from $-0.136\%$ to $1.051\%$ with $30\%$ censoring and from $-0.130\%$ to $2.638\%$ with $60\%$ censoring. The corresponding coverage ranges are $0.935$--$0.954$ and $0.927$--$0.957$. Under $\mathrm{MM}$, relative bias ranges from $9.341\%$ to $22.748\%$ with $30\%$ censoring and from $29.608\%$ to $79.167\%$ with $60\%$ censoring. Figure \ref{fig:irt-wo-all} summarizes these results.

\begin{table}[!htbp]
\centering
\caption{Doubly robust win odds results for IRTs with $n=500$ and $\tau_W=0.450$.}
\label{tab:irt-app-wo-low}
\begingroup
\begin{tabular*}{\textwidth}{@{\extracolsep{\fill}}lrrrr@{\hspace{14pt}}rrrr@{}}
\toprule
& \multicolumn{4}{c}{$c=0.300$} & \multicolumn{4}{c}{$c=0.600$} \\
\cmidrule(lr){2-5}\cmidrule(l){6-9}
Model & RB (\%) & MCSD & ASE & COV & RB (\%) & MCSD & ASE & COV \\
\midrule
\multicolumn{9}{@{}l}{$\tau=1.0$, $\mathrm{WO}_0=1.116$} \\
$\mathrm{CC}$ & 0.030 & 0.075 & 0.074 & 0.944 & 0.401 & 0.092 & 0.092 & 0.948 \\
$\mathrm{CM}$ & 0.233 & 0.086 & 0.086 & 0.948 & 0.581 & 0.104 & 0.103 & 0.952 \\
$\mathrm{MC}$ & 0.360 & 0.074 & 0.073 & 0.945 & 1.641 & 0.088 & 0.088 & 0.945 \\
$\mathrm{MM}$ & 9.710 & 0.085 & 0.087 & 0.822 & 30.714 & 0.105 & 0.103 & 0.279 \\
\addlinespace[5pt]
\multicolumn{9}{@{}l}{$\tau=1.5$, $\mathrm{WO}_0=1.146$} \\
$\mathrm{CC}$ & -0.136 & 0.080 & 0.079 & 0.935 & 0.723 & 0.108 & 0.105 & 0.954 \\
$\mathrm{CM}$ & 0.204 & 0.095 & 0.096 & 0.948 & 1.193 & 0.125 & 0.122 & 0.953 \\
$\mathrm{MC}$ & 0.303 & 0.080 & 0.078 & 0.941 & 2.282 & 0.099 & 0.100 & 0.957 \\
$\mathrm{MM}$ & 16.224 & 0.097 & 0.099 & 0.703 & 54.542 & 0.124 & 0.123 & 0.063 \\
\addlinespace[5pt]
\multicolumn{9}{@{}l}{$\tau=2.0$, $\mathrm{WO}_0=1.158$} \\
$\mathrm{CC}$ & -0.026 & 0.082 & 0.081 & 0.937 & 0.806 & 0.123 & 0.115 & 0.931 \\
$\mathrm{CM}$ & 0.469 & 0.103 & 0.102 & 0.947 & 1.655 & 0.148 & 0.138 & 0.938 \\
$\mathrm{MC}$ & 0.459 & 0.082 & 0.081 & 0.940 & 2.638 & 0.108 & 0.107 & 0.941 \\
$\mathrm{MM}$ & 22.748 & 0.107 & 0.106 & 0.544 & 79.167 & 0.143 & 0.139 & 0.016 \\
\bottomrule
\end{tabular*}
\endgroup
\end{table}

\begin{table}[!htbp]
\centering
\caption{Doubly robust win odds results for IRTs with $n=500$ and $\tau_W=0.700$.}
\label{tab:irt-app-wo-high}
\begingroup
\begin{tabular*}{\textwidth}{@{\extracolsep{\fill}}lrrrr@{\hspace{14pt}}rrrr@{}}
\toprule
& \multicolumn{4}{c}{$c=0.300$} & \multicolumn{4}{c}{$c=0.600$} \\
\cmidrule(lr){2-5}\cmidrule(l){6-9}
Model & RB (\%) & MCSD & ASE & COV & RB (\%) & MCSD & ASE & COV \\
\midrule
\multicolumn{9}{@{}l}{$\tau=1.0$, $\mathrm{WO}_0=1.116$} \\
$\mathrm{CC}$ & 0.068 & 0.072 & 0.073 & 0.954 & -0.031 & 0.090 & 0.090 & 0.952 \\
$\mathrm{CM}$ & 0.294 & 0.083 & 0.084 & 0.952 & 0.818 & 0.101 & 0.101 & 0.957 \\
$\mathrm{MC}$ & 0.427 & 0.071 & 0.072 & 0.952 & 1.236 & 0.087 & 0.087 & 0.955 \\
$\mathrm{MM}$ & 9.341 & 0.084 & 0.085 & 0.834 & 29.608 & 0.100 & 0.101 & 0.280 \\
\addlinespace[5pt]
\multicolumn{9}{@{}l}{$\tau=1.5$, $\mathrm{WO}_0=1.146$} \\
$\mathrm{CC}$ & 0.442 & 0.080 & 0.078 & 0.950 & -0.130 & 0.107 & 0.103 & 0.943 \\
$\mathrm{CM}$ & 0.777 & 0.095 & 0.095 & 0.947 & 1.310 & 0.124 & 0.120 & 0.954 \\
$\mathrm{MC}$ & 0.871 & 0.079 & 0.078 & 0.947 & 1.669 & 0.100 & 0.098 & 0.944 \\
$\mathrm{MM}$ & 16.282 & 0.097 & 0.097 & 0.683 & 53.063 & 0.122 & 0.121 & 0.063 \\
\addlinespace[5pt]
\multicolumn{9}{@{}l}{$\tau=2.0$, $\mathrm{WO}_0=1.161$} \\
$\mathrm{CC}$ & 0.559 & 0.083 & 0.080 & 0.949 & 0.078 & 0.123 & 0.112 & 0.927 \\
$\mathrm{CM}$ & 0.972 & 0.102 & 0.101 & 0.954 & 2.061 & 0.143 & 0.135 & 0.944 \\
$\mathrm{MC}$ & 1.051 & 0.082 & 0.080 & 0.950 & 2.230 & 0.109 & 0.105 & 0.943 \\
$\mathrm{MM}$ & 22.685 & 0.106 & 0.105 & 0.530 & 78.256 & 0.138 & 0.137 & 0.008 \\
\bottomrule
\end{tabular*}
\endgroup
\end{table}

\FloatBarrier

\begin{figure}[!htbp]
\centering
\includegraphics[width=0.97\textwidth,height=0.67\textheight,keepaspectratio]{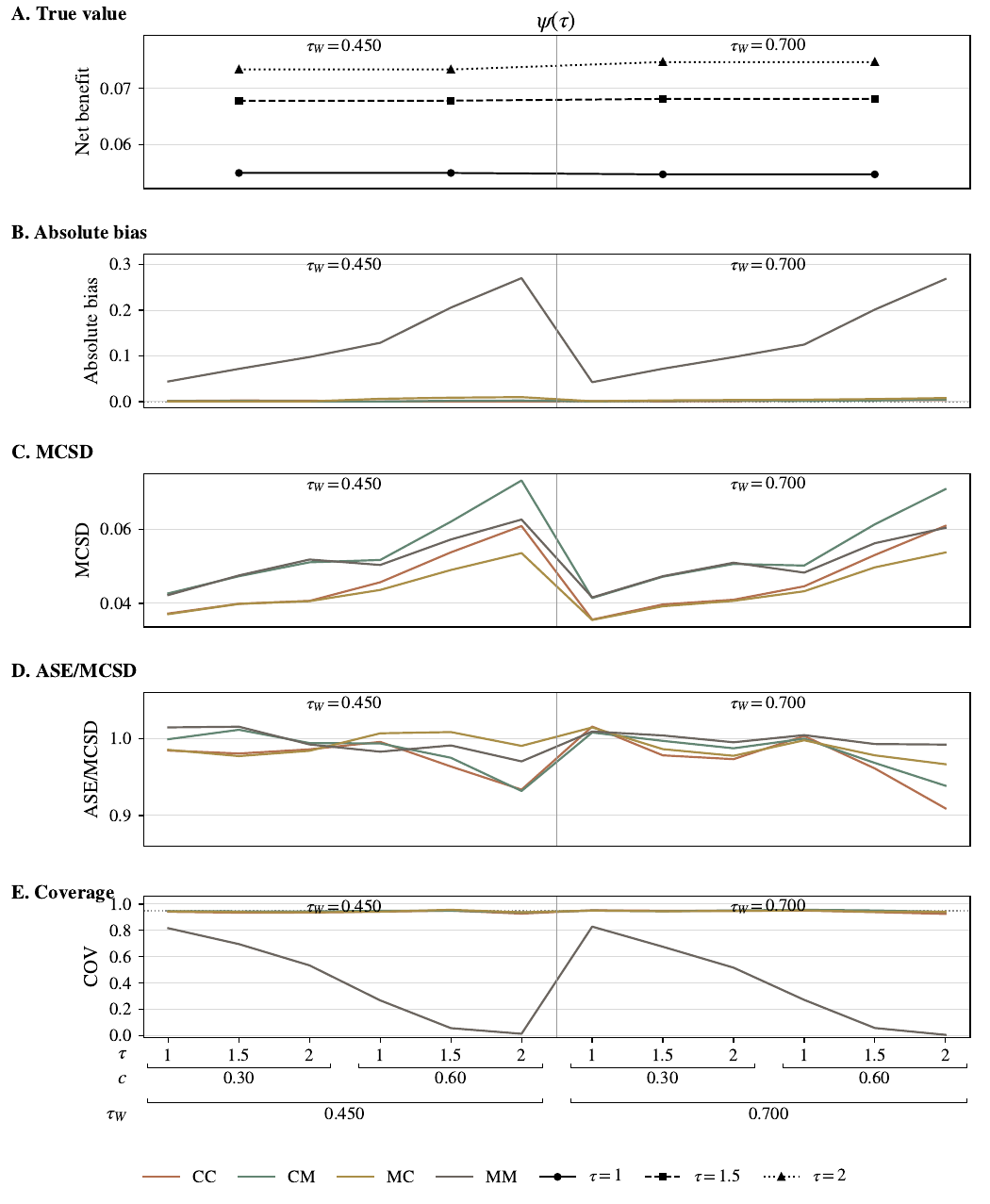}
\caption{Net benefit results for IRTs with $n=500$ and equal treatment allocation.}
\label{fig:irt-nb-all}
\end{figure}

Tables \ref{tab:irt-app-nb-low} and \ref{tab:irt-app-nb-high} report the net benefit results. Absolute bias under $\mathrm{MM}$ ranges from $0.0425$ to $0.0976$ with $30\%$ censoring and from $0.1248$ to $0.2703$ with $60\%$ censoring. Figure \ref{fig:irt-nb-all} summarizes these results.

\begin{table}[!htbp]
\centering
\caption{Doubly robust net benefit results for IRTs with $n=500$ and $\tau_W=0.450$.}
\label{tab:irt-app-nb-low}
\begingroup
\begin{tabular*}{\textwidth}{@{\extracolsep{\fill}}lrrrr@{\hspace{14pt}}rrrr@{}}
\toprule
& \multicolumn{4}{c}{$c=0.300$} & \multicolumn{4}{c}{$c=0.600$} \\
\cmidrule(lr){2-5}\cmidrule(l){6-9}
Model & AB & MCSD & ASE & COV & AB & MCSD & ASE & COV \\
\midrule
\multicolumn{9}{@{}l}{$\tau=1.0$, $\mathrm{NB}_0=0.0550$} \\
$\mathrm{CC}$ & 0.0013 & 0.037 & 0.037 & 0.942 & 0.0002 & 0.046 & 0.046 & 0.943 \\
$\mathrm{CM}$ & 0.0008 & 0.043 & 0.043 & 0.947 & 0.0001 & 0.052 & 0.051 & 0.949 \\
$\mathrm{MC}$ & 0.0003 & 0.037 & 0.037 & 0.945 & 0.0061 & 0.044 & 0.044 & 0.943 \\
$\mathrm{MM}$ & 0.0441 & 0.042 & 0.043 & 0.817 & 0.1286 & 0.050 & 0.050 & 0.268 \\
\addlinespace[5pt]
\multicolumn{9}{@{}l}{$\tau=1.5$, $\mathrm{NB}_0=0.0678$} \\
$\mathrm{CC}$ & 0.0024 & 0.040 & 0.039 & 0.935 & 0.0005 & 0.054 & 0.052 & 0.951 \\
$\mathrm{CM}$ & 0.0014 & 0.047 & 0.048 & 0.945 & 0.0018 & 0.062 & 0.061 & 0.950 \\
$\mathrm{MC}$ & 0.0002 & 0.040 & 0.039 & 0.940 & 0.0086 & 0.049 & 0.049 & 0.957 \\
$\mathrm{MM}$ & 0.0717 & 0.048 & 0.048 & 0.696 & 0.2057 & 0.057 & 0.057 & 0.057 \\
\addlinespace[5pt]
\multicolumn{9}{@{}l}{$\tau=2.0$, $\mathrm{NB}_0=0.0734$} \\
$\mathrm{CC}$ & 0.0019 & 0.041 & 0.040 & 0.936 & 0.0001 & 0.061 & 0.057 & 0.928 \\
$\mathrm{CM}$ & 0.0005 & 0.051 & 0.051 & 0.945 & 0.0023 & 0.073 & 0.068 & 0.937 \\
$\mathrm{MC}$ & 0.0005 & 0.041 & 0.040 & 0.938 & 0.0098 & 0.054 & 0.053 & 0.938 \\
$\mathrm{MM}$ & 0.0976 & 0.052 & 0.051 & 0.534 & 0.2703 & 0.063 & 0.061 & 0.014 \\
\bottomrule
\end{tabular*}
\endgroup
\end{table}

\begin{table}[!htbp]
\centering
\caption{Doubly robust net benefit results for IRTs with $n=500$ and $\tau_W=0.700$.}
\label{tab:irt-app-nb-high}
\begingroup
\begin{tabular*}{\textwidth}{@{\extracolsep{\fill}}lrrrr@{\hspace{14pt}}rrrr@{}}
\toprule
& \multicolumn{4}{c}{$c=0.300$} & \multicolumn{4}{c}{$c=0.600$} \\
\cmidrule(lr){2-5}\cmidrule(l){6-9}
Model & AB & MCSD & ASE & COV & AB & MCSD & ASE & COV \\
\midrule
\multicolumn{9}{@{}l}{$\tau=1.0$, $\mathrm{NB}_0=0.0547$} \\
$\mathrm{CC}$ & 0.0010 & 0.036 & 0.036 & 0.953 & 0.0023 & 0.045 & 0.045 & 0.951 \\
$\mathrm{CM}$ & 0.0004 & 0.041 & 0.042 & 0.951 & 0.0014 & 0.050 & 0.050 & 0.957 \\
$\mathrm{MC}$ & 0.0008 & 0.036 & 0.036 & 0.951 & 0.0041 & 0.043 & 0.043 & 0.953 \\
$\mathrm{MM}$ & 0.0425 & 0.042 & 0.042 & 0.829 & 0.1248 & 0.048 & 0.049 & 0.272 \\
\addlinespace[5pt]
\multicolumn{9}{@{}l}{$\tau=1.5$, $\mathrm{NB}_0=0.0681$} \\
$\mathrm{CC}$ & 0.0005 & 0.040 & 0.039 & 0.950 & 0.0037 & 0.053 & 0.051 & 0.939 \\
$\mathrm{CM}$ & 0.0015 & 0.047 & 0.047 & 0.946 & 0.0024 & 0.061 & 0.059 & 0.952 \\
$\mathrm{MC}$ & 0.0027 & 0.039 & 0.039 & 0.946 & 0.0055 & 0.050 & 0.049 & 0.942 \\
$\mathrm{MM}$ & 0.0719 & 0.047 & 0.048 & 0.676 & 0.2014 & 0.056 & 0.056 & 0.058 \\
\addlinespace[5pt]
\multicolumn{9}{@{}l}{$\tau=2.0$, $\mathrm{NB}_0=0.0747$} \\
$\mathrm{CC}$ & 0.0009 & 0.041 & 0.040 & 0.949 & 0.0036 & 0.061 & 0.055 & 0.926 \\
$\mathrm{CM}$ & 0.0020 & 0.051 & 0.050 & 0.952 & 0.0047 & 0.071 & 0.067 & 0.941 \\
$\mathrm{MC}$ & 0.0034 & 0.041 & 0.040 & 0.948 & 0.0078 & 0.054 & 0.052 & 0.943 \\
$\mathrm{MM}$ & 0.0974 & 0.051 & 0.051 & 0.517 & 0.2683 & 0.060 & 0.060 & 0.006 \\
\bottomrule
\end{tabular*}
\endgroup
\end{table}

\FloatBarrier
\subsection{Cluster-randomized trials}
\label{sec:app_simulation}
\label{subsec:app-crt-generation}
\label{subsec:app-crt-metrics}
\label{subsec:app-crt-results}

\begin{figure}[!htbp]
\centering
\includegraphics[width=0.95\textwidth]{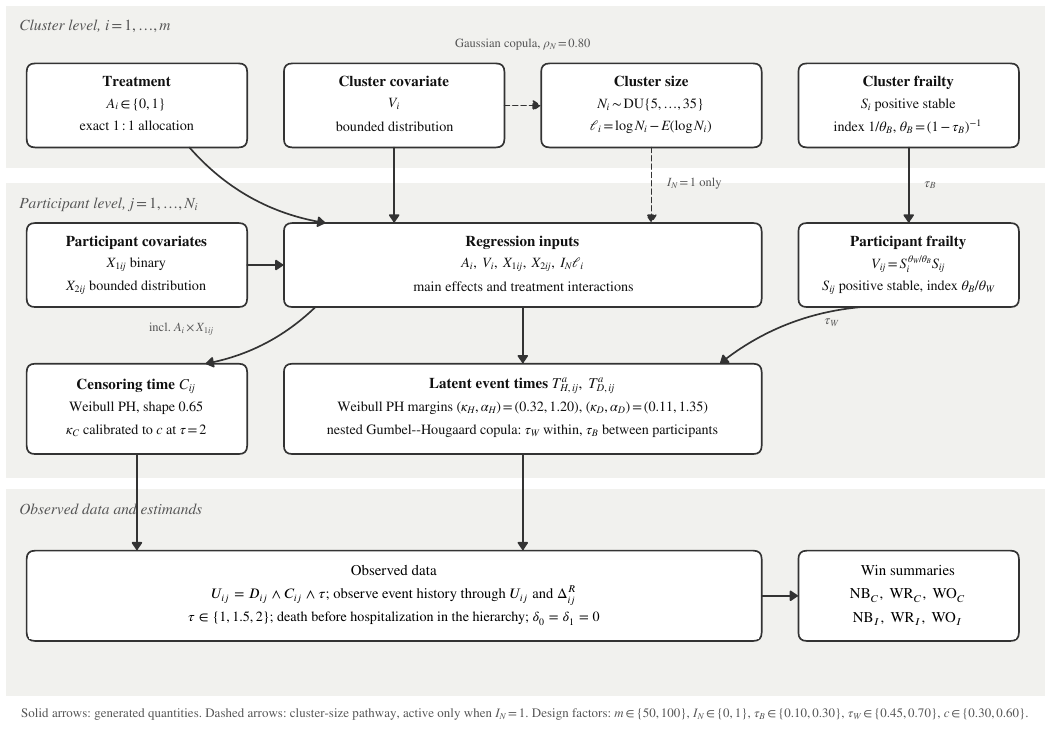}
\caption{Data-generating structure for the CRT simulations with equal treatment allocation and death prioritized over first hospitalization. Both indifference thresholds are zero.}
\label{fig:sim-dgp}
\end{figure}

Figure \ref{fig:sim-dgp} summarizes the data-generating process in Section \ref{subsec:sim-crt}. Shaded bands separate cluster-level variables, participant-level variables, and observed data and estimands. Solid arrows indicate the generation of subsequent variables; dashed arrows indicate associations present only under informative cluster size ($I_N=1$). The nested Gumbel copula determines between-participant and within-participant dependence through $\tau_B$ and $\tau_W$. The abbreviations DU and PH denote discrete uniform and proportional hazards, respectively.

Within each uncensored reference population in Section \ref{subsec:sim-crt}, treatment-control comparisons are averaged with equal cluster weights for the cluster-average target and equal participant weights for the individual-average target. The win and loss probabilities are averaged across the four reference populations before calculating the win statistics.

\begin{figure}[!htbp]
\centering
\includegraphics[width=0.99\textwidth,height=0.70\textheight,keepaspectratio]{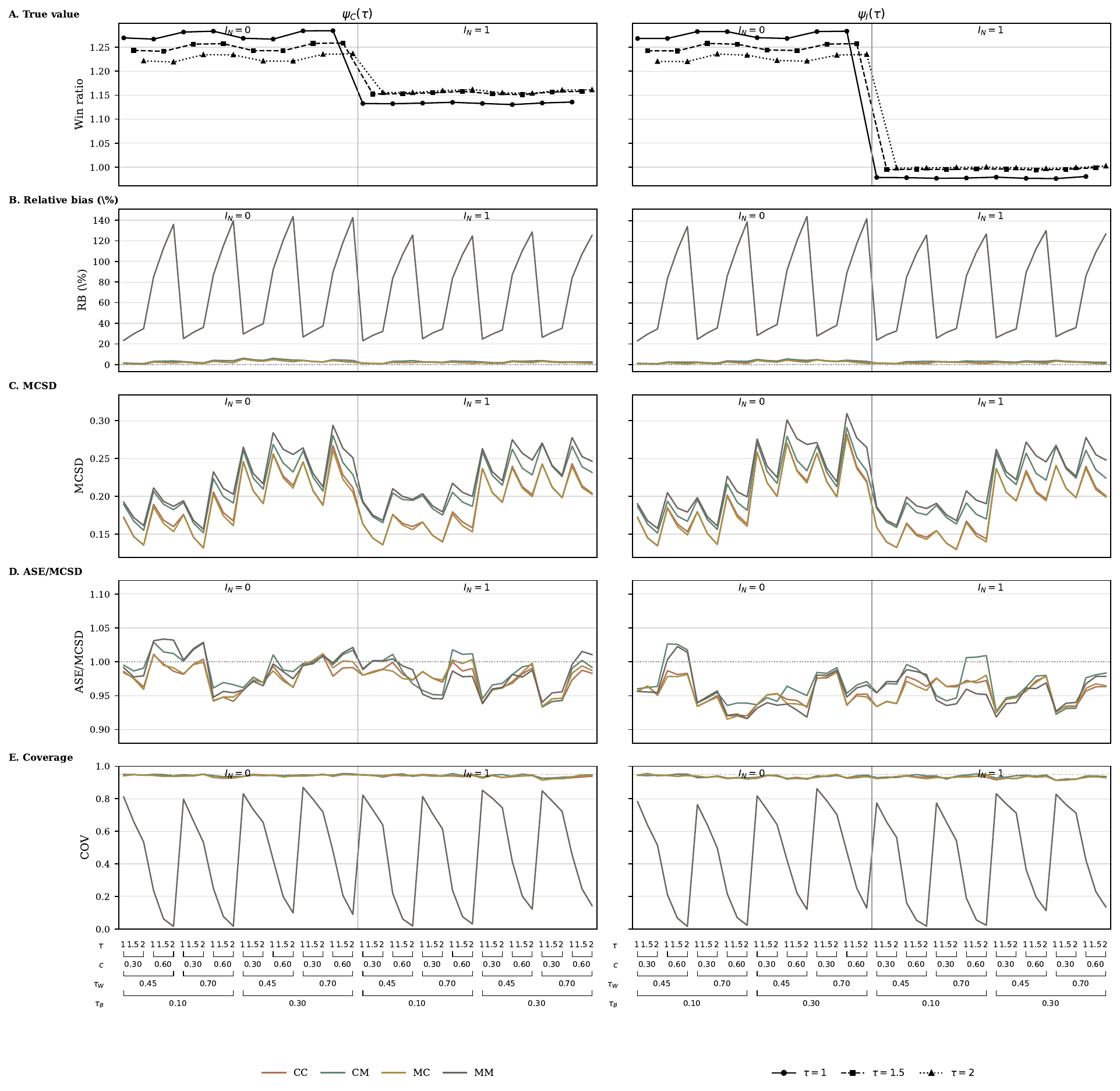}
\caption{Win ratio results for CRTs with $m=50$ and equal treatment allocation. The left and right columns show the cluster-average and individual-average targets.}
\label{fig:fig_appendix_01_wr_m50_loopstyle}
\end{figure}

\begin{figure}[!htbp]
\centering
\includegraphics[width=0.99\textwidth,height=0.70\textheight,keepaspectratio]{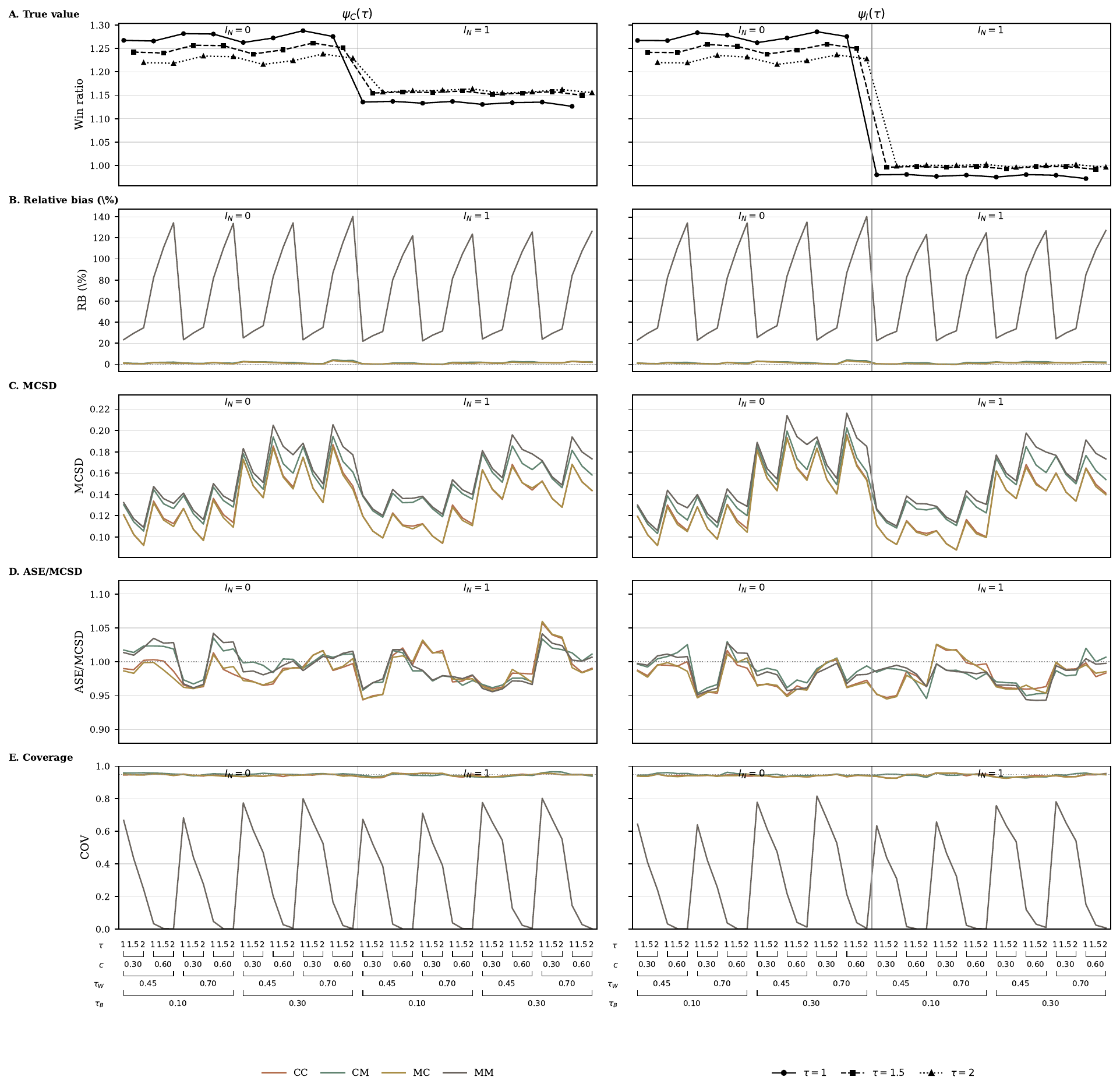}
\caption{Win ratio results for CRTs with $m=100$ and equal treatment allocation. The left and right columns show the cluster-average and individual-average targets.}
\label{fig:fig_appendix_02_wr_m100_loopstyle}
\end{figure}

\begin{figure}[!htbp]
\centering
\includegraphics[width=0.99\textwidth,height=0.70\textheight,keepaspectratio]{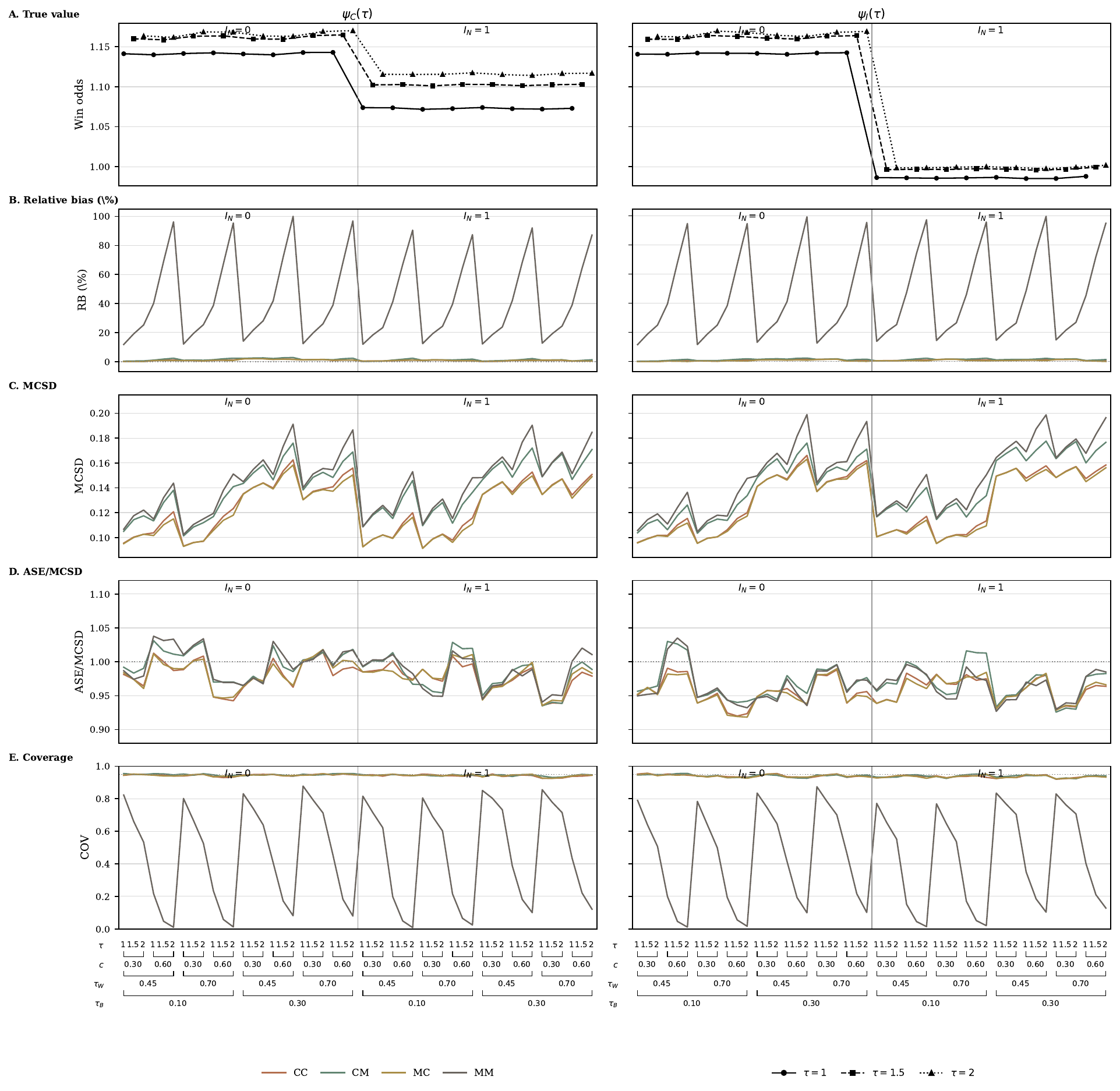}
\caption{Win odds results for CRTs with $m=50$ and equal treatment allocation. The left and right columns show the cluster-average and individual-average targets.}
\label{fig:fig_appendix_03_wo_m50_loopstyle}
\end{figure}

\begin{figure}[!htbp]
\centering
\includegraphics[width=0.99\textwidth,height=0.70\textheight,keepaspectratio]{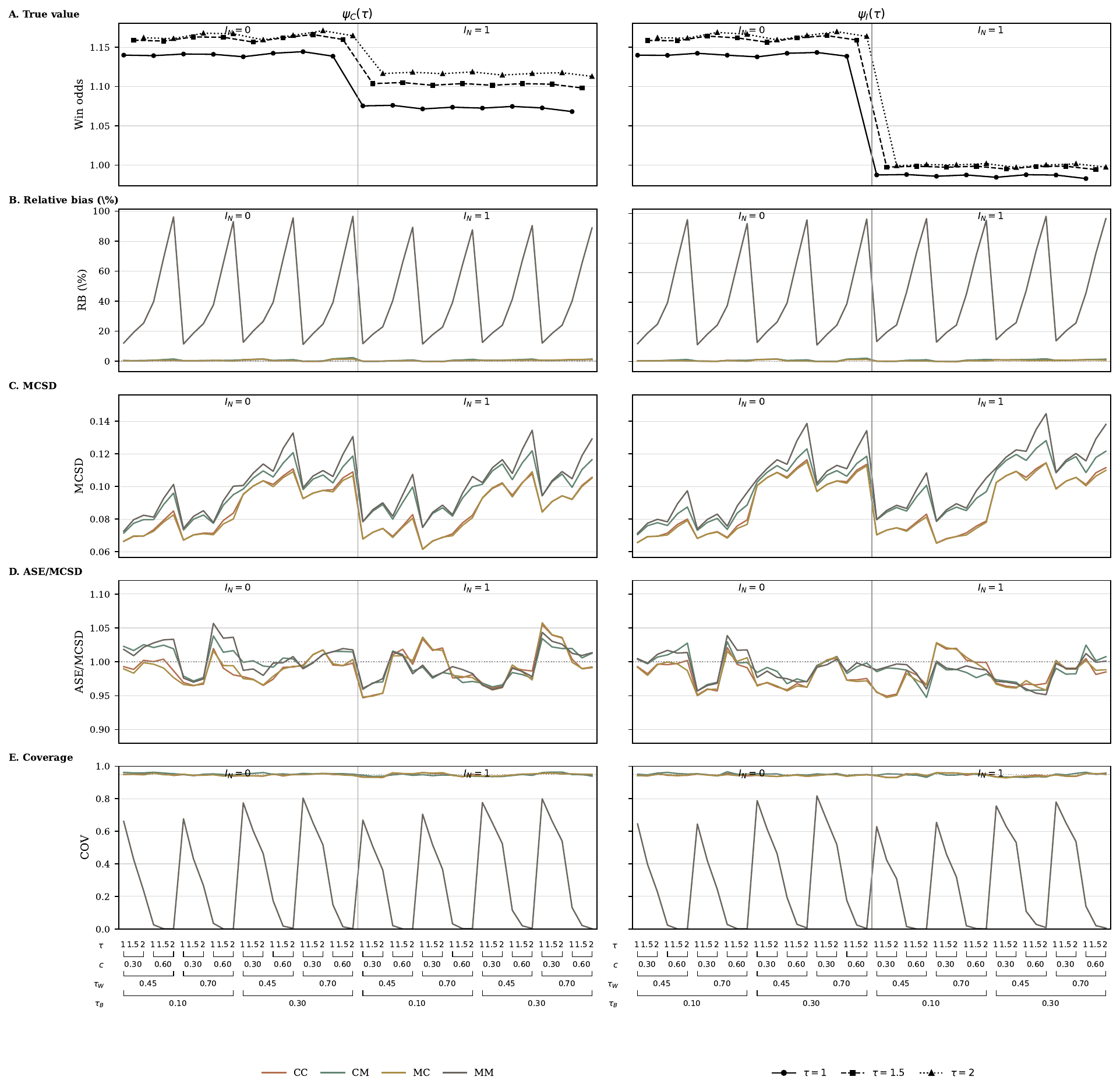}
\caption{Win odds results for CRTs with $m=100$ and equal treatment allocation. The left and right columns show the cluster-average and individual-average targets.}
\label{fig:fig_appendix_04_wo_m100_loopstyle}
\end{figure}

\begin{figure}[!htbp]
\centering
\includegraphics[width=0.99\textwidth,height=0.70\textheight,keepaspectratio]{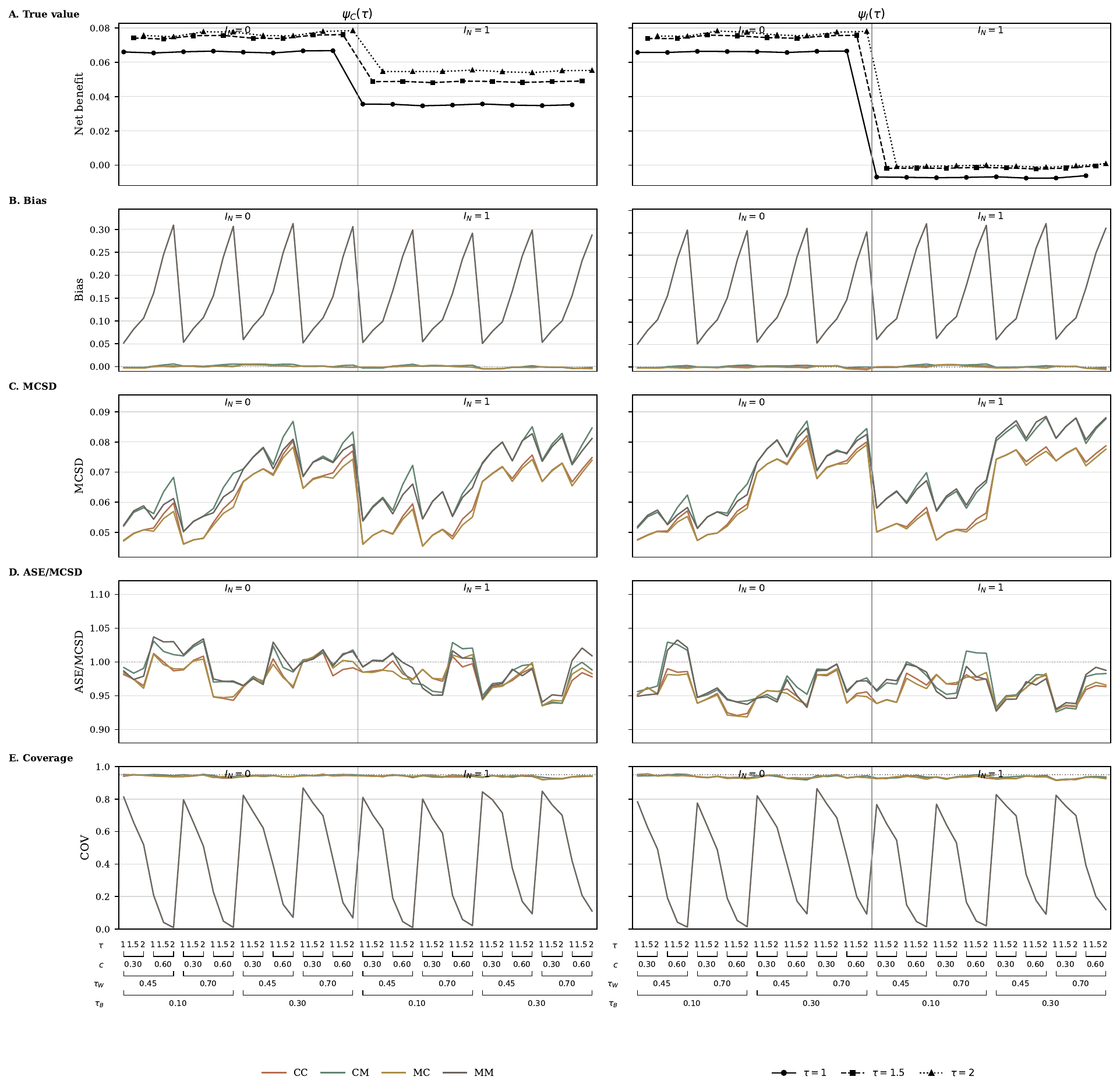}
\caption{Net benefit results for CRTs with $m=50$ and equal treatment allocation. The left and right columns show the cluster-average and individual-average targets.}
\label{fig:fig_appendix_05_nb_m50_loopstyle}
\end{figure}

\begin{figure}[!htbp]
\centering
\includegraphics[width=0.99\textwidth,height=0.70\textheight,keepaspectratio]{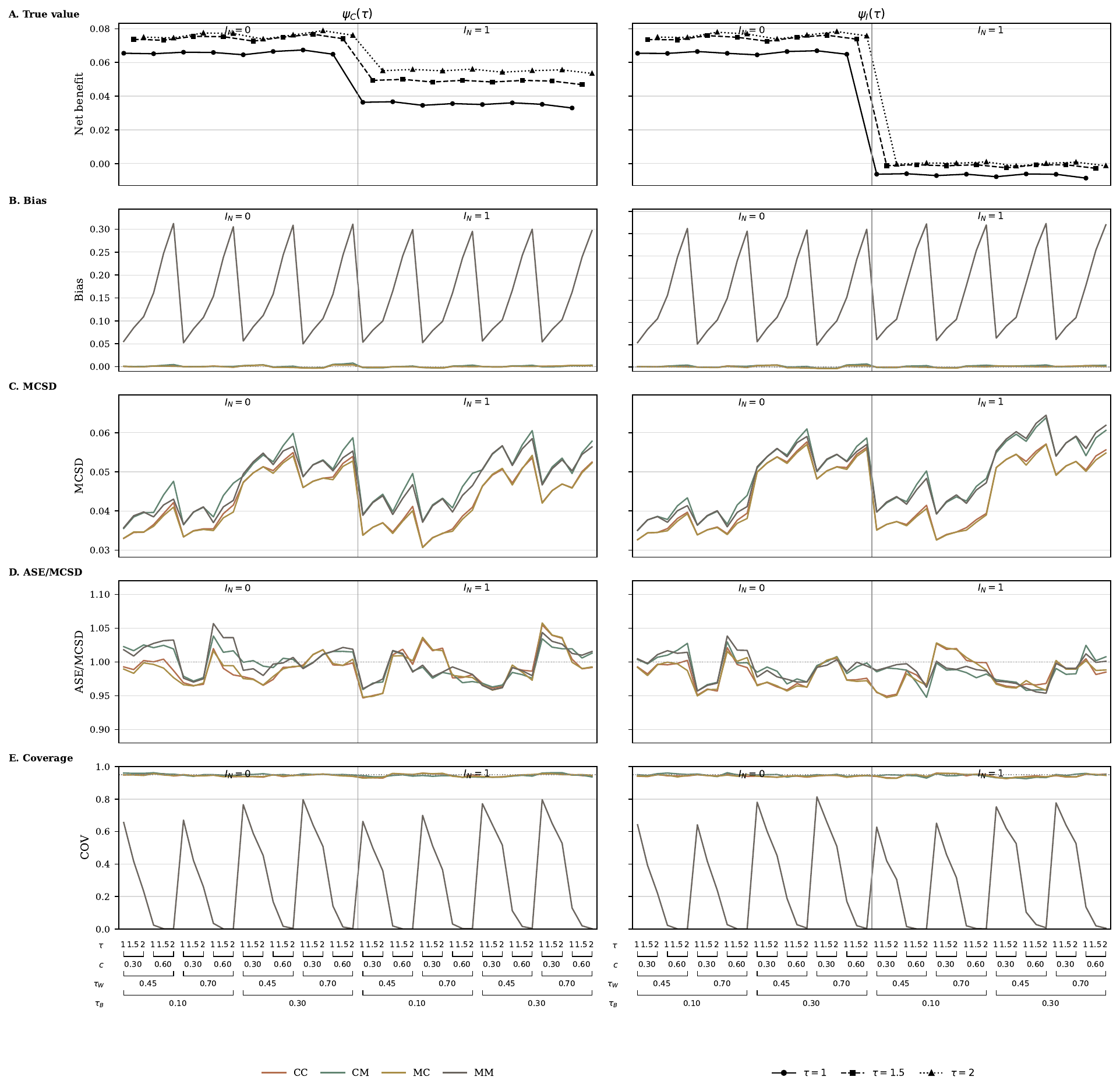}
\caption{Net benefit results for CRTs with $m=100$ and equal treatment allocation. The left and right columns show the cluster-average and individual-average targets.}
\label{fig:fig_appendix_06_nb_m100_loopstyle}
\end{figure}

Table \ref{tab:main-wr-c060} gives the win ratio results under $60\%$ censoring for the configuration in Table \ref{tab:main-wr-reference}. Figures \ref{fig:fig_appendix_01_wr_m50_loopstyle}--\ref{fig:fig_appendix_06_nb_m100_loopstyle} give all configurations. Under $\mathrm{MM}$, relative bias ranges from $21.896\%$ to $144.303\%$ for the win ratio and from $11.030\%$ to $100.089\%$ for the win odds; absolute bias ranges from $0.049$ to $0.323$ for net benefit. The largest biases occur with heavier censoring and longer restriction times.

\FloatBarrier
\begin{table}[!htbp]
\centering
\caption{Win ratio results for CRTs with $m=50$, $(\tau_B,\tau_W)=(0.100,0.450)$, and $c=0.600$.}
\label{tab:main-wr-c060}
\resizebox*{!}{0.7\textheight}{%
\begin{tabular}{@{}lrrrrrrrr@{}}
\toprule
& \multicolumn{4}{c}{$\mathrm{WR}_C(\tau)$} & \multicolumn{4}{c}{$\mathrm{WR}_I(\tau)$} \\
\cmidrule(lr){2-5}\cmidrule(l){6-9}
Model & RB (\%) & MCSD & ASE & COV & RB (\%) & MCSD & ASE & COV \\
\midrule
\multicolumn{9}{@{}l}{$I_N=0$} \\
\addlinespace[2pt]
$\tau=1.000$ & \multicolumn{4}{c}{$\mathrm{WR}_C=1.267$} & \multicolumn{4}{c}{$\mathrm{WR}_I=1.268$} \\
$\mathrm{CC}$ & 2.246 & 0.189 & 0.191 & 0.949 & 1.938 & 0.185 & 0.183 & 0.945 \\
$\mathrm{CM}$ & 3.035 & 0.206 & 0.212 & 0.950 & 2.450 & 0.194 & 0.199 & 0.944 \\
$\mathrm{MC}$ & 2.417 & 0.186 & 0.188 & 0.943 & 1.655 & 0.184 & 0.180 & 0.943 \\
$\mathrm{MM}$ & 85.039 & 0.211 & 0.218 & 0.235 & 83.996 & 0.205 & 0.206 & 0.210 \\
\addlinespace[2pt]
$\tau=1.500$ & \multicolumn{4}{c}{$\mathrm{WR}_C=1.241$} & \multicolumn{4}{c}{$\mathrm{WR}_I=1.242$} \\
$\mathrm{CC}$ & 2.097 & 0.168 & 0.168 & 0.944 & 1.677 & 0.163 & 0.160 & 0.949 \\
$\mathrm{CM}$ & 3.287 & 0.190 & 0.192 & 0.949 & 2.398 & 0.174 & 0.178 & 0.952 \\
$\mathrm{MC}$ & 1.918 & 0.164 & 0.163 & 0.937 & 1.079 & 0.160 & 0.157 & 0.937 \\
$\mathrm{MM}$ & 112.972 & 0.193 & 0.200 & 0.061 & 111.610 & 0.185 & 0.189 & 0.066 \\
\addlinespace[2pt]
$\tau=2.000$ & \multicolumn{4}{c}{$\mathrm{WR}_C=1.219$} & \multicolumn{4}{c}{$\mathrm{WR}_I=1.220$} \\
$\mathrm{CC}$ & 1.911 & 0.160 & 0.158 & 0.937 & 1.442 & 0.153 & 0.151 & 0.942 \\
$\mathrm{CM}$ & 3.481 & 0.182 & 0.185 & 0.941 & 2.445 & 0.167 & 0.170 & 0.952 \\
$\mathrm{MC}$ & 1.469 & 0.153 & 0.152 & 0.939 & 0.576 & 0.149 & 0.146 & 0.943 \\
$\mathrm{MM}$ & 136.033 & 0.187 & 0.193 & 0.015 & 134.432 & 0.179 & 0.182 & 0.015 \\
\addlinespace[5pt]
\multicolumn{9}{@{}l}{$I_N=1$} \\
\addlinespace[2pt]
$\tau=1.000$ & \multicolumn{4}{c}{$\mathrm{WR}_C=1.132$} & \multicolumn{4}{c}{$\mathrm{WR}_I=0.978$} \\
$\mathrm{CC}$ & 1.966 & 0.176 & 0.176 & 0.945 & 1.717 & 0.165 & 0.161 & 0.943 \\
$\mathrm{CM}$ & 3.091 & 0.204 & 0.207 & 0.948 & 2.768 & 0.192 & 0.191 & 0.940 \\
$\mathrm{MC}$ & 2.285 & 0.176 & 0.173 & 0.950 & 1.427 & 0.163 & 0.159 & 0.939 \\
$\mathrm{MM}$ & 83.756 & 0.210 & 0.211 & 0.217 & 85.275 & 0.199 & 0.197 & 0.160 \\
\addlinespace[2pt]
$\tau=1.500$ & \multicolumn{4}{c}{$\mathrm{WR}_C=1.153$} & \multicolumn{4}{c}{$\mathrm{WR}_I=0.996$} \\
$\mathrm{CC}$ & 1.885 & 0.164 & 0.161 & 0.942 & 1.622 & 0.150 & 0.146 & 0.935 \\
$\mathrm{CM}$ & 3.291 & 0.196 & 0.193 & 0.953 & 2.910 & 0.179 & 0.177 & 0.949 \\
$\mathrm{MC}$ & 2.074 & 0.162 & 0.158 & 0.945 & 1.084 & 0.148 & 0.143 & 0.931 \\
$\mathrm{MM}$ & 106.925 & 0.200 & 0.199 & 0.061 & 108.402 & 0.187 & 0.185 & 0.053 \\
\addlinespace[2pt]
$\tau=2.000$ & \multicolumn{4}{c}{$\mathrm{WR}_C=1.156$} & \multicolumn{4}{c}{$\mathrm{WR}_I=0.998$} \\
$\mathrm{CC}$ & 1.973 & 0.160 & 0.156 & 0.941 & 1.604 & 0.146 & 0.141 & 0.932 \\
$\mathrm{CM}$ & 3.711 & 0.195 & 0.188 & 0.938 & 3.178 & 0.176 & 0.172 & 0.941 \\
$\mathrm{MC}$ & 1.974 & 0.156 & 0.152 & 0.943 & 0.757 & 0.143 & 0.137 & 0.923 \\
$\mathrm{MM}$ & 125.505 & 0.196 & 0.194 & 0.018 & 126.160 & 0.184 & 0.181 & 0.017 \\
\bottomrule
\end{tabular}%
}
\end{table}
\FloatBarrier

Informative cluster size also affects net benefit. With $(\tau_B,\tau_W)=(0.100,0.450)$ and $\tau=2$, the reference cluster-average and individual-average net benefits are $0.0546$ and $-0.0008$. The treatment-win probability therefore exceeds the control-win probability by approximately $5.46$ percentage points under equal cluster weighting, whereas the probabilities are approximately equal under equal participant weighting. This reflects the association of cluster size with prognosis and treatment effect in the generating model.

The left and right columns of the CRT figures show the cluster-average and individual-average targets. Their horizontal axes group noninformative ($I_N=0$) and informative ($I_N=1$) cluster size, between-participant dependence $\tau_B$, within-participant dependence $\tau_W$, censoring probability $c$, and restriction time $\tau$.

\FloatBarrier
\section{Illustrative data applications}
\label{sec:app-applications}

This section gives event-history reconstruction, supplementary results, and sensitivity analyses for Section \ref{sec:application}. The working models are collected in Appendix \ref{sec:app-working-models}. All displayed confidence intervals are pointwise. Net benefit and its intervals are expressed in percentage points; intervals for win ratios and win odds are calculated on the logarithmic scale and exponentiated.

\subsection{HF-ACTION}
\label{sec:app-hf-action}
\label{subsec:hf-reconstruction}
\label{subsec:hf-results-details}

The analysis uses the version 2015a noncommercial data release described in Section \ref{subsec:hf-action}.\cite{biolinccHFAction} Death times were reconciled across the analysis dataset, clinical events committee records, and deaths confirmed after final study contact. First HF hospitalization was identified from committee records and the nonfatal component of the HF composite in the release. A composite event was classified as hospitalization only when it preceded the reconciled death time and occurred within the supported follow-up. Worsening HF without hospital admission was excluded. This reconstruction produced 530 first HF hospitalizations through four years.

HF hospitalization was treated as a nonterminal event and death as the terminal event. Before the first HF hospitalization, follow-up was subject to the earlier of the hospitalization and vital status ascertainment limits. After hospitalization, follow-up for death continued until death or the end of vital status ascertainment. The resulting histories contain 349 deaths through four years. Participants without observed follow-up were retained as censored at enrollment. Table \ref{tab:hf-descriptive} gives baseline characteristics and observed outcomes before imputation. Continuous variables are reported as means and standard deviations (SD), and percentages use the randomized-arm total. Estimated glomerular filtration rate (GFR) is missing for 218 participants and peak oxygen uptake for 52. NYHA, KCCQ, and US denote New York Heart Association functional class, Kansas City Cardiomyopathy Questionnaire, and United States. Etiology uses release code 2.

Event counts are strictly before two years. Hospitalization person-time ends at first HF hospitalization, death, observation end, or two years; mortality person-time continues after hospitalization. Frequencies and rates are descriptive, and unresolved censoring is not a disjoint event category.

\begin{table}[!htbp]
\centering
\caption{Baseline characteristics and recorded two-year outcomes in the HF-ACTION analysis cohort: 1,070 usual care and 1,060 exercise participants.}
\label{tab:hf-descriptive}
\begin{tabular}{p{0.62\textwidth}cc}
\toprule
Characteristic & Usual care & Exercise\\
\midrule
\multicolumn{3}{l}{\textit{Baseline characteristics}}\\
Participants & 1,070 & 1,060\\
Age, years & 58.5 (13.0) & 58.6 (12.4)\\
Left ventricular ejection fraction, \% & 25.2 (7.4) & 25.2 (7.6)\\
Systolic blood pressure, mm Hg & 114.0 (18.6) & 113.6 (17.9)\\
Estimated GFR, mL/min/1.73 m$^2$ & 67.3 (24.2) & 68.2 (24.7)\\
Peak oxygen uptake, mL/kg/min & 15.1 (4.8) & 14.8 (4.5)\\
KCCQ overall score & 66.5 (21.1) & 66.2 (19.9)\\
Recorded female sex, $n$ (\%) & 281 (26.3) & 318 (30.0)\\
NYHA class III or IV, $n$ (\%) & 379 (35.4) & 400 (37.7)\\
Etiology (release code 2), $n$ (\%) & 524 (49.0) & 510 (48.1)\\
Diabetes, $n$ (\%) & 335 (31.3) & 343 (32.4)\\
Atrial fibrillation or flutter, $n$ (\%) & 219 (20.5) & 224 (21.1)\\
Prior HF hospitalization within 6 months, $n$ (\%) & 295 (27.6) & 269 (25.4)\\
Non-US participant, $n$ (\%) & 124 (11.6) & 126 (11.9)\\
\addlinespace
\multicolumn{3}{l}{\textit{Follow-up and observed outcomes before two years}}\\
Recorded observation time, years (median) & 2.52 & 2.54\\
Recorded all-cause deaths, $n$ (\%) & 127 (11.9) & 103 (9.7)\\
First HF hospitalization, $n$ (\%) & 221 (20.7) & 204 (19.2)\\
Mortality rate per 100 person-years & 6.82 & 5.51\\
First HF hospitalization rate per 100 person-years & 13.23 & 12.10\\
Observation ends before outcome resolution, $n$ (\%) & 247 (23.1) & 268 (25.3)\\
Alive and observed at two years, $n$ (\%) & 696 (65.0) & 689 (65.0)\\
\bottomrule
\end{tabular}
\end{table}

Inference uses the participant influence contributions in Section \ref{subsec:irt-theory}, centered within treatment arms and including nuisance estimation. The IPCW, outcome regression, and unadjusted estimates in the main paper use the same restricted outcomes and clinical comparison.

Table \ref{tab:hf-probabilities} gives the estimated win and loss probabilities and their residual at two years. For the doubly robust, IPCW, and outcome regression estimators, the residual estimates the tie probability under the corresponding assumptions. For the complete-resolution comparator, it does not represent a censoring-adjusted tie probability. These entries describe pairwise comparisons, not participant event risks, and sum to one before rounding.

\begin{table}[!htbp]
\centering
\caption{Two-year win and loss probabilities and residual $1-\widehat\lambda^+-\widehat\lambda^-$ in HF-ACTION, comparing exercise training with usual care.}
\label{tab:hf-probabilities}
\begin{tabular*}{\textwidth}{@{\extracolsep{\fill}}lrrr@{}}
\toprule
Estimator & $\widehat\lambda^+$ & $\widehat\lambda^-$ & Residual\\
\midrule
Doubly robust & 0.2539 & 0.2192 & 0.5268\\
IPCW & 0.2518 & 0.2150 & 0.5331\\
Outcome regression & 0.2561 & 0.2244 & 0.5195\\
Complete-resolution comparator & 0.1537 & 0.1325 & 0.7138\\
\bottomrule
\end{tabular*}
\end{table}

\begin{figure}[!htbp]
\centering
\includegraphics[width=0.88\textwidth]{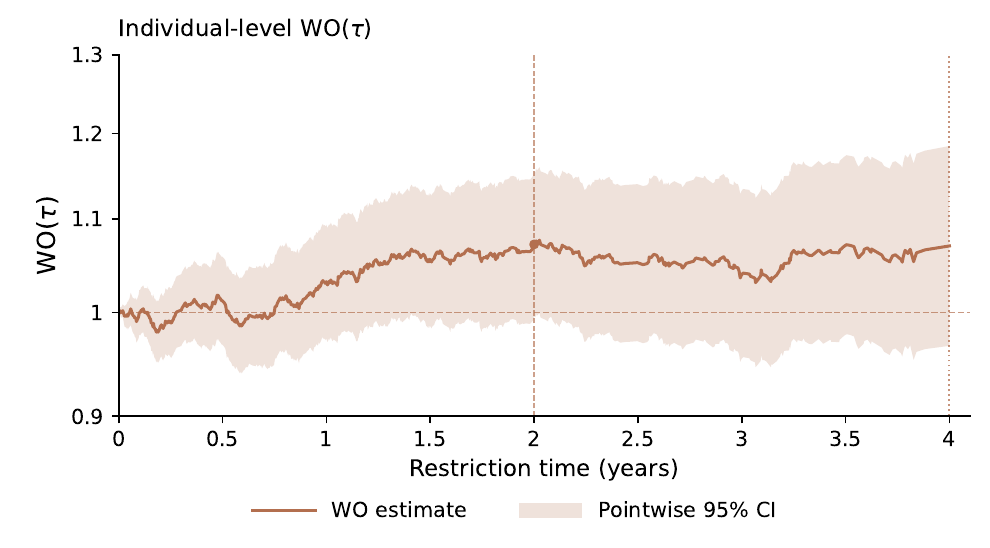}
\caption{Doubly robust win odds trajectory for HF-ACTION with death prioritized over first HF hospitalization among 2,130 participants. The solid curve is the estimate, and the shaded region gives pointwise 95\% confidence intervals.}
\label{fig:hf-wo}
\end{figure}

\begin{figure}[!htbp]
\centering
\includegraphics[width=0.88\textwidth]{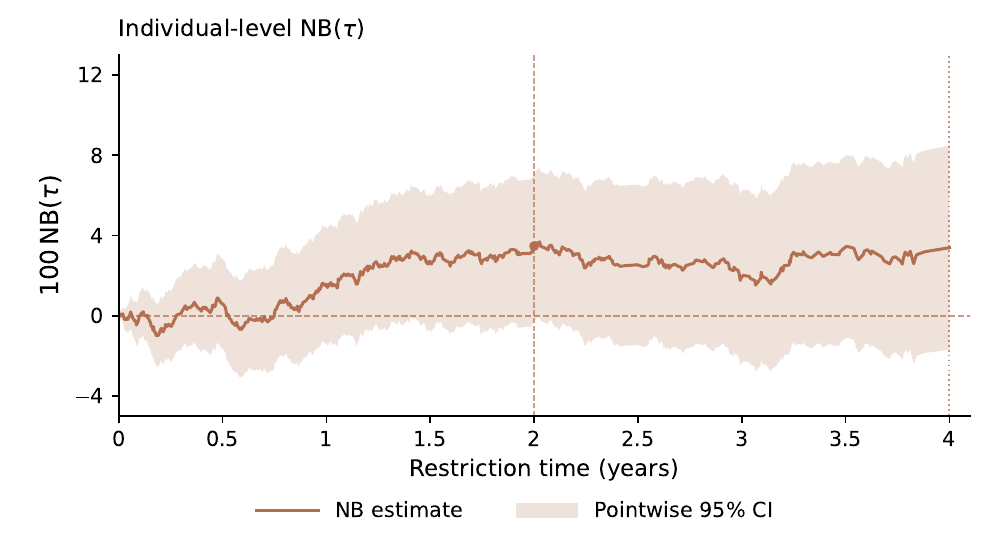}
\caption{Doubly robust net benefit trajectory for HF-ACTION with death prioritized over first HF hospitalization. Estimates and pointwise 95\% confidence intervals are expressed in percentage points.}
\label{fig:hf-nb}
\end{figure}

Figures \ref{fig:hf-wo} and \ref{fig:hf-nb} show the supplementary trajectories. At four years, the win odds is 1.070 (95\% CI 0.967 to 1.185), and net benefit is 3.39 percentage points (95\% CI $-1.69$ to 8.47). The point estimates favor exercise training over intermediate and later follow-up, with uncertainty compatible with the null comparison. Follow-up becomes less complete near four years, so the principal interpretation concerns the prespecified two-year comparison. At the earliest restriction times, a logarithmic win ratio requires positive estimated win and loss probabilities; win odds and net benefit remain defined when one is zero. The dashed and dotted vertical lines mark two and four years. Values above one for win odds and above zero for net benefit favor exercise training.

\FloatBarrier

\FloatBarrier
\subsection{STRIDE}
\label{sec:app-stride}
\label{subsec:stride-probabilities}
\label{subsec:stride-sensitivity}

Table \ref{tab:stride-descriptive} reports characteristics and recorded outcomes for 5,451 participants from 86 practices in 10 health care systems. Counts refer to practices or participants as indicated, and percentages use the corresponding randomized-arm total. Practice-size tertiles are the supplied prerandomization categories. Event counts are strictly before two years. Adjudicated injury denotes serious fall injury; the self-reported endpoint includes the broader category of reported fall injury.

First-injury person-time ends at first injury, death, observation end, or two years. Mortality follow-up continues after injury, so injury and death counts can overlap. Frequencies and rates are descriptive, and unresolved censoring is not a disjoint event category.

\begin{table}[!htbp]
\centering
\caption{Practice and participant characteristics and recorded two-year outcomes in STRIDE.}
\label{tab:stride-descriptive}
\resizebox{0.8\textwidth}{!}{%
\begin{tabular}{lcc}
\toprule
Characteristic & Enhanced usual care & Intervention\\
\midrule
\multicolumn{3}{l}{\textit{Practice characteristics}}\\
Practices & 43 & 43\\
Enrolled participants per practice, mean (SD) & 61.6 (27.4) & 65.2 (37.6)\\
Enrolled participants per practice, range & 10 to 129 & 10 to 199\\
Urban practices, n (\%) & 39 (90.7) & 39 (90.7)\\
Predominantly White practices, n (\%) & 35 (81.4) & 35 (81.4)\\
Practice-size tertile 1, n (\%) & 14 (32.6) & 14 (32.6)\\
Practice-size tertile 2, n (\%) & 15 (34.9) & 14 (32.6)\\
Practice-size tertile 3, n (\%) & 14 (32.6) & 15 (34.9)\\
\addlinespace
\multicolumn{3}{l}{\textit{Participant characteristics}}\\
Participants & 2,649 & 2,802\\
Age in years, mean (SD) & 79.5 (5.8) & 79.9 (5.7)\\
Recorded female sex, n (\%) & 1,629 (61.5) & 1,752 (62.5)\\
Recorded White race, n (\%) & 2,394 (90.4) & 2,571 (91.8)\\
Unknown or missing race category, n (\%) & 18 (0.7) & 15 (0.5)\\
Surrogate consent, n (\%) & 76 (2.9) & 86 (3.1)\\
Chronic conditions, mean (SD) & 2.14 (1.29) & 2.08 (1.34)\\
Recorded follow-up in years, median & 2.35 & 2.35\\
\addlinespace
\multicolumn{3}{l}{\textit{Observed outcomes before two years}}\\
Recorded deaths, n (\%) & 72 (2.7) & 75 (2.7)\\
Recorded death rate per 100 person-years & 1.49 & 1.49\\
First adjudicated serious injury, n (\%) & 248 (9.4) & 215 (7.7)\\
Adjudicated injury rate per 100 person-years & 5.36 & 4.44\\
First self-reported injury, n (\%) & 1,090 (41.1) & 1,004 (35.8)\\
Self-reported injury rate per 100 person-years & 29.50 & 25.02\\
Observation ends before resolution, n (\%) & 572 (21.6) & 600 (21.4)\\
Observed alive at two years, n (\%) & 2,005 (75.7) & 2,127 (75.9)\\
\bottomrule
\end{tabular}}
\end{table}

Tables \ref{tab:stride-probabilities} and \ref{tab:stride-wo-nb} report the two-year probabilities and win summaries. Each row of estimated probabilities sums to $100\%$ before rounding. These are adjusted pairwise probabilities; unresolved comparisons are not classified as ties. With death prioritized over first self-reported fall injury, the cluster-average and individual-average win odds are 1.089 (95\% CI 1.023 to 1.159) and 1.103 (95\% CI 1.047 to 1.162). The corresponding net benefits are 4.26 (95\% CI 1.15 to 7.37) and 4.90 (95\% CI 2.31 to 7.48) percentage points. Both summaries favor intervention at two years under each target population.

\begin{table}[!htbp]
\centering
\caption{Doubly robust win, loss, and tie probabilities at two years in STRIDE, expressed as percentages, with death prioritized over first adjudicated serious fall injury or first self-reported fall injury.}
\label{tab:stride-probabilities}
\resizebox{0.8\textwidth}{!}{%
\begin{tabular}{lcccc}
\toprule
Lower-priority endpoint & Target population & Win (\%) & Loss (\%) & Tie (\%)\\
\midrule
Adjudicated serious fall injury & Cluster-average ($C$) & 12.00 & 10.82 & 77.18\\
 & Individual-average ($I$) & 12.17 & 10.57 & 77.26\\
\addlinespace
Self-reported fall injury & Cluster-average ($C$) & 36.14 & 31.87 & 31.99\\
 & Individual-average ($I$) & 36.59 & 31.69 & 31.72\\
\bottomrule
\end{tabular}}
\end{table}

\begin{table}[!htbp]
\centering
\caption{Two-year win odds and net benefit in STRIDE under the Gumbel outcome model. Parentheses give 95\% confidence intervals; net benefit is expressed in percentage points.}
\label{tab:stride-wo-nb}
\resizebox{0.8\textwidth}{!}{%
\begin{tabular}{lccc}
\toprule
Lower-priority endpoint & Target population & Win odds (95\% CI) & Net benefit (95\% CI)\\
\midrule
Adjudicated serious fall injury & Cluster-average ($C$) & 1.024 (0.985, 1.064) & 1.17 (-0.75, 3.10)\\
 & Individual-average ($I$) & 1.032 (0.999, 1.067) & 1.60 (-0.04, 3.24)\\
\addlinespace
Self-reported fall injury & Cluster-average ($C$) & 1.089 (1.023, 1.159) & 4.26 (1.15, 7.37)\\
 & Individual-average ($I$) & 1.103 (1.047, 1.162) & 4.90 (2.31, 7.48)\\
\bottomrule
\end{tabular}}
\end{table}

\begin{figure}[!htbp]
\centering
\includegraphics[width=\textwidth]{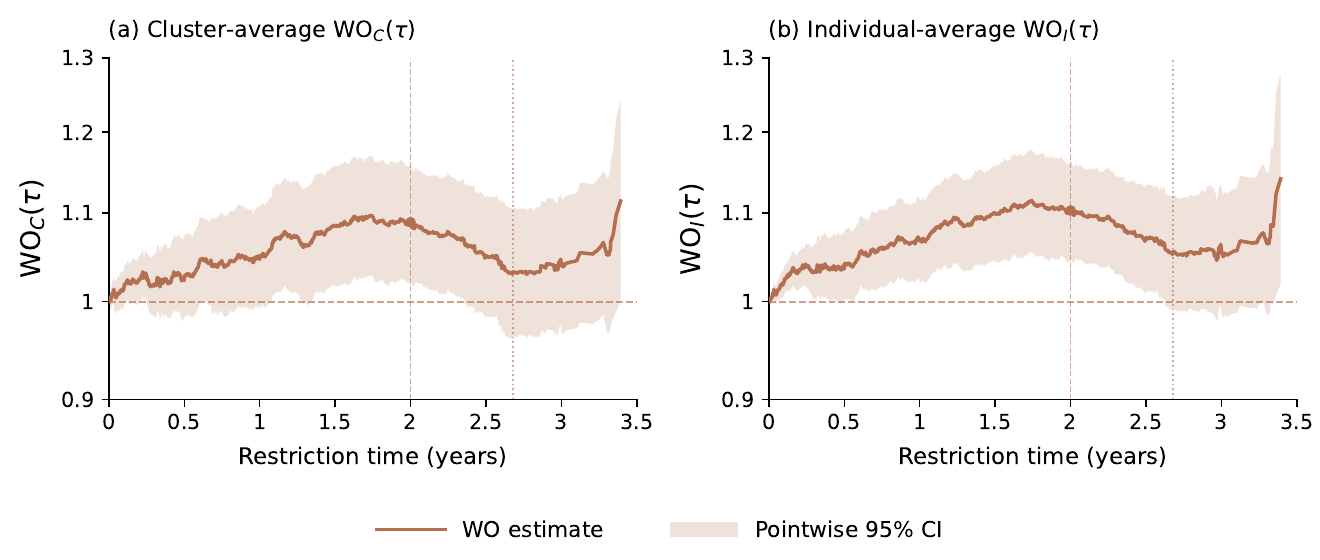}
\caption{Doubly robust win odds trajectories for STRIDE with death prioritized over first self-reported fall injury. Panels (a) and (b) show the cluster-average and individual-average targets; shaded regions give pointwise 95\% confidence intervals.}
\label{fig:stride-wo-self-reported}
\end{figure}

\begin{figure}[!htbp]
\centering
\includegraphics[width=\textwidth]{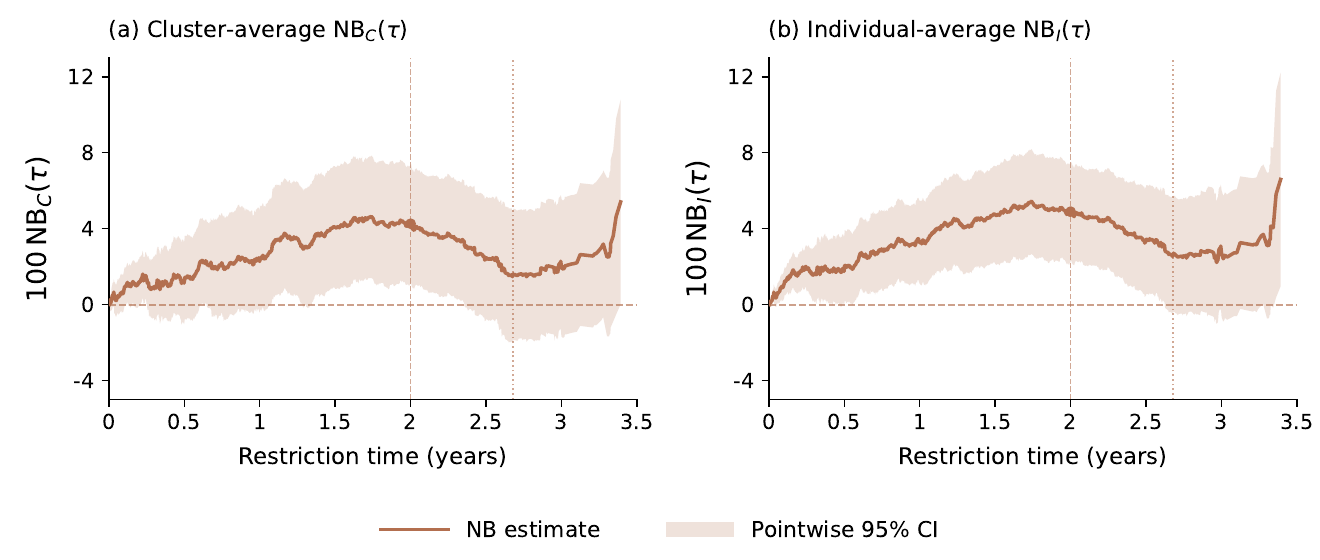}
\caption{Doubly robust net benefit trajectories for STRIDE with death prioritized over first self-reported fall injury. Panels (a) and (b) show the cluster-average and individual-average targets. Estimates and pointwise 95\% confidence intervals are expressed in percentage points.}
\label{fig:stride-nb-self-reported}
\end{figure}

Figures \ref{fig:stride-wo-self-reported} and \ref{fig:stride-nb-self-reported} show the win odds and net benefit trajectories. The principal covariance is centered within health care system and treatment arm, with standard normal critical values. The dashed vertical line marks two years. The dotted line marks 2.678 years, when the minimum fitted probability of remaining uncensored falls below $0.05$. Follow-up is less complete beyond approximately 2.68 years; interpretation therefore emphasizes the prespecified two-year comparisons. Both indifference thresholds are zero, and values above one for win odds and above zero for net benefit favor intervention over enhanced usual care.

\FloatBarrier

Under the independence outcome model in Appendix \ref{subsec:stride-models}, the two-year cluster-average and individual-average win ratios are 1.134 (95\% CI 1.033 to 1.244) and 1.154 (95\% CI 1.069 to 1.247). With the Gumbel model and Student $t$ critical values with 66 degrees of freedom, the win ratio intervals are 1.031 to 1.246 and 1.068 to 1.248. With covariance centered within treatment arms and normal critical values, the intervals are 1.038 to 1.239 and 1.073 to 1.242. Each interval excludes one and favors intervention at two years.

\FloatBarrier
\section{Using the DRwinStats package}
\label{sec:app-package}

The \texttt{DRwinStats} package implements the proposed estimators for IRTs and CRTs through the function \texttt{drwins}. The same function fits the outcome and censoring models, estimates the win, loss, and tie probabilities and the corresponding win statistics, and calculates analytic standard errors. We illustrate its use with the simulated datasets \texttt{example\_irt}, containing 500 participants, and \texttt{example\_crt}, containing 334 participants in 30 clusters. Both datasets prioritize death over first hospitalization, with follow-up measured in years.

\subsection{Installation and data format}

The package can be installed from GitHub and the example datasets loaded as follows.
\begin{verbatim}
install.packages("remotes")
remotes::install_github("fancy575/DRwinStats")
library(DRwinStats)
data(example_irt, package = "DRwinStats")
data(example_crt, package = "DRwinStats")
head(example_irt)
\end{verbatim}

The input is a data frame of longitudinal event records. The columns \texttt{id}, \texttt{treatment}, \texttt{time}, and \texttt{status} identify the participant, randomized treatment, observation time, and event type. Code 1 denotes the terminal event, code 2 denotes the next outcome in the clinical hierarchy, and subsequent positive codes identify additional nonterminal outcomes. Code 0 records the end of observation without the terminal event. Each participant has exactly one final record with code 0 or 1, together with any preceding nonterminal event records. For example, hospitalization followed by death is represented by a code-2 record followed by a code-1 record. Baseline covariates and treatment assignment are repeated across the participant's records. CRT data additionally contain a cluster identifier, and all participants in a cluster have the same assignment.

The interpretation of code 0 depends on the restriction time. An observation ending before $\tau$ indicates censoring before resolution, whereas an observation ending at or after $\tau$ provides complete follow-up for the restricted outcome. An observed death before $\tau$ also resolves the restricted outcome. The reported censoring rate therefore refers to censoring before each specified restriction time. For a nonterminal outcome with repeated records, the default \texttt{recurrent = "first"} uses its first occurrence.

\subsection{Individual-randomized trials}

The following analysis uses the Gumbel outcome model and Cox censoring model described in Appendix \ref{sec:app-working-models}. The formulas specify the baseline covariates, and models are fitted separately in each treatment arm. The argument \texttt{treatment\_levels = c(0, 1)} specifies treatment 1 versus control 0, and \texttt{priorities} specifies death before hospitalization. The default indifference thresholds are zero. Because the example trial assigns a fixed number of participants to each arm, \texttt{allocation = "complete"} is used.
\begin{verbatim}
fit_irt <- drwins(
  data = example_irt,
  outcome = ~ age + sex + x,
  censoring = ~ age + sex + x,
  id = "id", time = "time", status = "status",
  treatment = "treatment", treatment_levels = c(0, 1),
  priorities = c(Death = 1L, Hospitalization = 2L),
  outcome_model = "cox", censoring_model = "cox",
  copula = "gumbel", times = c(1, 1.5, 2),
  allocation = "complete"
)

irt_summary <- summary(fit_irt)
plot(fit_irt, measure = "WR")
\end{verbatim}

Omitting \texttt{cluster} treats each participant as an independent randomization unit. The fitted object contains the results at all requested restriction times. The \texttt{summary} method reports the estimates, standard errors, confidence intervals, and censoring percentages by treatment arm and overall. The returned \texttt{censoring} table stores these rates as fractions. The \texttt{plot} method displays estimates and pointwise confidence intervals; \texttt{measure = "WO"} or \texttt{measure = "NB"} selects the win odds or net benefit. The win ratio and win odds intervals are calculated on the logarithmic scale and exponentiated. If \texttt{times} is omitted, the function uses the 25th, 50th, and 75th percentiles of the final observed follow-up times, taking one value per participant. Duplicate times are removed, and the times used are stored in \texttt{fit\_irt\$times}. For a prespecified trial analysis, the intended restriction times should be supplied explicitly. Under independent randomization, use \texttt{allocation = "independent"} and specify the known treatment probability through \texttt{treatment\_probability}.

\subsection{Cluster-randomized trials}

For a CRT, \texttt{cluster} identifies the randomization unit, and \texttt{target = "both"} requests the cluster-average and individual-average estimands. The formulas may contain participant-level and cluster-level baseline covariates. In the following example, cluster size is explicitly included in the outcome model.
\begin{verbatim}
fit_crt <- drwins(
  data = example_crt,
  outcome = ~ age + sex + x + cluster_cov + log(cluster_size),
  censoring = ~ age + sex + x + cluster_cov,
  id = "id", time = "time", status = "status",
  treatment = "treatment", treatment_levels = c(0, 1),
  cluster = "cluster",
  priorities = c(Death = 1L, Hospitalization = 2L),
  outcome_model = "cox", censoring_model = "cox",
  copula = "gumbel", times = c(1, 1.5, 2),
  target = "both", allocation = "complete"
)

crt_summary <- summary(fit_crt, measure = c("WR", "WO", "NB"))
plot(fit_crt, measure = "WR")
\end{verbatim}

The two estimands are displayed together by default. The arguments \texttt{target = "cluster"} and \texttt{target = "individual"} select either target in \texttt{summary} or \texttt{plot}. Variance estimation uses the cluster as the independent unit and includes estimation of the working models and, for the individual-average probabilities, mean cluster size. At each restriction time, the individual-average censoring rate is the proportion of participants censored before resolution, and the cluster-average rate is the average of the within-cluster proportions. These summaries use the same population weights as their corresponding estimands.

\end{document}